\documentstyle{article}

\def\noheaderplainsetup{%

\topmargin=0pt \headheight=0pt \headsep=0pt  \oddsidemargin=0pt \evensidemargin=0pt  \textheight=8.9truein \textwidth=6.5truein}

\noheaderplainsetup

\begin{document}


\newcommand{\chess}{\mbox{\em Chess}}
\newcommand{\checkers}{\mbox{\em Checkers}}
\newcommand{\elzz}[2]{ \langle #1\rangle\hspace{-2pt} \downarrow\hspace{-2pt} #2} 
\newcommand{\elz}[1]{\mbox{$\parallel\hspace{-3pt} #1 \hspace{-3pt}\parallel$}} 
\newcommand{\propel}{\mbox{\bf CL1}}
\newcommand{\propell}{\mbox{\bf CL2}}
\newcommand{\propellc}{\mbox{\bf CL2}^\circ}
\newcommand{\propseqc}{\mbox{\bf CL9}^\circ}
\newcommand{\propseq}{\mbox{\bf CL9}}
\newcommand{\predseq}{\mbox{\bf CL11}}
\newcommand{\propseqe}{\mbox{\bf CL10}}
\newcommand{\propseqec}{\mbox{\bf CL10}^\circ}
\newcommand{\propseqecc}{\overline{\mbox{\bf CL10}^\circ}}
\newcommand{\propseqecn}{\overline{\mbox{\bf CL10}^\circ}}
\newcommand{\emptyrun}{\langle\rangle} 
\newcommand{\oo}{\bot}            
\newcommand{\pp}{\top}            
\newcommand{\xx}{\wp}               
\newcommand{\legal}[2]{\mbox{\bf Lr}^{#1}_{#2}} 
\newcommand{\Legal}[1]{\mbox{\bf LR}^{#1}} 
\newcommand{\win}[2]{\mbox{\bf Wn}^{#1}_{#2}} 
\newcommand{\seq}[1]{\langle #1 \rangle}           


\newcommand{\intimpl}{\mbox{\hspace{2pt}$\circ$\hspace{-0.14cm} \raisebox{-0.043cm}{\Large --}\hspace{2pt}}}
\newcommand{\sintimpl}{\mbox{\hspace{2pt}\raisebox{0.033cm}{\tiny $ | \hspace{-4pt} >$}\hspace{-0.14cm} \raisebox{-0.039cm}{\large --}\hspace{2pt}}}
\newcommand{\ade}{\mbox{\Large $\sqcup$}}      
\newcommand{\ada}{\mbox{\Large $\sqcap$}}      
\newcommand{\sst}{\mbox{\raisebox{-0.07cm}{\scriptsize $-$}\hspace{-0.2cm}$\pst$}}
\newcommand{\scost}{\mbox{\raisebox{0.20cm}{\scriptsize $-$}\hspace{-0.2cm}$\pcost$}}
\newcommand{\sqc}{\mbox{\small \raisebox{0.0cm}{$\bigtriangleup$}}}
\newcommand{\sqci}{\mbox{\tiny \raisebox{0.0cm}{$\bigtriangleup$}}}
\newcommand{\sqd}{\mbox{\small \raisebox{0.06cm}{$\bigtriangledown$}}}
\newcommand{\sqdi}{\mbox{\tiny \raisebox{0.05cm}{$\bigtriangledown$}}}
\newcommand{\sqe}{\mbox{\large \raisebox{0.07cm}{$\bigtriangledown$}}}
\newcommand{\sqa}{\mbox{\large \raisebox{0.0cm}{$\bigtriangleup$}}}
\newcommand{\mld}{\vee}    
\newcommand{\mlc}{\wedge}  
\newcommand{\mle}{\mbox{\Large $\vee$}}    
\newcommand{\mla}{\mbox{\Large $\wedge$}}  
\newcommand{\add}{\sqcup}                      
\newcommand{\adc}{\sqcap}                      
\newcommand{\gneg}{\neg}                  
\newcommand{\rneg}{\neg}               
\newcommand{\pneg}{\neg}               
\newcommand{\mli}{\rightarrow}                     
\newcommand{\intf}{\$}               
\newcommand{\tlg}{\bot}               
\newcommand{\twg}{\top}               

\newcommand{\pst}{\mbox{\raisebox{-0.01cm}{\scriptsize $\wedge$}\hspace{-4pt}\raisebox{0.16cm}{\tiny $\mid$}\hspace{2pt}}}
\newcommand{\cla}{\mbox{\large $\forall$}}      
\newcommand{\cle}{\mbox{\large $\exists$}}        
\newcommand{\pintimpl}{\mbox{\hspace{2pt}\raisebox{0.033cm}{\tiny $>$}\hspace{-0.18cm} \raisebox{-0.043cm}{\large --}\hspace{2pt}}}
\newcommand{\pcost}{\mbox{\raisebox{0.12cm}{\scriptsize $\vee$}\hspace{-4pt}\raisebox{0.02cm}{\tiny $\mid$}\hspace{2pt}}}
\newcommand{\st}{\mbox{\raisebox{-0.05cm}{$\circ$}\hspace{-0.13cm}\raisebox{0.16cm}{\tiny $\mid$}\hspace{2pt}}}
\newcommand{\cost}{\mbox{\raisebox{0.12cm}{$\circ$}\hspace{-0.13cm}\raisebox{0.02cm}{\tiny $\mid$}\hspace{2pt}}}


\newtheorem{theoremm}{Theorem}[section]
\newtheorem{conjecturee}[theoremm]{Conjecture}
\newtheorem{exercisee}[theoremm]{Exercise}
\newtheorem{definitionn}[theoremm]{Definition}
\newtheorem{lemmaa}[theoremm]{Lemma}
\newtheorem{propositionn}[theoremm]{Proposition}
\newtheorem{conventionn}[theoremm]{Convention}
\newtheorem{examplee}[theoremm]{Example}
\newtheorem{remarkk}[theoremm]{Remark}
\newtheorem{factt}[theoremm]{Fact}

\newenvironment{conjecture}{\begin{conjecturee}}{\end{conjecturee}}
\newenvironment{definition}{\begin{definitionn} \em}{ \end{definitionn}}
\newenvironment{theorem}{\begin{theoremm}}{\end{theoremm}}
\newenvironment{lemma}{\begin{lemmaa}}{\end{lemmaa}}
\newenvironment{proposition}{\begin{propositionn} }{\end{propositionn}}
\newenvironment{convention}{\begin{conventionn} \em}{\end{conventionn}}
\newenvironment{remark}{\begin{remarkk} \em}{\end{remarkk}}
\newenvironment{proof}{ {\bf Proof.} }{\  $\Box$ \vspace{.1in} }
\newenvironment{example}{\begin{examplee} \em}{\end{examplee}}
\newenvironment{exercise}{\begin{exercisee} \em}{\end{exercisee}}
\newenvironment{fact}{\begin{factt} \em}{\end{factt}}

\title{Sequential operators in computability logic}
\author{Giorgi Japaridze\thanks{This material is based upon work supported by the National Science Foundation under Grant No. 0208816}   \\  
\\ {\footnotesize Department of Computing Sciences, Villanova University, 800 Lancaster Avenue, Villanova, PA 19085, USA.}\\
{\footnotesize Email: giorgi.japaridze@villanova.edu
 \ URL: http://www.csc.villanova.edu/$^\sim$japaridz/}}
\date{}
\maketitle
\begin{abstract} 
{\em Computability logic} (CL) is a semantical platform and research program for redeveloping logic as a formal theory of computability, as opposed to the formal theory of truth which it has more traditionally been. Formulas in CL stand for (interactive) computational problems, understood as games between a machine and its environment; logical operators represent operations on such entities; and ``truth'' is understood as existence of an effective solution, i.e.,  of an algorithmic  winning strategy. 

The formalism of CL is open-ended, and may undergo series of extensions as the study of the subject advances. The main groups of operators on which CL has been focused so far are the {\em parallel}, {\em choice}, {\em branching}, and {\em blind} operators, with the logical behaviors of the first three groups 
resembling those of the multiplicatives, additives and exponentials of linear logic, respectively. The present paper introduces a new important group of operators, called {\em sequential}. The latter come in the form of sequential conjunction and disjunction, sequential quantifiers, and sequential recurrences (``exponentials''). As the name may suggest, the algorithmic intuitions associated with this group are those of sequential computations, as opposed to the intuitions of parallel computations associated with the parallel group of operations. Specifically, while playing a parallel combination of games means playing all components of the combination simultaneously, playing a sequential combination means playing the components in a sequential fashion, one after one.

The main technical result of the present paper is a sound and complete axiomatization of the propositional fragment of computability logic whose vocabulary, together with negation, includes all three --- parallel, choice and sequential --- sorts of conjunction and disjunction. An extension of this result to the first-order level is also outlined. 

 \end{abstract}

\noindent {\em MSC}: primary: 03B47; secondary: 03F50; 03B70; 68Q10; 68T27; 68T30; 91A05 

\

\noindent {\em Keywords}: Computability logic; Interactive computation; Game semantics; Linear logic; Constructive logics

\

\section{Introduction}\label{sintr}

This article is yet another addition to the evolving list \cite{Jap03}-\cite{Japdeep} of papers devoted to developing {\em computability logic} (CL).  Baptized so in \cite{Jap03}, in a broad sense, CL  is not a particular syntactic system or a particular semantics for a particular collection of logical operators, but rather a general platform and an ambitious program for redeveloping logic as a formal theory of computability, as opposed to 
the formal theory of truth which it has more traditionally been. Formulas in CL stand for computational problems,  logical operators represent operations on problems, ``truth'' is understood as existence of an algorithmic solution, and proofs encode such solutions. Among the main goals of CL at the present stage of development is finding axiomatizations for incrementally expressive fragments of it. Considerable advances have already been made in this direction, and the present paper tells one more success story. 

The traditional theory of computation has been primarily evolving around batch computation, despite the fact of life that most tasks performed by computers and computer networks (as well as humans in everyday life) are interactive. Aiming at being a comprehensive formal theory of computation, CL understands computational problems and computability in their most general --- {\em interactive} --- sense. And interactive problems are formalized as games played by a {\em machine} (computer, robot) against its {\em environment} (user, nature), with {\em computability} meaning existence of a machine that wins the game against any possible (behavior of the) environment. 

Technically, the semantics of CL is thus a game semantics. Among the features distinguishing it from other game-semantical approaches, including Blass's approach \cite{Bla72,Bla92} which is the closest precursor of computability logic,
one should point out the following. 

First of all, in CL, machine's (proponent's, $\exists$-player's) strategies are limited to algorithmic ones. This is a minimal condition that a game semantics should satisfy if it is meant to find applications in computer science. Due to the same condition, CL has good --- semantically rather than syntactically justified --- claims to be a {\em constructive logic}.  
 
Second, players' strategies are no longer considered as functions from positions (the sequences of the previously made moves) to moves. Rather, they are defined in terms of interactive machines, where computation is one continuous process interspersed with --- and influenced by --- multiple ``input'' (environment's moves) and ``output'' (machine's moves) events. A good game semantics is or should be  about interaction, while functions are inherently non-interactive. The traditional {\em strategies-as-functions} approach misses this very important point and creates an unnatural hybrid of interactive (games) and non-interactive (functions) entities. To appreciate the difference, it would be sufficient to reflect on the behavior of one's personal computer. The job of your computer is to play one long --- potentially infinite --- game against you. Now, have you noticed your computer getting slower every time you use it? Probably not. That is because the computer is smart enough to follow a non-functional strategy in this game. If its strategy was a function from positions (interaction histories) to moves, the response time would inevitably keep worsening due to the need to read the entire --- continuously lengthening and, in fact, practically infinite --- interaction history every time before responding. Defining strategies as functions of only the latest moves (rather than entire interaction histories) in Abramsky and Jagadeesan's \cite{Abr94} tradition  is also not a way out, as typically more than just the last move matters. Back to your personal computer, its actions certainly depend on more than 
your last keystroke. Thus, the difference between the traditional {\em functional} strategies and the {\em post-functional} strategies of CL is not just a matter of taste or convenience. It will become especially important when it comes to (yet to be developed) interactive complexity theory: hardly any meaningful interactive complexity theory can be done with the strategies-as-functions approach. And complexity issues will inevitably come forward when computability logic or similar approaches achieve a certain degree of maturity: nowadays, 95\% of the theory of computation is about complexity rather than just computability.

Third, the concept of games that CL deals with is more general than the traditional concepts. Among the distinguishing features of CL games is the absence of {\em procedural rules} --- rules regulating which player can or should move in any given position, the most typical procedural rule being the one according to which the players take turns in an alternating order. In CL games, both players may have legal moves in a given situation.  It has been repeatedly argued that only this flexible approach allows us to adequately model truly interactive real-life computational tasks and account for phenomena such as asynchronous communication, concurrency and parallelism. So, again, this difference is not just a difference of tastes, and will certainly play a crucial role when it comes to  interactive computational complexity and various flavors of it.  

Time has not yet matured for seriously addressing complexity issues though, and CL, including the present paper, continues to be focused on just computability, where there still are too many open questions calling for answers.  
 
The formalism of CL is open-ended, and is expected to undergo series of extensions as the study of the subject advances. The main groups of operations studied so far 
are:
\begin{itemize}
\item {\em Constant elementary games} ($0$-ary operations): $\twg$, $\tlg$.
\item {\em Negation}: $\gneg$.
\item {\em Choice operations}: $\adc$ (conjunction), $\add$ (disjunction), $\ada$ (universal quantifier), $\ade$ (existential quantifier).
\item {\em Parallel operations}: $\mlc$ (conjunction), $\mld$ (disjunction), $\mla$ (universal quantifier), $\mle$ (existential quantifier), $\pst$ (recurrence), $\pcost$ (corecurrence).
\item {\em Blind operations}: $\cla$ (universal quantifier), $\cle$ (existential quantifier).
\item {\em Branching operations}: $\st$ (recurrence), $\cost$ (corecurrence) and a series of their restricted versions  such as $\st^{\aleph_0}$ (countable recurrence), $\cost^{\aleph_0}$ (countable corecurrence).
\end{itemize} 
There are also various {\em reduction} operations: $\mli$, defined by $A\mli B=\gneg A\mld B$; $\pintimpl$, defined by $A\pintimpl B=\pst A\mli B$; $\intimpl$, defined by $A\intimpl B=\st A\mli B$; etc.

The present paper introduces the following new group:
\begin{itemize}
\item {\em Sequential operations}:  $\sqc$ (conjunction), $\sqd$ (disjunction), $\sqa$ (universal quantifier), $\sqe$ (existential quantifier), $\sst$ (recurrence), $\scost$ (corecurrence),
\end{itemize}
which also induces the reduction operation $\sintimpl$ defined by $A\sintimpl B=\sst A\mli B$.

The main technical result of this paper is constructing a sound and complete axiomatization for the propositional fragment of CL whose logical vocabulary consists of $\twg$, $\tlg$, $\gneg$, $\mlc$, $\mld$, $\adc$, $\add$, $\sqc$, $\sqd$. An extension of this result to the first-order level that additionally includes the quantifiers $\ada,\ade,\cla,\cle$ is also outlined. 

\section{A tour of the zoo}\label{s2}

In this section we give a very brief and informal overview of the language of computability logic and the game-semantical meanings of its main operators  for those unfamiliar with the subject. In what follows, $\pp$ and $\oo$ are symbolic names for the players to which we referred as the machine and the environment, respectively. 

First of all, it should be noted that computability logic is a conservative extension of classical logic. Classical propositions --- as well as predicates as generalized propositions --- are viewed as special, {\em elementary} sorts of games that have no moves and are automatically won by the machine if true, and lost if false. The languages of various reasonably expressive fragments of computability logic would typically include two sorts of atoms: {\em elementary} atoms $p$, $q$, $r(x)$, $s(x,y)$, \ldots to represent elementary games, and {\em general atoms} $P$, $Q$, $R(x)$, $S(x,y)$, \ldots to represent any, not-necessarily-elementary, games. The classically-shaped operators $\gneg,\mlc,\mld,\cla,\cle$ are conservative generalizations of the corresponding classical operations from elementary games to all games. That is in the sense that, when applied to elementary games, they again produce elementary games, and their meanings happen to coincide with the classical meanings. 

\subsection{Constant elementary games} These are two $0$-ary ``operations'', for which we use the same symbols $\twg$ and $\tlg$ as for the two players. $\twg$ is an elementary game automatically won by $\pp$, and $\tlg$ is an elementary game won by $\oo$. Just as classical logic, computability logic sees no difference between two true or two false propositions, so that we have ``Snow is white''=``$0=0$''=$\twg$ and ``Snow is black''=``$0=1$''=$\tlg$.    

\subsection{Negation}
Negation $\gneg$ is a role-switch operation: $\gneg A$ is obtained from $A$ by turning $\pp$'s (legal) moves and wins into $\oo$'s (legal) moves and wins, and vice versa. For example, if {\em Chess} means the game of chess from the point of view of the white player, then $\gneg${\em Chess} is the same game from the point of view of the black player. And where $0=0$ is an elementary game automatically won by $\pp$, $\gneg 0=0$ is an elementary game automatically won by $\oo$ --- there are no moves to interchange here, so only the winners are interchanged. From this explanation it must be clear that $\gneg$, when applied to elementary games
(propositions or predicates), indeed acts like classical negation, as promised.

\subsection{Choice operations}
The choice operations model decision steps in the course of interaction, with disjunction 
and existential quantifier meaning $\pp$'s choices, and conjunction and universal quantifier 
meaning choices by $\oo$. For instance, where $f(x)$ is a function, $\ada x\ade y \bigl(y=f(x)\bigr)$ is a game in which the first move/choice is by the environment, consisting in 
specifying a particular value $m$ for $x$. Such a move, which intuitively can be seen as asking the machine the question ``{\em what is the value of $f(m)$?}\hspace{1pt}'' brings the game down to the position $\ade y \bigl(y=f(m)\bigr)$. 
The next step is by the machine, which should specify a value 
$n$ for $y$, further bringing the game down to the elementary game $n=f(m)$,  won by the machine if true and lost if false. $\pp$'s move $n$ can thus be seen as answering/claiming that $n$ is the value of $f(m)$.
From this explanation it must be clear that   
$\ada x\ade y \bigl(y=f(x)\bigr)$ represents the problem of computing $f$, with $\pp$ having an algorithmic winning strategy for this game iff $f$ is a computable function. Similarly, where $p(x)$ is a predicate, $\ada x\bigl(p(x)\add \gneg p(x)\bigr)$ represents the problem of deciding $p(x)$: here, again, the first move is by the environment, consisting in choosing a value $m$ for $x$ (asking whether $p(m)$ is true); and the next step is by the machine which, in order to win, should 
choose the true disjunct of  $p(m)\add \gneg p(m)$, i.e. correctly answer the question. Formally, $A\add B$  can be defined as $\gneg(\gneg A\adc \gneg B)$, or 
$A\adc B$ can be defined as $\gneg(\gneg A\add \gneg B)$; furthermore, assuming that the universe of discourse is $\{1,2,3,\ldots\}$,  $\ada xA(x)$ can be defined as 
$A(1)\adc A(2)\adc A(3)\adc\ldots$ and $\ade xA(x)$ as $A(1)\add A(2)\add A(3)\add\ldots$. It should be mentioned that making an initial choice of a component by the corresponding player in a choice combination of games is not only that player's privilege, but also an obligation: the player will be considered the loser if it fails to make a choice.

\subsection{Parallel operations}
The parallel operations combine games in a way that corresponds to the intuition of concurrent computations. Playing $A\mlc B$ or $A\mld B$ means playing, in parallel, the two games $A$ and $B$. In $A\mlc B$, $\pp$ is considered the winner if it wins in both of the components, while in $A\mld B$ it is sufficient to win in one of the components. Then the parallel quantifiers and recurrences are defined by: 
\[\begin{array}{rcl}
\mla xA(x) & = & A(1)\mlc A(2)\mlc A(3)\mlc\ldots\vspace{3pt}\\
\mle xA(x) & = & A(1)\mld A(2)\mld A(3)\mld\ldots\vspace{3pt}\\ 
\pst A & = & A\mlc A\mlc A\mlc\ldots\vspace{3pt}\\
\pcost A & = & A\mld A\mld A\mld\ldots
\end{array}\] 
To appreciate the difference between choice operations and their parallel counterparts, let us compare the games ${\em Chess}\mld\gneg \mbox{\em Chess}$ and $\mbox{\em Chess} \add \gneg \mbox{\em Chess}$. 
The former  is, in fact, a simultaneous play on two boards, where on the left board $\pp$ plays white, and on the right  board plays black. There is a simple strategy for $\pp$ that guarantees success against any adversary. All that $\pp$ needs to do is to mimic, in {\em Chess}, the moves made by $\oo$ in $\gneg \mbox{\em Chess}$, and vice versa. On the other hand, to win the game $\mbox{\em Chess}\hspace{0.02in} \add\gneg \mbox{\em Chess}$ is not easy: here, at the very beginning, $\pp$ has to choose between {\em Chess} and $\gneg \mbox{\em Chess}$ and then win the chosen one-board game. 

While all classical tautologies automatically hold when the classically-shaped operators are applied to elementary games, in the general (nonelementary) case the class of valid principles shrinks. For example,  $\gneg P\mld (P\mlc P)$ is no longer valid. The above ``mimicking strategy" would obviously fail in the three-board game \[\gneg \mbox{\em Chess}\mld (\mbox{\em Chess}\mlc \mbox{\em Chess}),\] for here the best that $\pp$ can do is to pair $\gneg \mbox{\em Chess}$ with one of the two conjuncts of $\mbox{\em Chess}\mlc \mbox{\em Chess}$. It is possible that then $\gneg \mbox{\em Chess}$ and the unmatched {\em Chess} are both lost, in which case the whole game will be lost.  As much as this example may remind us of linear logic, it should be noted that the class of principles with parallel connectives validated by computability logic is not the same as the class of multiplicative formulas provable in linear or affine logic. An example separating CL from both linear and affine logics is Blass's \cite{Bla92} principle \[\bigl((\gneg P\mld\gneg Q)\mlc(\gneg R\mld\gneg S)\bigr) \mld \bigl((P\mld R)\mlc(Q\mld  S)\bigr),\] not provable in affine logic but valid in CL.
The same applies to principles containing choice (``additive'') and recurrence (``exponential'') operators.
 
\subsection{Reduction}
The operation $\mli$, defined in the standard way by $A\mli B=\gneg A\mld B$, is perhaps most  interesting from the 
computability-theoretic point of view. Intuitively, $A\mli B$ is the problem of {\em reducing} $B$ to $A$. Putting it in other words, solving $A\mli B$ means solving $B$ having $A$ as an (external) {\em computational resource}.\label{0cr} ``Computational resource" is symmetric to ``computational problem": what is a problem (task) for the machine, is a resource for the environment, and vice versa.
To get a feel of $\mli$ as a problem reduction operator, let us look at  reducing 
the acceptance problem to the halting problem. The halting problem  can be expressed by 
\[\ada x\ada y \bigl(\mbox{\em Halts}(x,y) \add \gneg \mbox{\em Halts}(x,y)\bigr),\]
 where $\mbox{\em Halts}(x,y)$ is the predicate  
``Turing machine  (encoded by) $x$ halts on input $y$". And the acceptance problem can be expressed by 
\[\ada x\ada y \bigl(\mbox{\em Accepts}(x,y) \add \gneg \mbox{\em Accepts}(x,y)\bigr),\] 
with $\mbox{\em Accepts}(x,y)$ meaning 
``Turing machine  $x$ accepts input $y$". While the acceptance problem is not decidable, it is algorithmically reducible to the halting problem. In particular, there is a machine that always wins the game
\[
\ada x\ada y \bigl(\mbox{\em Halts}(x,y)\add \gneg \mbox{\em Halts}(x,y)\bigr) \mli\ \ada x\ada y \bigl(\mbox{\em Accepts}(x,y)\add \gneg \mbox{\em Accepts}(x,y)\bigr).
\]
A strategy for solving this problem is to wait till the environment specifies values $m$ and $n$ for $x$ and $y$ in the consequent, thus asking $\pp$ the question ``does machine $m$ accept input $n$?''. In response, $\pp$ selects the same values $m$ and $n$ for  $x$ and $y$ in the antecedent (where the roles of $\pp$ and $\oo$ are switched), thus asking the counterquestion ``does $m$ halt on $n$?''. The environment will have to correctly answer this counterquestion, or else it loses. If it answers ``No'', then $\pp$ also says ``No'' in the consequent, i.e., selects the right disjunct there, as not halting implies not accepting. Otherwise, if the environment's response in the antecedent is ``Yes'',
$\pp$ simulates machine $m$ on input $n$ until it halts and then selects, in the consequent, the left or the right disjunct depending on whether the simulation accepted or rejected. 

\subsection{Blind operations}
The blind group of operations comprises $\cla$  and its dual $\cle$ ($\cle x =\gneg\cla x\gneg$). The meaning of $\cla xA(x)$ is similar to that of $\ada xA(x)$, with the difference that the particular value of $x$ that the environment ``selects" is invisible to the machine, so that it has to play blindly in a way that guarantees success no matter what that value is. This way, $\cla$ and $\cle$  produce games with {\em imperfect information}. 

Compare the problems
\[\ada x\bigl(\mbox{\em Even$(x)$}\add \mbox{\em Odd$(x)$}\bigr)\] and  \[\cla x\bigl(\mbox{\em Even$(x)$}\add \mbox{\em Odd$(x)$}\bigr).\] 
Both of them are about telling whether a given number is even or odd; the difference is only in whether that ``given number" is communicated to the machine or not. The first problem is an easy-to-win, two-move-deep game of a structure that we have already seen.  The second game, on the other hand, is one-move deep with only by the machine to make a move --- select the ``true"  disjunct, which is hardly possible to do as the value of $x$ remains unspecified. 

As an example of a solvable nonelementary $\cla$-problem, let us look at
\[\cla x\Bigl(\mbox{\em Even$(x)$}\add \mbox{\em Odd$(x)$}\ \mli\ \ada y\bigl(\mbox{\em Even$(x+y)$}\add
\mbox{\em Odd$(x+y)$}\bigr)\Bigr),\]
solving which means solving what follows ``$\cla x$" without knowing the value of $x$. Unlike $\cla x\bigl(\mbox{\em Even$(x)$}\add \mbox{\em Odd$(x)$}\bigr)$, this game is certainly winnable: The machine waits till the environment selects  a value $n$ for $y$ in the consequent and also selects one of the $\add$-disjuncts in the antecedent (if either selection is never made, the machine automatically wins). Then: If $n$ is even, in the consequent the machine makes the same selection 
{\em left}  or {\em right} as the environment made in the antecedent, and otherwise, if $n$ is odd, it reverses the environment's selection. 

\subsection{Sequential operations} \label{s2seq}
The new, sequential group of operations forms another natural phylum in this zoo of game operations. 
The sequential conjunction $A\sqc B$ is a game that starts and proceeds as a play of $A$; it will also end as an ordinary play of $A$ unless, at some point, 
$\oo$ decides --- by making a special {\em switch} move --- to abandon $A$ and switch to $B$. In such a case the play restarts, continues and ends as an ordinary play of $B$ without the possibility to go back to $A$. $A\sqd B$ is the same, only here it is $\pp$ who decides whether and when to switch from $A$ to $B$. These generalize to the infinite cases $A_0\sqc A_1\sqc A_2\sqc\ldots$ and $A_0\sqd A_1\sqd A_2\sqd\ldots$: here the corresponding player can make any finite number $n$ of switches, in which case the winner in the play will be the player who wins in $A_n$; and if an infinite number of switches are made, then the player responsible for this is considered the loser.  The sequential quantifiers, as we may guess, are defined by \[\sqa xA(x)=A(1)\sqc A(2)\sqc A(3)\sqc\ldots\] and \[\sqe xA(x)=A(1)\sqd A(2)\sqd A(3)\sqd\ldots,\] and the sequential recurrence and corecurrence are defined by 
\[\sst A=A\sqc A\sqc A\sqc\ldots\] and \[\scost A=A\sqd A\sqd A\sqd\ldots.\] Below are a few examples providing insights into the computational intuitions and motivations associated with the sequential operations. 

Let $p(x)$ be any predicate. Remember that the game $\ada x\bigl(\gneg p(x)\add p(x)\bigr)$ represents the problem of deciding $p(x)$. Then what is  represented by $\ada x\bigl(\gneg p(x)\sqd p(x)\bigr)$? If you guessed that this is the problem of {\em semideciding} $p(x)$, you have guessed right. It is not hard to see that this game has an effective winning strategy by $\pp$ iff $p(x)$ is semidecidable (recursively enumerable). Indeed, if $p(x)$ is semidecidable, a winning strategy is to wait until $\oo$ selects a particular $m$ for $x$, thus bringing the game down to  $\gneg p(m)\sqd p(m)$. After that, $\pp$ starts looking for a certificate of $p(m)$'s being true. If and when such a certificate is found (meaning that $p(m)$ is indeed true), $\pp$ makes a switch move turning $\gneg p(m)\sqd p(m)$ into the true and hence $\pp$-won $p(m)$; and if no certificate exists (meaning that $p(m)$ is false), then  $\pp$ keeps looking for a non-existent certificate forever and thus never makes any moves, meaning that the game ends as $\gneg p(m)$, which, again, is a true and hence $\pp$-won elementary game. And vice versa: any effective winning strategy for $\ada x\bigl(\gneg p(x)\sqd p(x)\bigr)$ can obviously be seen as a semidecision procedure for $p(x)$, which accepts an input $m$ iff the strategy ever makes a switch move in the scenario where $\oo$'s initial choice of a value for $x$ is $m$.       

 Algorithmic solvability (computability) of games has been shown to be closed under modus ponens, as well as the rules --- along with a number of other rules --- ``from $A$ and $B$ conclude $A\mlc B$'', ``from $A$ conclude $\ada x A$'', ``from $A$ conclude $\pst A$''. In view of these closures, the validity (= ``always computability'') of the principles discussed below implies  certain 
known facts from the theory of computation. Needless to say, those examples demonstrate how CL can be used as a systematic tool for defining new interesting properties and relations between computational problems, and not only reproducing already known theorems but also discovering an infinite variety of new facts.  

The following formula, later proven --- in a stronger form --- to be a theorem of our presumably sound and complete (with respect to validity) first-order system $\predseq$, implies --- in a sense, ``expresses'' --- the well known fact that, if both a predicate $p(x)$ and its negation $\gneg p(x)$ are recursively enumerable, then $p(x)$ is decidable:

\begin{equation}\label{nov16}
\ada x\bigl(\gneg p(x)\sqd p(x)\bigr)\mlc\ada x\bigl(p(x)\sqd\gneg p(x)\bigr)\mli \ada x\bigl(p(x)\add\gneg p(x)\bigr).
\end{equation}
Actually, the validity of the above formula means something more than just noted: it means that the problem of deciding $p(x)$ is reducible to the ($\mlc$-conjunction of) the problems of semideciding $p(x)$ and $\gneg p(x)$. In fact, a reducibility in an even stronger sense (in a sense that has no name) holds, expressed by the following valid formula:
\begin{equation}\label{nov16a}
\ada x\Bigl(\bigl(\gneg p(x)\sqd p(x)\bigr)\mlc\bigl((p(x)\sqd\gneg p(x)\bigr)\mli \bigl(p(x)\add\gneg p(x)\bigr)\Bigr).
\end{equation}

Computability logic defines computability of a game $A(x)$ as computability of its $\ada$-closure, so the prefix $\ada x$ can be safely removed in the above formula and, after writing simply ``$p$'' instead of ``$p(x)$'', the validity of (\ref{nov16a}) means the same as the validity of the following propositional-level formula, provable in our sound and complete propositional system $\propseq$:
 
\begin{equation}\label{nov16b}
(\gneg p\sqd p)\mlc(p\sqd\gneg p)\mli p\add\gneg p.
\end{equation}

Furthermore, the above principle is valid not only for predicates (elementary games) but also for all games that we consider, as evidenced by the provability of the following formula in (the sound) $\propseq$:

\begin{equation}\label{nov16c}
(\gneg P\sqd P)\mlc(P\sqd\gneg P)\mli P\add\gneg P.
\end{equation}
Similarly, formula (\ref{nov16}) remains provable in $\predseq$ and hence valid with $P(x)$ instead of $p(x)$:
\begin{equation}\label{nov16d}
\ada x\bigl(\gneg P(x)\sqd P(x)\bigr)\mlc\ada x\bigl(P(x)\sqd\gneg P(x)\bigr)\mli \ada x\bigl(P(x)\add\gneg P(x)\bigr).
\end{equation}

For our next example, remember the relation of {\em mapping reducibility} (more often  called {\em many-one reducibility}) of a predicate $q(x)$ to a predicate $p(x)$, defined as existence of an effective function $f$ such that, for any $n$, $q\bigl(n\bigr)$ is equivalent to $p\bigl(f(n)\bigr)$. It is not hard to see that this relation holds if and only if the game \[\ada x\ade y\Bigl(\bigl(q(x)\mli p(y)\bigr)\mlc \bigl(p(y)\mli q(x)\bigr)\Bigr),\]
which we abbreviate as  $\ada x\ade y\bigl(q(x)\leftrightarrow p(y)\bigr)$, has an algorithmic winning strategy by $\pp$. In this sense, $\ada x\ade y\bigl(q(x)\leftrightarrow p(y)\bigr)$  expresses the problem of mapping reducing $q(x)$ to $p(x)$. Then the  validity (that can be established through  $\predseq$-provability) of the following formula implies the known fact that, if $q(x)$ is mapping reducible to $p(x)$ and $p(x)$ is recursively enumerable, then so is $q(x)$:\footnote{By the way, the same principle does not hold with ``Turing reducible'' instead of ``mapping reducible''.} 

\begin{equation}\label{nov16e}
\ada x\ade y\bigl(q(x)\leftrightarrow p(y)\bigr)\mlc \ada x\bigl(\gneg p(x)\sqd p(x)\bigr)\mli \ada x\bigl(\gneg q(x)\sqd q(x)\bigr).\end{equation}
As in the earlier examples, the validity of (\ref{nov16e}), in fact, means something even more: it means that the problem of semideciding $q(x)$ is reducible to the ($\mlc$-conjunction of the) problems of mapping reducing $q(x)$ to $p(x)$ and semideciding $p(x)$. 

Certain other reducibilities hold only in a sense weaker than the sense captured by $\mli$. We characterized $A\mli B$ as a game where $\pp$  can use $A$ as a computational resource: playing in the role of $\oo$ in $A$, $\pp$ can observe how the adversary is solving $A$ and employ that information in its solving $B$. It is however important to note that only one ``copy'' of $A$ is available to $\pp$ as a resource in $A\mli B$. In many cases, however, more than one runs of $A$ may be necessary. An example of a reduction of this sort is Turing reduction, where the oracle (resource $A$) can be queried an unlimited number of times. A way  to account for the possibility of repeated usage of $A$ is prefixing it with a recurrence operation. In the following two examples a recurrence that suffices is $\pst$, which (just as the other types of recurrences) induces the weak reduction operation $\pintimpl$ defined by $A\pintimpl B=\pst A\mli B$.  

The following formula is valid (and remains so with $P(x,y)$ instead of $p(x,y)$):
\begin{equation}\label{xc}
\ada x\ada y\bigl(\gneg p(x,y)\add p(x,y)\bigr)\pintimpl \ada x\bigl(\gneg \cle yp(x,y)\sqd \cle yp(x,y)\bigr),\end{equation}
meaning that the problem of semideciding a predicate $\cle y p(x,y)$ is $\pintimpl$-reducible to the problem of deciding $p(x,y)$. This, in turn, implies the known fact that if $p(x,y)$ is decidable, then $\cle y p(x,y)$ is recursively enumerable. Unlike the earlier cases where we appealed to provability in (the sound) $\propseq$ or $\predseq$ in claiming validity, (\ref{xc}) is not a formula of the languages of those systems because it contains $\pintimpl$. So, let us verify its validity directly. 

Here is $\pp$'s strategy for (\ref{xc}), equally good for (and not depending on) any predicate $p(x,y)$. Wait till $\oo$ specifies a value $m$ for $x$ in the consequent, thus bringing the game down to 
\[\ada x\ada y\bigl(\gneg p(x,y)\add p(x,y)\bigr)\pintimpl \bigl(\gneg \cle yp(m,y)\sqd \cle yp(m,y)\bigr).\]
Then, initialize $i$ to $1$ and do the following. Specify $x$ and $y$ in the $i$th copy of the antecedent as $m$ and $i$, respectively. $\oo$ will have to respond by choosing one of the $\add$-disjuncts in that copy, which now looks like $\gneg p(m,i)\add p(m,i)$, or else it loses. If $\oo$ chooses $\gneg p(m,i)$, increment $i$ to $i+1$ and repeat the step. Otherwise, if $\oo$ chooses $p(m,i)$, make a switch move in the consequent and rest your case.

Let us see one more example with $\pintimpl$-reducibility. Let $\mbox{\em NEQ}(x,y)$ be the predicate ``Turing machines (encoded by) $x$ and $y$ are not equivalent'', with equivalence meaning that the two machines accept exactly the same inputs.  This predicate is neither semidecidable nor co-semidecidable. However, the problem of its semideciding $\pintimpl$-reduces to the halting problem.  Specifically, $\pp$ has an algorithmic winning strategy for the following game:

\begin{equation}\label{po}
\ada z\ada t \bigl(\gneg \mbox{\em Halts}(z,t)\add \mbox{\em Halts}(z,t)\bigr)\pintimpl \ada x\ada y\bigl(\gneg \mbox{\em NEQ}(x,y)\sqd \mbox{\em NEQ}(x,y)\bigr).
\end{equation}
A strategy here is to wait till $\oo$ specifies some values $m$ and $n$ for $x$ and $y$ in the consequent, respectively. 
Then, initialize $i$ to $1$ and do the following. Specify $z$ and $t$  as $m$ and $i$ in one yet-unused copy of the antecedent, and as $n$ and $i$ in another yet-unused copy. That is, ask $\oo$ whether $m$ halts on input $i$ and whether $n$ halts on the same input. $\oo$ will have to provide the correct pair of answers, or else it loses. 
\begin{enumerate}
\item If the answers are ``No,No", increment $i$ to $i+1$ and repeat the step. 
\item If the answers are ``Yes,Yes'', then simulate both $m$ and $n$ on input $i$ until they halt. If both machines accept or both reject, increment $i$ to $i+1$ and repeat the step. Otherwise, if one accepts and one rejects, make a switch move in the consequent and celebrate victory.
\item If the answers are ``Yes,No'', then simulate $m$ on $i$ until it halts. If $m$ rejects $i$, increment $i$ to $i+1$ and repeat the step. Otherwise, if $m$ accepts $i$, make a switch move in the consequent and you win.
\item If the answers are ``No,Yes'', then simulate $n$ on $i$ until it halts. If $n$ rejects $i$, increment $i$ to $i+1$ and repeat the step. Otherwise, if $n$ accepts $i$, make a switch move in the consequent and you win.
\end{enumerate}

For our last example, remember the concept of the Kolmogorov complexity of a given number $m$, which can be defined as the size of the smallest Turing machine that returns $m$ on input $1$. We denote the Kolmogorov complexity of $x$ by $k(x)$. The latter is known to be bounded, not exceeding $x$ itself.\footnote{Well, strictly speaking, this is so only for sufficiently large numbers $x$. But since only for finitely many (very small) numbers $x$ do we have $k(x)>x$, we may ignore this minor technicality and assume in our treatment that $k(x)$ never exceeds $x$.} Function $k(x)$ is not computable, meaning that   $\pp$ has no algorithmic winning strategy in \[\ada x\ade y \bigl(y=k(x)\bigr).\] In contrast, the problem  
\[\ada x\scost\ade y \bigl(y=k(x)\bigr)\]
does have an algorithmic solution. Here is one: Wait till $\oo$ specifies a value $m$ for $x$, thus asking ``what is the Kolmogorov complexity of $m$?'' and bringing the game down to $\scost \ade y \bigl(y=k(m)\bigr)$. Answer that it is $m$, i.e. specify $y$ as $m$, and after that start simulating, in parallel, all machines $n$ with $n<m$ on input $1$. Whenever you find a machine $n$ that returns $m$ on input $1$ and is smaller than any of the previously found such machines, make a switch move and, in the new copy of  $\ade y \bigl(y=k(m)\bigr)$, specify $y$ as the size (=logarithm) $|n|$ of $n$. This obviously guarantees success: sooner or later the real Kolmogorov complexity $c$ of $m$ will be reached and named; and, even though the strategy will never be sure that $k(m)$ is not something  yet smaller than $c$, it will never really find a reason to further reconsider its latest claim that $c=k(m)$.  

The following game also has an algorithmic winning strategy, describing which is left as an exercise for the reader: 
\[\ada x\sqe y \bigl(k(x)=(x-y)\bigr).\]

\subsection{Branching operations}
The branching operations come in the form of branching recurrence $\st$ and its dual branching corecurrence $\cost$, which can be defined by $\cost A=\gneg\st\gneg A$. The two other --- parallel and sequential --- sorts of recurrences we have already seen, and it might be a good idea to explain $\st$ by comparing it with them.

What is common to all members of the family of (co)recurrence operations is that, when applied to $A$, they turn it into a game playing which means repeatedly playing $A$. In terms of resources, recurrence operations generate multiple ``copies'' of $A$, thus making $A$ a reusable/recyclable resource. The difference between the various sorts of recurrences is how ``reusage'' is exactly understood.

Imagine a computer that has a program successfully playing $\chess$. The resource that such a computer provides is obviously something stronger than just $\chess$, for it permits to play $\chess$ as many times as the user wishes, while $\chess$, as such, only assumes one play. The simplest operating system would allow to start a session of $\chess$, then --- after finishing or abandoning and destroying it --- start a new play again, and so on. The game that such a system plays --- i.e. the resource that it supports/provides --- 
is the already known to us sequential recurrence $\sst \chess$, which assumes an unbounded number of plays of $\chess$ in a sequential fashion.  A more advanced operating system, however, would not require to destroy the old sessions before starting new ones; rather, it would allow to run as many parallel sessions as the user needs. This is what is captured by the parallel recurrence $\pst\chess$. As a resource, $\pst\chess$ is obviously stronger than $\sst\chess$ as it gives the user more flexibility. But $\pst$ is still not the strongest form of reusage. A really good operating system would not only allow the user to start new sessions of $\chess$ without destroying old ones; it would also make it possible to branch/replicate each particular stage of each particular session, i.e. create any number of ``copies" of any already reached position
of the multiple parallel plays of $\chess$, thus giving the user 
the possibility to try different continuations from the same position. What corresponds to this intuition is the branching recurrence $\st\chess$.

Thus, the user of the resource $\st A$  does not have to restart $A$ from the very beginning every time it wants to reuse it; rather, it is (essentially) allowed to backtrack to any of the previous --- not necessarily starting --- positions and try a new continuation from there, thus depriving the adversary of the possibility to reconsider the moves it has already made in that position. This is in fact the type of reusage every purely software resource allows or would allow in the presence of an advanced operating system and unlimited memory:
one can start running process $A$; then fork it  
at any stage  thus creating two threads  that have a common past but possibly diverging futures  (with the possibility to treat one of the threads as 
a ``backup copy'' and preserve it for backtracking purposes); then further fork any of the branches at any time; and so on. The less flexible type of reusage of $A$ assumed by $\pst A$, on the other hand, is closer to what infinitely many autonomous 
physical resources would naturally offer, such as an unlimited number of independently acting robots each performing task $A$, or an unlimited number of computers with limited memories, each one only capable of and responsible for running a single thread 
of process $A$. Here  the effect of replicating/forking an advanced stage of $A$ cannot be achieved unless, by good luck, 
there are two identical copies of the stage, meaning that the corresponding two robots or computers have so far acted in precisely the same ways. As for $\sst A$, it models the task performed by a single reusable physical resource --- the resource that can perform  task $A$ over and over again any number of times.  

A formal definition of branching recurrence is more complicated than the definitions of its parallel and sequential counterparts. For this reason, in our present relaxed tour we refrain from going into more technical details of how, exactly, games of the form $\st A$ are played. Such details, together with formal definitions and additional explanations, can be found, for example, in \cite{Japfin}. $\st$ also has a series of weaker versions obtained by imposing various restrictions on the quantity and form of reusages. Among the interesting and natural weakenings of $\st$ is the {\em countable branching recurrence} $\st^{\aleph_0}$ in the style of Blass's \cite{Bla72,Bla92} {\em repetition operation} $R$. See \cite{Japfour} for a discussion of such operations. 

Branching recurrence $\st$ stands out as the strongest of all recurrence operations, allowing to reuse $A$ (in $\st A$) in the strongest algorithmic sense possible. 
This makes the associated reduction operation $\intimpl$, defined by $A\intimpl B=\st A\mli B$, the weakest and hence most general form of algorithmic reduction. 
The well known concept of {\em Turing reduction} has the same claims. The latter, however, is only defined for the traditional, non-interactive sorts of computational problems --- two-step, input-output, question-answer sorts of problems that in our terms are written as $\ada x\bigl(p(x)\add\gneg p(x)\bigr)$ (the problem of deciding predicate $p$) or 
 $\ada x\ade y\bigl(y=f(x)\bigr)$ (the problem of computing function $f$). And it is no surprise that our $\intimpl$, when restricted to such problems, turns out to be equivalent to Turing reduction. Furthermore, when $A$ and $B$ are traditional sorts of problems, $A\intimpl B$ further turns out to be equivalent to $A\pintimpl B$ (but not $A\sintimpl B$), as the differences between 
$A\pintimpl B$ and $A \intimpl B$, while substantial in the general (truly interactive) case, turn out to be  too subtle to   be relevant when $A$ is a game that models only a very short and simple potential dialogue between the interacting parties, consisting in just 
asking a question and giving an answer.  The benefits from the greater degree of resource-reusage flexibility offered by $A\intimpl B$ (as opposed to $A\pintimpl B$) are related to the possibility for the machine to try different reactions to the same action(s) by the environment in $A$. But such potential benefits cannot be realized when $A$ is, say, $\ada x\bigl(p(x)\add \gneg p(x)\bigr)$. Because here a given individual session of $A$ immediately ends with an environment's move, to which the machine simply has no legal or meaningful responses at all, let alone having multiple possible responses to experiment  with. 

Thus, both $\intimpl$ and $\pintimpl$ are conservative extensions of Turing reduction from traditional sorts of problems to problems of arbitrary degrees and forms of interactivity. Of these two operations, however, only $\intimpl$ has the moral right to be called a legitimate successor of Turing reducibility, in the sense that, just like Turing reducibility (in its limited context), $\intimpl$ rather than $\pintimpl$ is an ultimate formal counterpart of our most general intuition of algorithmic reduction. And perhaps it is no accident that, as shown in \cite{Japjsl,Propint}, its logical behavior --- along with the choice operations --- is precisely captured by Heyting's intuitionistic calculus. As an aside, this means that CL offers a good justification  --- in the form of a mathematically strict and intuitively convincing semantics --- of the constructivistic claims of intuitionistic logic, and a materialization of Kolmogorov's \cite{Kol32} well known yet  so far rather abstract thesis, according to which intuitionistic logic is a logic of problems. 

Our recurrence operations, in their logical spirit, are reminiscent of the exponential operators of linear logic. It should be noted that, as shown in \cite{Japfin}, linear --- in fact, affine --- logic turns out to be sound but incomplete when its additives are read as our choice operators, multiplicatives as parallel operators, and exponentials as either parallel or branching recurrences. Here the sequential sort of recurrences stands out in that linear logic becomes simply unsound if its exponentials $!,?$ are interpreted as our $\sst,\scost$. 

Just like the acceptance problem $\ada x\ada y \bigl(\mbox{\em Accepts}(x,y)\add \gneg \mbox{\em Accepts}(x,y)\bigr)$, the Kolmogorov complexity problem 
$\ada x\ade y \bigl(y=k(x)\bigr)$ is known to be algorithmically reducible --- specifically, Turing reducible --- to the halting problem. Unlike the former case, however, the reduction in the latter case essentially requires repeated usage of the halting problem as a resource. That is, the reducibility holds only in the sense of $\intimpl$ or $\pintimpl$ but not in the sense of $\mli$. As an exercise, the reader may try to come up with an informal description of an algorithmic winning strategy for either one of the following games:
\[\begin{array}{cc}
\ada x\ada y \bigl(\mbox{\em Halts}(x,y)\add \gneg \mbox{\em Halts}(x,y)\bigr)\pintimpl \ada x\ade y \bigl(y=k(x)\bigr);\\
\ada x\ada y \bigl(\mbox{\em Halts}(x,y)\add \gneg \mbox{\em Halts}(x,y)\bigr)\intimpl \ada x\ade y \bigl(y=k(x)\bigr).
\end{array}\]

\section{Sequential operators in CL-based applied systems}

As we had a chance to see, CL offers a flexible and convenient formalism for specifying and studying computational problems and relations between them.  It is a formal theory of computability in the same sense as classical logic is a formal theory of truth, and axiomatizations of various fragments of it provide a systematic way to answer the fundamental question  `what can be computed'. CL also takes us one step closer to developing the long-overdue comprehensive theories of interactive computation and interactive complexity. In fact, if and when further advanced  and sufficiently developed (but certainly not in its present, embryonic  form), computability logic itself can be considered such a theory or, at least, an integral part of it.

The significance of CL, however, is not limited to theory of computing or pure logic. All of the soundness results for the known axiomatizations of CL come in the strong form that we call {\em uniform-constructive soundness}. The uniform-constructive soundness of a deductive system means that: ({\em uniform soundness}:) for every provable formula $F$, there is a {\em uniform}, meaning-independent strategy in the sense that it wins the game represented by $F$ no matter how its atoms are interpreted,\footnote{As opposed simple soundness which means existence of a winning strategy for each particular interpretation, so that different interpretations may require different strategies.} and
({\em constructive soundness}:) such a strategy can  be effectively extracted from a proof of $F$.  This is good news, signifying that CL is not only about ``{\em what} can be computed'', but also equally about ``{\em how} can be computed'', opening various application areas, such as (constructive) applied theories, (interactive) knowledgebase systems, or (resource-oriented) AI systems for planning. All such systems would follow the same general scheme. One takes a basic set of formulas expressing problems (computational, informational or physical resources) whose solutions are available (known, maintainable, providable). To such a set $S$, depending on the context, we may refer as the set of {\em axioms}, or the {\em knowledgebase}, or the {\em resourcebase}. Provability of a formula $F$ in the system can be defined as provability --- in pure CL --- of the formula $S\mli F$, with $S$ here identified with the $\mlc$-conjunction of its elements.\footnote{Of course, this is not the only way to construct CL-based systems. But we are trying to keep things as simple as possible in this brief discussion.} Such a system becomes a problem-solving tool: all one needs for solving a problem is to express it in the language of the system and then find a proof of it. In view of  the uniform-constructive soundness of the underlying axiomatization of CL and the closure of computability under modus ponens, a solution for the problem can be automatically obtained from its proof. This is a very brief summary. See, for example, Section 10 of \cite{Japfin} for an extended discussion of CL-based applied systems. In this section we only outline --- very briefly and informally --- some intuitions associated with sequential operators that are relevant to potential applications in knowledgebase and planning systems.

In knowledgebase systems, sequential operators can be used to express dynamic or unstable knowledge. Imagine a knowledgebase system that maintains information on all people. Part of the information provided by such a system is knowledge of whether any given person is dead or alive. What the old (sequential-operator-free) language  of CL could offer to express such knowledge, with $x$ ranging over people, is 
\begin{equation}\label{tom}
\pst\ada x\bigl(\mbox{\em Alive}(x)\add \mbox{\em Dead}(x)\bigr):
\end{equation}
a system containing such a formula/resource in its knowledgebase is able to repeatedly tell us, for any person, whether he or she is alive or dead. This is sufficient to represent a snapshot of some stage of the knowledge(base). Facts change over time though and, in particular, so does the alive/dead status of a person. Yet, once a system asserts    $\mbox{\em Alive}(\mbox{\em Tom})$ in the process of playing (\ref{tom}), it cannot take it back later, specifically, when Tom dies. So, (\ref{tom}) is not an adequate way to express a dynamic informational resource of people's alive/dead status. What does fit the bill is  
\[\pst \ada x\bigl(\mbox{\em Alive}(x)\sqd \mbox{\em Dead}(x)\bigr)\]
instead (but, note, by no means $\pst\ada x\bigl(\mbox{\em Dead}(x)\sqd \mbox{\em Alive}(x)\bigr)$). 
 
Imagine further that the system maintains dynamic information on everybody's marital status. Unlike the alive/dead status, the marital status may change many times through a person's lifetime. To account for having this informational resource, we would include the following formula in the system's knowledgebase:
 
\[\pst \ada x\scost\bigl(\mbox{\em Single}(x)\sqd \mbox{\em Married}(x)\bigr).\]

Of course, in both of the above examples, the prefix $\pst\ada x$ can be replaced by $\st \ada x$ or just $\mla x$. This would not change the strength of the knowledgebase, but taking $\mla x$ instead of $\pst \ada x$ or $\st \ada x$ could negatively affect its efficiency, obligating the system to resolve each and every person's status no matter whether so requested or not.   

Having sequential operators in CL-based planning systems could be even more imperative, as such systems are inherently dynamic, where the truth status of various facts keeps changing from situation to situation, and is affected not only by the environment (as in dynamic knowledgebase systems) but also by actions of the agent. Here we restrict ourselves to just one simple and naive example to provide some insights. 

Imagine a controller for the outside front entrance light, whose job is to turn the light on at night and --- to save energy --- turn it off during the daytime. The controller has the capability to repeatedly turn the light on and off. This capability, as a (physical) resource, can be expressed by the formula  $\sst (\mbox{\em Off} \sqc \mbox{\em On})$, with {\em Off} expressing the fact ``the light is off'', and {\em On} expressing ``the light is on''. It further has a bright-light sensor, reporting whether it is day or night (or rather whether it is bright enough or not quite so). Let {\em Day} mean ``it is bright enough'' and {\em Night} mean 
the opposite. The resource provided by such a sensor can then be expressed by  $\scost (\mbox{\em Day} \sqd \mbox{\em Night})$. And the goal of the controller is to maintain the truth of $(\mbox{\em Day} \mlc \mbox{\em Off})\mld (\mbox{\em Night}\mlc \mbox{\em On})$. The overall planning/maintainance problem can then be expressed by
\begin{equation}\label{day} 
\scost (\mbox{\em Day}\sqd \mbox{\em Night})\mlc \sst (\mbox{\em Off}\sqc \mbox{\em On})\mli (\mbox{\em Day} \mlc \mbox{\em Off})\mld (\mbox{\em Night}\mlc \mbox{\em On}).\end{equation}
Can the controller, with perfect knowledge of CL and without any other specific knowledge of the world (namely, without knowledge of the meanings of the atoms of (\ref{day})), successfully perform its job? With a little thought, this question can be seen to be equivalent to whether there is an effective winning strategy for 
(\ref{day}). Sure there is one: every time the environment switches from {\em Day} to {\em Night}, switch from {\em Off} to {\em On}; and every time the environment switches to the next $\scost$-component in the left conjunct of the antecedent, switch to the next $\sst$-component in the right conjunct.  In this example, 
the {\em resourcebase} of the planning agent is $\{\scost (\mbox{\em Day}\sqd \mbox{\em Night}), \sst (\mbox{\em Off}\sqc \mbox{\em On})\}$, and the {\em goal} task is $(\mbox{\em Day} \mlc \mbox{\em Off})\mld (\mbox{\em Night}\mlc \mbox{\em On})$. 

Abstract resource semantics, briefly discussed in Appendix A, potentially offers an alternative (similar but not the same) way of using the formalism of computability logic in planning systems.

\section{Formal definitions}\label{sdef}
Here we only provide formal definitions for sequential operations, because they have never been defined before. Strict definitions of the other game operations, as well as definitions of games and all related basic concepts can be found, for example, in \cite{Japfin}. In fact, in what follows we rely on \cite{Japfin} as an external source. Although long, the latter is very easy to read and has a convenient glossary to look up any unfamiliar terms and symbols. A reader not familiar with \cite{Japfin} or unwilling to do some parallel reading, may want to either stop here or just browse the rest of the paper without attempting to go into the technical details of formal definitions and proofs. Due to the very dynamic recent development, computability logic has already reached a point where it is no longer feasible to reintroduce all relevant concepts all over again in each new paper on the subject. 

We fix $\S$ as  a special-meaning symbol, and say that a run $\Gamma$ is {\bf presequential} iff every move of $\Gamma$ is either $\S$ (we call such moves {\bf switch moves},  or simply {\bf switches}) or $.\alpha$ (we call such moves {\bf non-switch moves}) for some string $\alpha$. Where $\xx\in\{\pp,\oo\}$, by the {\bf $\xx$-degree} of such a run we mean the number of switch moves made by player $\xx$ in it, if this number is finite; if $\Gamma$ has infinitely many switches by $\xx$, then we say that its $\xx$-degree is infinite. 
When $\Gamma=\seq{\Phi,\xx .\alpha,\Delta}$, by the {\bf degree} of the (indicated occurrence of the) non-switch labmove\footnote{Remember from \cite{Japfin} that {\em labmove} means ``labeled move'', i.e., a move prefixed with $\pp$ or $\oo$, with such a prefix indicating who has made the move. Terminologically we are not always strict about differentiating between moves and labmoves, and often say ``move'' where, strictly speaking, we should have said ``labmove''.} $\xx.\alpha$ we mean the $\xx$-degree of $\Phi$. Where $i\geq 0$, by $\Gamma^{\#i}$ we mean the result of deleting from $\Gamma$ all labmoves except the non-switch labmoves of degree $i$, and then further deleting 
the prefix ``$.$'' in each such labmove. For example, we have:
\[\begin{array}{ccl}
\seq{\pp .\alpha,\ \oo .\beta,\ \pp \S,\ \pp .\gamma, \ \oo .\delta,\ \oo \S,\ \pp .\sigma,\ \oo .\omega,\ \pp \S,\ \pp .\psi }^{\#0} & = & \seq{\pp\alpha,\oo\beta, \oo\delta};\\
\seq{\pp .\alpha,\ \oo .\beta,\ \pp \S,\ \pp .\gamma, \ \oo .\delta,\ \oo \S,\ \pp .\sigma,\ \oo .\omega,\ \pp \S,\ \pp .\psi }^{\#1} & = & \seq{\pp\gamma,\pp\sigma, \oo\omega};\\
\seq{\pp .\alpha,\ \oo .\beta,\ \pp \S,\ \pp .\gamma, \ \oo .\delta,\ \oo \S,\ \pp .\sigma,\ \oo .\omega,\ \pp \S,\ \pp .\psi }^{\#2} & = & \seq{\pp\psi};\\
\seq{\pp .\alpha,\ \oo .\beta,\ \pp \S,\ \pp .\gamma, \ \oo .\delta,\ \oo \S,\ \pp .\sigma,\ \oo .\omega,\ \pp \S,\ \pp .\psi }^{\#3} & = & \seq{}.\end{array}\]
 
\begin{definition}\label{nov10}
Let $A_0,\ldots,A_n$ ($n\geq 1$) be any constant games. We define the games $A_0\sqc\ldots\sqc A_n$ and $A_0\sqd\ldots\sqd A_n$ as follows:

\begin{enumerate}
\item  
\begin{itemize}
\item A position $\Phi$ is a legal position of $A_0\sqc\ldots\sqc A_n$ iff $\Phi$ is presequential, the $\oo$-degree of $\Phi$ does not exceed $n$, the $\pp$-degree of $\Phi$ or any of its initial segments does not exceed the $\oo$-degree of the same position and, for each $i\in \{0,\ldots,n\}$, $\Phi^{\#i}$ is a legal position of $A_i$.
\item Let $\Gamma$ be a legal run of $A_0\sqc\ldots\sqc A_n$, and $k$ be the $\oo$-degree of $\Gamma$. Then $\Gamma$ is  a $\oo$-won run of $A_0\sqc\ldots\sqc A_n$ iff $\Gamma^{\#k}$ is a $\oo$-won run of $A_k$.
\end{itemize}
\item  
\begin{itemize}
\item A position $\Phi$ is a legal position of $A_0\sqd\ldots\sqd A_n$ iff $\Phi$ is presequential, the $\pp$-degree of $\Phi$ does not exceed $n$, the $\oo$-degree of $\Phi$ or any of its initial segments does not exceed the $\pp$-degree of the same position and, for each $i\in \{0,\ldots,n\}$, $\Phi^{\#i}$ is a legal position of $A_i$.
\item Let $\Gamma$ be a legal run of $A_0\sqd\ldots\sqd A_n$, and $k$ be the $\pp$-degree of $\Gamma$. Then $\Gamma$ is  a $\pp$-won run of $A_0\sqc\ldots\sqc A_n$ iff $\Gamma^{\#k}$ is a $\pp$-won run of $A_k$.
\end{itemize}
\end{enumerate}
\end{definition}

Thus, whenever $\oo$ wants to switch from a given component $A_i$ to $A_{i+1}$ in $A_0\sqc\ldots\sqc A_n$, it makes the move $\S$. But, as we see, $\pp$, too, is expected to make switch moves in a $\sqc$-game to ``catch up" with $\oo$.\footnote{This arrangement is necessary to ensure that the sequential operators do not violate the static property of games.} The switches made by $\oo$ in a $\sqc$-game we call {\bf leading switches}, and the switches made by $\pp$ in a $\sqc$-game  we call {\bf catch-up switches}.  As for a non-switch move $.\alpha$ by either player $\xx$, its effect is making move $\alpha$ in $A_k$, where $k$ is the number of switch moves made by $\xx$ so far. $A_0\sqd\ldots\sqd A_n$, of course, is symmetric. Specifically, here it is $\pp$ who makes leading switches, while the switches by $\oo$ are catch-up switches. In either case, the number of leading switches cannot exceed $n$, and the number of catch-up switches cannot exceed the number of leading switches.

Intuitively, in a play (run) $\Gamma$ over a sequential combination of games, $\Gamma^{\#i}$ is the sequence of moves made within the $i$th component (starting the count from $0$ rather than $1$) of the combination. Each switch move by a player $\xx$ ``activates'' the next component for that player, in the sense that every subsequent non-switch move $.\alpha$ (until the next switch) by $\xx$ will {\bf signify} making move $\alpha$ in that component; we also say that the {\bf effect} of $.\alpha$ is making move $\alpha$ in the corresponding component. This intuitive and semiformal terminology, on which we will heavily rely in our further treatments, extends to more complex situations and other types of moves as well. Consider, for example, the game $(A\add B)\mlc \bigl((C\adc D)\sqc (E\adc F)\bigr)$ and the legal run $\seq{\pp 1.1, \oo 2.\S,\oo 2..2}$ of it. We say that:
\begin{itemize}
\item The effect of the move $1.1$ by $\pp$ is (or such a move signifies) choosing $A$ within the $A\add B$ component. Indeed, notice that after this move is made in 
$(A\add B)\mlc \bigl((C\adc D)\sqc (E\adc F)\bigr)$, the game is brought down to --- in the sense that it continues as --- $A\mlc \bigl((C\adc D)\sqc (E\adc F)\bigr)$.
\item The effect of the next move $2.\S$ by $\oo$ is switching from $C\adc D$ to $E\adc F$ in the  $(C\adc D)\sqc (E\adc F)$ component.
\item The effect of the last move $2..2$ by $\oo$ is choosing $F$ in the $(E\adc F)$ component. If this move was made before the switch, then its effect would be choosing $D$ in the $(C\adc D)$ component. 
\end{itemize}

Definition \ref{nov10} extends from finite cases to the infinite case as follows: 
\begin{definition}\label{nov101}
Let $A_0,A_1,A_2,\ldots$  be any constant games. We define the games $A_0\sqc A_1\sqc A_2\sqc\ldots$ and $A_0\sqd A_1\sqd A_2\sqd \ldots$ as follows:

\begin{enumerate}
\item  
\begin{itemize}
\item A position $\Phi$ is a legal position of $A_0\sqc A_1\sqc A_2\sqc \ldots$ iff $\Phi$ is presequential, the $\pp$-degree of $\Phi$ or any of its initial segments does not exceed the $\oo$-degree of the same position and, for each $i\geq 0$,  $\Phi^{\#i}$ is a legal position of $A_i$.
\item Let $\Gamma$ be a legal run of $A_0\sqc A_1\sqc A_2\sqc \ldots$. Then $\Gamma$ is  a $\oo$-won run of $A_0\sqc A_1\sqc A_2\sqc \ldots$ iff the $\oo$-degree of $\Gamma$ is finite and, where $k$ is that $\oo$-degree,  $\Gamma^{\#k}$ is a $\oo$-won run of $A_k$.
\end{itemize}
\item  
\begin{itemize}
\item A position $\Phi$ is a legal position of $A_0\sqd A_1\sqd A_2\sqd \ldots$ iff $\Phi$ is presequential, the $\oo$-degree of $\Phi$ or any of its initial segments does not exceed the $\pp$-degree of the same position and, for each $i\geq 0$,  $\Phi^{\#i}$ is a legal position of $A_i$.
\item Let $\Gamma$ be a legal run of $A_0\sqd A_1\sqd A_2\sqd \ldots$. Then $\Gamma$ is  a $\pp$-won run of $A_0\sqd A_1\sqd A_2\sqd \ldots$ iff the $\pp$-degree of $\Gamma$ is finite and, where $k$ is that $\pp$-degree,  $\Gamma^{\#k}$ is a $\pp$-won run of $A_k$.
\end{itemize}
\end{enumerate}
\end{definition}

Even though the above definitions officially define $\sqc$ and $\sqd$ only for constant games, they extend to all games in the standard way, as explained in the second paragraph of Section 4 of \cite{Japfin}. Specifically, for any not-necessarily-constant  games $A_0,\ldots,A_n$, $A_0\sqc\ldots\sqc A_n$ is the unique game such that, for any valuation (assignment of constants to variables) $e$, we have  $e[A_0\sqc\ldots\sqc A_n]=e[A_0]\sqc\ldots\sqc e[A_n]$. Similarly for $\sqd$ and the infinite cases of $\sqc,\sqd$. (The meaning of the notation $e[\ldots]$, just as the meanings of any other unfamiliar terms or notations, as already noted, can  and should be looked up in \cite{Japfin}.) 

The remaining sequential operations, as we already know from Section \ref{s2seq}, are defined as follows:
\begin{definition}\label{nov102}
For any games $A$ or $A(x)$:
\begin{enumerate}
\item $\sst A\ =\ A\sqc A\sqc A\sqc\ldots$
\item $\scost A\ =\ A\sqd A\sqd A\sqd\ldots$
\item $\sqa xA(x)\ =\ A(1)\sqc A(2)\sqc A(3)\sqc\ldots$
\item $\sqe xA(x)\ =\ A(1)\sqd A(2)\sqd A(3)\sqd\ldots$
\end{enumerate} 
\end{definition}

It is not hard to see that the DeMorgan dualities hold for the sequential operations, just as they do for all other groups of operations (parallel, choice, branching, blind). Namely, we have:
\[\begin{array}{l}
\gneg(A_0\sqc\ldots\sqc A_n)\ =\ \gneg A_0\sqd\ldots\sqd \gneg A_n,\\
\gneg(A_0\sqd\ldots\sqd A_n)\ =\ \gneg A_0\sqc\ldots\sqc \gneg A_n,
\end{array}\]
and similarly for infinite sequential conjunctions and disjunctions, including $\sst A$, $\scost A$, $\sqa xA(x)$, $\sqe xA(x)$.

Whenever new game operations are introduced, one needs to make sure that they preserve the static property of games, for otherwise many things can go wrong: 
\begin{theorem}\label{static}
The class of static games is closed under our sequential operations.
\end{theorem}
\begin{proof} Given in Appendix B. \end{proof}

\section{Logic $\propseq$}\label{intr}
In this section we introduce the propositional system $\propseq$.  The building blocks of its language are:
\begin{itemize}
\item Infinitely many nonlogical {\bf elementary atoms}, for which we use the metavariables $p,q,r,s$;
\item Infinitely many nonlogical {\bf general atoms}, for which we use the metavariables $P,Q,R,S$;
\item The 0-ary operators $\twg$ and $\tlg$. They can as well be called {\bf logical atoms};
\item The unary operator $\gneg$;
\item The operators $\mlc,\mld,\adc,\add,\sqc,\sqd$. Their arities are not fixed and can be any $n\geq 2$.   
\end{itemize}

{\bf Formulas}, to which we refer as {\bf $\propseq$-formulas}, are built from atoms and operators in the standard way, with the requirement (yielding no loss of expressiveness) that $\gneg$ can only be applied to nonlogical atoms. A {\bf literal} means $L$ or $\gneg L$, where $L$ is an atom. Such a literal is said to be elementary, general, nonlogical or logical if $L$ is so. When $F$ is not a nonlogical atom, $\gneg F$ is understood as an abbreviation defined by:
\[\begin{array}{rcl}
\gneg \twg & = & \tlg\\
\gneg \tlg & = & \twg\\ 
\gneg\gneg E & = & E\\
\gneg(E_1\mlc\ldots\mlc E_n) & = & \gneg E_1\mld\ldots\mld \gneg E_n\\
\gneg(E_1\mld\ldots\mld E_n) & = & \gneg E_1\mlc\ldots\mlc \gneg E_n\\  
\gneg(E_1\adc\ldots\adc E_n) & = & \gneg E_1\add\ldots\add \gneg E_n\\
\gneg(E_1\add\ldots\add E_n) & = & \gneg E_1\adc\ldots\adc \gneg E_n\\  
\gneg(E_1\sqc\ldots\sqc E_n) & = & \gneg E_1\sqd\ldots\sqd \gneg E_n\\  
\gneg(E_1\sqd\ldots\sqd E_n) & = & \gneg E_1\sqc\ldots\sqc \gneg E_n
\end{array}\]
Also, if we write $E\mli F$, it is to be understood as an abbreviation of $\gneg E\mld F$.  

The formulas that do not contain elementary nonlogical atoms we call {\bf general-base}, and the formulas that do not contain general atoms 
we call {\bf elementary-base}. This terminology also extends to the corresponding two fragments of $\propseq$; in particular,  the {\em general-base fragment} of $\propseq$ is the set of all general-base theorems of $\propseq$, and the {\em elementary-base fragment} of $\propseq$ is the set of all elementary-base theorems of $\propseq$. 

An {\bf interpretation}\label{z6} for the language of $\propseq$ a function that sends each nonlogical elementary atom  to an elementary game, and sends each general atom to any, not-necessarily-elementary, static game. This mapping extends to all formulas by letting it respect all logical operators as the corresponding game operations. That is, $\twg^*=\twg$, $(E\sqc F)^*=E^*\sqc F^*$, etc. When $F^*=A$, we say that {\bf $^*$ interprets $F$ as $A$}.

A formula $F$ is said to be {\bf valid} iff, for every interpretation $^*$, the game $F^*$ is computable. And $F$ is {\bf uniformly valid} iff there is an HPM $\cal H$, called a {\bf uniform solution} for $F$, such that $\cal H$ wins (computes) $F^*$ for every interpretation $^*$. 

A {\bf sequential (sub)formula} is one of the form $F_0\sqc\ldots\sqc F_n$ or $F_0\sqd\ldots\sqd F_n$. We say that $F_0$ is the {\bf head} of such a (sub)formula, 
and $F_1,\ldots, F_n$ form its {\bf tail}. 

The {\bf capitalization} of a formula is the result of replacing in it every sequential subformula by its head.
 
 A formula is said to be  {\bf elementary} iff it is a formula of classical propositional logic, i.e., contains no general atoms and no operators other than $\twg,\tlg,\gneg,\mlc,\mld$.  

An occurrence of a subformula in a formula is {\bf positive} iff it is not in the scope of $\gneg$. Otherwise it is {\bf negative}. According to our conventions regarding the usage of $\gneg$, only atoms may have negative occurences.
 
A {\bf surface occurrence} is an occurrence that is not in the scope of a choice connective and not in the tail of any sequential subformula. 

The {\bf elementarization} of a $\propseq$-formula $F$ means the result of replacing in the capitalization of $F$ every surface occurrence of the form $G_1\adc\ldots\adc G_n$ by $\twg$, every surface occurrence of the form $G_1\add\ldots\add G_n$ by $\tlg$, and every positive surface occurrence of each general literal by $\tlg$. 

Finally,  a formula is said to be {\bf stable}\label{z14} iff its elementarization is a classical tautology; otherwise it is {\bf instable}.\vspace{10pt}

\begin{definition}\label{defcl9}
With ${\cal P}\mapsto F$ meaning ``from premise(s) $\cal P$ conclude $F$'', $\propseq$ is the system given by the following four rules of inference:\footnote{There are no axioms, but the rule of Wait can act as such when the set of its premises is empty.} 

\begin{description}
\item[Wait:]  $\vec{H}\mapsto F$, where $F$ is stable and $\vec{H}$ is the smallest set of formulas satisfying the following two conditions: 
\begin{enumerate} 
\item whenever $F$ has a surface occurrence of a subformula $G_1\adc\ldots\adc G_n$, for each 
$i\in\{1,\ldots,n\}$, $\vec{H}$ contains the result of replacing that occurrence in $F$ by $G_i$;
\item whenever $F$ has a surface occurrence of a subformula $G_0\sqc G_1\sqc\ldots\sqc G_n$, $\vec{H}$ contains the result of replacing that occurrence in $F$ by $G_1\sqc\ldots\sqc G_n$.\footnote{In this definition, if $n=1$, $G_1\sqc\ldots\sqc G_n$ or $G_1\sqd\ldots\sqd G_n$ is simply understood as $G_1$.} 
\end{enumerate}
\item[Choose:]  $H\mapsto F$, where $H$ is the result of replacing in $F$ a surface occurrence of a subformula $G_1\add\ldots\add G_n$  by $G_i$ for some $i\in\{1,\ldots, n\}$.
\item[Switch:]  $H\mapsto F$, where $H$ is the result of replacing in $F$ a surface occurrence of a subformula $G_0\sqd G_1\sqd\ldots\sqd G_n$  by 
$G_1\sqd\ldots\sqd G_n$.$^8$
\item[Match:] $H\mapsto F$, where $H$ is the result of replacing in $F$ two --- one positive and one negative ---
surface occurrences of some general atom by a nonlogical elementary atom that does not occur in $F$.\vspace{10pt}
\end{description}
\end{definition}

\begin{example}\label{oct17} The following is a $\propseq$-proof of $P\add Q \mli P\sqd Q$:\vspace{3pt}

$\begin{array}{ll}
1.\ \gneg q \mld q & \mbox{(from $\{\}$ by Wait)};\\
2.\ \gneg Q \mld Q & \mbox{(from 1 by Match)};\\
3.\ \gneg Q \mld (P\sqd Q) & \mbox{(from 2 by Switch)};\\
4.\ \gneg p \mld (p\sqd Q) & \mbox{(from $\{\}$ by Wait)};\\
5.\ \gneg P \mld (P\sqd Q) & \mbox{(from 4 by Match)};\\
6.\ (\gneg P\adc \gneg Q) \mld (P\sqd Q) & \mbox{(from \{3,5\} by Wait)}.
\end{array}$\vspace{3pt}

On the other hand, $P\sqd Q \mli P\add Q$ is not provable. Indeed,  $(\gneg P\sqc \gneg Q) \mld (P\add Q)$ is instable, so Wait cannot be used at the last step of its derivation. It contains no $\sqd$, so Switch cannot be used, either. And it has only one surface occurrence of an atom, so Match does not apply. This leaves us with Choose. Then the premise is  $(\gneg P\sqc\gneg Q) \mld P$ or $(\gneg P\sqc\gneg Q) \mld Q$. In either case, Choose no longer applies. If we are dealing with $(\gneg P\sqc\gneg Q) \mld Q$, evidently the other three rules do not apply either. And if we are dealing with  $(\gneg P\sqc\gneg Q) \mld P$, then only Match applies, which takes us to the premise  $(\gneg p\sqc\gneg Q) \mld p$. This formula could only be the conclusion of Wait, where the set of premises consists of  
$\gneg Q \mld p$. Now we are stuck with this premise, as it cannot be derived by any of the four rules of $\propseq$.

With about an equal amount of effort, where $\propseq\vdash F$ means ``$F$ is provable in $\propseq$'' and $\propseq\not\vdash F$ means ``$F$ is not provable in $\propseq$'', one can further verify that:
\begin{itemize}
\item $\propseq\vdash P\sqd Q\mli P\mld Q$;
\item $\propseq\not\vdash P\mld Q\mli P\sqd Q$.
\end{itemize}
\end{example}

\begin{example}\label{jan11} The following is a $\propseq$-proof of $(P\mlc Q)\mld (\gneg P\sqc \gneg R)\mld (\gneg Q\sqc \gneg S)\mld (R\add S)$:\vspace{3pt}

$\begin{array}{ll}
1.\ (p\mlc q)\mld (\gneg p\sqc \gneg R)\mld \gneg s\mld s & \mbox{(from $\{\}$ by Wait)};\\
2.\ (p\mlc q)\mld (\gneg p\sqc \gneg R)\mld \gneg S\mld S & \mbox{(from 1 by Match)};\\
3.\ (p\mlc q)\mld (\gneg p\sqc \gneg R)\mld \gneg S\mld (R\add S) & \mbox{(from 2 by Choose)};\\
4.\ (p\mlc q)\mld \gneg r \mld (\gneg q\sqc \gneg S)\mld r & \mbox{(from $\{\}$ by Wait)};\\
5.\ (p\mlc q)\mld \gneg R \mld (\gneg q\sqc \gneg S)\mld R & \mbox{(from 4 by Match)};\\
6.\ (p\mlc q)\mld \gneg R \mld (\gneg q\sqc \gneg S)\mld (R\add S) & \mbox{(from 5 by Choose)};\\
7.\ (p\mlc q)\mld (\gneg p\sqc \gneg R) \mld (\gneg q\sqc \gneg S)\mld (R\add S) & \mbox{(from \{3,6\} by Wait)};\\
8.\ (p\mlc Q)\mld (\gneg p\sqc \gneg R)\mld (\gneg Q\sqc \gneg S)\mld (R\add S) & \mbox{(from 7 by Match)};\\
9.\ (P\mlc Q)\mld (\gneg P\sqc \gneg R)\mld (\gneg Q\sqc \gneg S)\mld (R\add S) & \mbox{(from 8 by Match)}.
\end{array}$
\end{example}

\begin{exercise} Verify that:\label{jan31}

1. $\propseq\vdash P\mld\gneg P$ 

2. $\propseq\not\vdash P\add\gneg P$ 

3. $\propseq\not\vdash P\sqd\gneg P$

4. $\propseq\not\vdash P\mli P\mlc P$ 

5. $\propseq\vdash P\mli P\adc P$ 

6. $\propseq\not\vdash P\mli P\sqc P$ 

7. $\propseq\vdash P\mlc Q\mli Q\mlc P$

8. $\propseq\vdash P\adc Q\mli Q\adc P$

9. $\propseq\not\vdash P\sqc Q\mli Q\sqc P$ 

\end{exercise}

\begin{exercise}\label{jan32}
Check the $\propseq$-provability status of the formulas of Exercise \ref{jan31} with $p,q$ instead of $P,Q$. Where are you getting differences?
\end{exercise}

\begin{exercise}\label{jan327}
Construct $\propseq$-proofs of formulas (\ref{nov16b}) and (\ref{nov16c}) of Section \ref{s2seq}. Try to extract winning strategies from your proofs.
\end{exercise}

Below comes our main theorem. It is simply the later-proven Lemmas \ref{sound} and \ref{compl} put together:
 
\begin{theorem}\label{thcl2}
$\propseq\vdash F$ iff $F$ is valid $($any $\propseq$-formula $F$$)$.
Furthermore:

a) There is an effective procedure that takes a $\propseq$-proof of an arbitrary formula $F$ and 
constructs an HPM $\cal H$ such that,  for every interpretation $^*$, \ $\cal H$ computes $F^*$.

b) If $\propseq\not\vdash F$, then $F^*$ is not computable  for some  interpretation 
$^*$ that interprets all elementary atoms of $F$ as finitary predicates of arithmetical complexity $\Delta_2$, and interprets all general atoms of $F$ as problems of the form
\((A^{1}_{1}\add\ldots\add A_{m}^{1})\adc\ldots\adc (A_{1}^{m}\add\ldots\add A_{m}^{m}),\)
where each $A_{i}^{j}$ is a finitary predicate of arithmetical complexity $\Delta_2$.
\end{theorem}

The following facts are immediate corollaries of Theorem \ref{thcl2}, so we state them without proofs:

\begin{fact}\label{fact1}
A $\propseq$-formula is valid iff it is uniformly valid.
\end{fact}

\begin{fact}\label{fact2}
$\propseq$ is a conservative extension of classical logic. That is, an elementary formula is provable in $\propseq$ iff it is a classical tautology. 
\end{fact}

It is also worth noting that $\propseq$ is decidable, with a brute force decision algorithm 
obviously running in at most polynomial space. Whether there are more efficient algorithms is unknown.

\section{Preliminaries for the soundness proof}\label{shf}
Our proof of Theorem \ref{thcl2} starts here and ends in Section \ref{s9}. It closely follows the soundness and completeness proof given in \cite{Japtocl1,Japtocl2} for the less expressive, $\sqc,\sqd$-free logic {\bf CL2}, but the style here is much more relaxed and schematic, often relying on arguments such as ``with some thought, we can see that...'' in places where \cite{Japtocl1,Japtocl2} would provide a full technical elaboration of such ``some thought''.  The present section contains certain necessary 
preliminaries for the soundness part of the proof.

\subsection{Hyperformulas}
In the bottom-up (from conclusion to premises) view, the Match rule introduces two occurrences of some new 
nonlogical elementary atom. For technical convenience, we want to differentiate elementary atoms introduced this way
from all other elementary atoms, and also somehow keep track of the exact origin of each such elementary atom $q$ --- that is, remember what general atom $P$ was replaced by $q$ when Match was applied. For this purpose, 
we extend the language of $\propseq$ by  adding to it a new sort of nonlogical atoms, called  {\bf hybrid}.\label{z3} In particular, each hybrid atom is a pair consisting of a general atom $P$, called its {\bf general component}, 
and a nonlogical elementary atom $q$, called its {\bf elementary component}. We denote such a pair by $P_q$. 
As we are going to see later, the presence of $P_q$ in a (modified $\propseq$-) proof will be an indication of the fact that, in the bottom-up 
view of proofs, $q$ has been introduced by Match and that, when this happened, the general atom that $q$ replaced was $P$. 

For similar reasons, we do not want to fully forget the earlier components of sequential subformulas when Switch or Wait are applied. Hence,   
we further modify the language of $\propseq$ by requiring that, in every sequential (sub)formula, one of the components be {\bf underlined}. With such 
formulas, applying Match (in the bottom up view) simply moves the underline to the next component of a $\sqd$-subformula without deleting the ``abandoned'' component as was the case in $\propseq$.  Similarly for the effect of (clause 2 of) Wait on $\sqc$-subformulas. The role of an underline is thus to indicate which component of the sequential subformula would be the head of the corresponding subformula in the corresponding $\propseq$-proof. 

The formulas of this extended and modified language we call {\bf hyperformulas}. We understand each $\propseq$-formula $F$ as the hyperformula (and identify $F$ with such) obtained from  $F$ by underlining the first component of every sequential subformula. $\propseq$-formulas are thus special cases of hyperformulas where the underlined component of a sequential subformula is always the leftmost component, and where no hybrid atoms are present. 

By the {\bf general dehybridization} of a hyperformula $F$ we mean the $\propseq$-formula that results from $F$ by replacing in the latter every hybrid atom by its general component, and  removing all underlines in sequential subformulas. Where $^*$ is an interpretation and $F$ is a hyperformula, we define the game \[F^*\] as $G^*$, where $G$ is the general dehybridization of $F$. 
Extending the earlier-established lingo to hyperformulas, for a hyperformula $F$ and an 
interpretation $^*$, whenever $F^*=A$, we say that $^*$ {\bf interprets} $F$ as $A$.

By a {\bf surface occurrence} of a subexpression in a given hyperformula $F$ we mean an occurrence that is not in the scope of a choice operator, such that, if the subexpression occurs within a component of a sequential subformula, that component is underlined or occurs earlier than (is to the left of) the underlined component.  

An {\bf active occurrence} is an occurrence such that, whenever it happens to be within a component of a sequential subformula, that component is underlined. If an occurrence is within a component of a sequential subformula which (the component) is to the left of the underlined component of the same subformula,  then we say that the occurrence is {\bf abandoned}.

The terms {\bf positive occurrence} and {\bf negative occurrence} have the same meanings for hyperformulas as for $\propseq$-formulas. 
 
As in the case of $\propseq$-formulas, an {\bf elementary hyperformula} is one not containing choice and sequential operators, underlines, and general and hybrid atoms. Thus,  `elementary hyperformula' and `elementary $\propseq$-formula'
(as well as `formula of classical propositional logic') mean the same. 

We define the {\bf capitalization} of a hyperformula $F$ as the result of replacing in it every sequential subformula by its underlined component, after which all underlines are removed. 

The {\bf elementarization} \[\elz{F}\label{z25}\] of a hyperformula $F$ is 
the result of replacing, in the capitalization of $F$, every surface occurrence of the form $G_1\adc\ldots\adc G_n$ by $\twg$, every surface occurrence of the form $G_1\add\ldots\add G_n$
by  $\tlg$, every  positive surface occurrence of each general literal by $\tlg$, and every surface occurrence 
of each hybrid atom by the elementary component of that atom. 

As in the case of $\propseq$-formulas, we say that a hyperformula $F$ is {\bf stable} 
iff its elementarization $\elz{F}$ is a classical tautology; otherwise it is {\bf instable}.

A hyperformula $F$ is said to be {\bf balanced} iff,  for every hybrid atom $P_q$ occurring in $F$, the following two conditions are satisfied: 
\begin{enumerate}
\item $F$ has exactly two occurrences of $P_q$, where one occurrence is positive and the other occurrence is negative, and both occurrences are surface occurrences;
\item the elementary atom $q$ does not occur in $F$, nor is it the elementary component of any hybrid atom occurring in $F$ other than $P_q$.  
\end{enumerate}

We say that an active occurrence of a hybrid atom (or the corresponding literal) in a balanced hyperformula is {\bf widowed} iff the other occurrence of the same hybrid atom is abandoned.

\subsection{Logic $\propseqc$}
In our soundness proof for $\propseq$ we will employ a ``version" of $\propseq$ called $\propseqc$. Unlike $\propseq$ whose language consists only of formulas, the language of $\propseqc$ allows any balanced hyperformulas, which we also refer to as {\bf $\propseqc$-formulas}. 

\begin{definition}\label{nov23}
Logic $\propseqc$ is given by the following rules for balanced hyperformulas (below simply referred to as ``(sub)formulas''): 
\begin{description}
\item[Wait$^\circ$:]  $\vec{H}\mapsto F$, where $F$ is stable and $\vec{H}$ is the smallest set of formulas satisfying the following two conditions: 
\begin{enumerate} 
\item whenever $F$ has an active surface occurrence of a subformula $G_1\adc\ldots\adc G_n$, for each 
$i\in\{1,\ldots,n\}$, $\vec{H}$ contains the result of replacing that occurrence in $F$ by $G_i$;
\item whenever $F$ has an active surface occurrence of a subformula $G_0\sqc \ldots\sqc \underline{G_m}\sqc G_{m+1}\sqc\ldots\sqc G_n$, $\vec{H}$ contains the result of replacing that occurrence in $F$ by $G_0\sqc \ldots\sqc G_m\sqc \underline{G_{m+1}}\sqc\ldots\sqc G_n$.
\end{enumerate}
\item[Choose$^\circ$:]  $H\mapsto F$, where $H$ is the result of replacing in $F$ an active surface occurrence of a subformula $G_1\add\ldots\add G_n$  by $G_i$ for some $i\in\{1,\ldots, n\}$.
\item[Switch$^\circ$:]  $H\mapsto F$, where $H$ is the result of replacing in $F$ an active surface occurrence of a subformula 
\(G_0\sqd\ldots\sqd \underline{G_m}\sqd G_{m+1}\sqd\ldots\sqd G_{n}\)  
by \(G_0\sqd\ldots\sqd {G_m}\sqd \underline{G_{m+1}}\sqd\ldots\sqd G_{n}.\)
\item[Match$^\circ$:] $H\mapsto F$, where $H$ has two --- a positive and a negative --- active surface occurrences of some hybrid atom $P_q$, and $F$ is the result of replacing in $H$ both occurrences by $P$.\vspace{10pt}  
\end{description}
\end{definition}

\begin{lemma}\label{fdc}
For any $\propseq$-formula $G$, \ if $\propseq\vdash G$, then $\propseqc\vdash G$.

Furthermore, there is an effective procedure that converts any $\propseq$-proof of any formula $G$ into a $\propseqc$-proof of $G$. 
\end{lemma}

\begin{proof} 
Consider any $\propseq$-proof tree $T$ for $G$, i.e. a tree every node of which is labeled with a $\propseq$-formula that follows by one of the rules of $\propseq$ from the set of (the labels of) its children, with $G$ sitting at the root. For safety and without loss of generality, we assume that, in the bottom-up view of this proof, Match never introduces an elementary atom that had occurrences in some earlier formulas (but such occurrences later disappeared due to cutting off the heads of sequential subformulas).

We modify $T$ as follows. 
First, we underline the heads of all sequential subformulas of all formulas of $T$. 
Next, for each node $F$ of the tree that is derived from its child $H$ by Match --- in particular, where $H$ is the result of replacing in $F$ a positive and a negative active surface occurrences of a general atom $P$ by a nonlogical elementary atom $q$ --- we replace both occurrences of $q$ by the hybrid atom $P_q$ in $H$ as well as in all of its descendants in the tree. In other words, we turn each application of Match into the corresponding application of Match$^\circ$.  
Next, we similarly turn each application of Switch into an application of Switch$^\circ$. That is, we modify Switch so that this rule, when moving from a conclusion $F$ to the premise (child) $H$, simply moves the underline from a given component to the next component without otherwise deleting any components in the corresponding $\sqd$-subformula (and, of course, the undeleted old components should also be added to the corresponding $\sqd$-subformulas in all descendants of $H$ in the tree). Finally, we do the same for Wait whenever it modifies a $\sqc$-subformula.  It is not hard to see that the resulting tree $T^\circ$ is a $\propseqc$-proof of $G$. 
\end{proof}

\subsection{Perfect interpretations}

An interpretation $^*$ is said to be {\bf perfect} iff it interprets every atom as 
a constant game. All of our game operations can be easily seen to preserve the constant property of games (for the non-sequential operations, this was officially established in \cite{Jap03}, Theorem 14.1), which means that perfect interpretations interpret all (hyper)formulas as constant games. This fact may be worth marking as we will often implicitly rely on it.  
For an interpretation $^*$ and valuation $e$, the {\bf perfect interpretation induced by $(^*,e)$} is the interpretation $^\dagger$ that interprets each (elementary or general) atom $L$ as the constant game $e[L^*]$.

\begin{lemma}\label{perf}
Assume $F$ is any $\propseqc$-formula, $e$ any valuation, $^*$ any interpretation and $^\dagger$ the perfect interpretation induced by $(^*,e)$. Then $e[F^*]=F^\dagger$.
\end{lemma}

\begin{proof} Induction on the complexity of $F$. For an atomic $F$, $e[F^*]=F^\dagger$ is immediate. And 
the inductive step is also straightforward, taking into account that the operations $e[\ldots]$,  $^*$,  $^\dagger$ 
commute with $\gneg,\mlc,\mld,\adc,\add,\sqc,\sqd$. \end{proof}

\subsection{Manageability}
We continue using our informal jargon introduced in Section \ref{sdef}. Let us further agree, context permitting, to see no distinction between (hyper)formulas and the games that they would represent once some interpretation was applied to them. We can say, for example, that ``$\pp$ makes a choice move within the subformula 
$E\add F$ of $(E\add F)\mlc G$'' to mean that $\pp$ makes the move $1.1$ or $1.2$ (thus choosing $E$ or $F$ in the first $\mlc$-conjunct of the formula/game). 
For terminological simplicity, we may also not always be careful about distinguishing between a subformula and a particular occurrence of it and, for example,  say ``active (surface, etc.) subformula $E$ of $F$'' instead of ``active occurrence of $E$ in $F$''. When we use such terminology, we usually have some run $\Gamma$ in mind --- the run about the moves of which we are talking. Let us call such a run the {\bf contextual run}. 
  
\begin{definition}\label{qint}
Let $F$ be a $\propseqc$-formula. We say that a run $\Gamma$ is  {\bf $F$-manageable} iff, with $\Gamma$ being the contextual run,  the following five conditions are satisfied:
\begin{enumerate}
\item No (choice) moves have been made within active choice subformulas.
\item If $E_0\sqc\ldots\sqc \underline{E_{i}}\sqc\ldots\sqc E_n$ is an active subformula of $F$, both of the players have made exactly $i$ switches in this subformula.\footnote{According to our terminological conventions, making moves in (a given occurrence of) subformula $E_0\sqc\ldots\sqc E_n$ means making moves in the (sub)game represented by this (sub)formula. Note that such a (sub)game does not depend on which of the components of the formula is underlined. So, in the present context the phrase ``this subformula'' should be understood as referring to $E_0\sqc\ldots\sqc E_n$ without regard to where the underline goes. The same applies to clause 3 of the definition. And a similar comment should be made for clause 5, where ``within \ldots $P_q$'' can or should be understood as ``within \ldots $P$'', as the game represented by $P_q$ does not depend on $q$ or the presence/absence of it.}
\item If $E_0\sqd\ldots\sqd \underline{E_{i}}\sqd\ldots\sqd E_n$ is an active subformula of $F$, $\pp$ has made exactly $i$ switches in this subformula, and $\oo$ has made $\leq i$ switches. 
\item No moves have been made by $\pp$ within general atoms. 
\item If, for some hybrid atom $P_q$ both of whose occurrences are active,  $\Sigma^+$ and $\Sigma^-$ are the sequences of the moves that have been made within the positive and the negative occurrence of $P_q$ in $F$, respectively, then $\Sigma^+$ is a $\pp$-delay (see \cite{Japfin}, Section 5) of $\rneg\Sigma^-$.
\end{enumerate}  
\end{definition}

The above technical concept will play a central role in our soundness proof for $\propseq$. The main intuition here is that, when $\Gamma$ is $F$-manageable, playing it in no way affects/modifies the choice subgames of the game (clause 1), guarantees that at any time the underline in a sequential subformula accurately indicates the number of leading switches made therein (clauses 2 and 3),  lets the subgames in the ``matched" occurrences of 
atoms evolve to --- in a sense --- the same games  (clause 5), and makes sure that $\pp$ does not make any hasty moves in unmatched atoms (clause 4), so that, if and when at some later point such an atom finds a match, $\pp$ will still have a chance to ``even out" the corresponding two subgames.   

\begin{lemma}\label{feb21}
Let $E$ be any $\propseqc$-formula, $^*$ any perfect interpretation, and $\Gamma$ an infinite run 
with arbitrarily long finite initial segments that are $E$-manageable legal positions of $E^*$. Then $\Gamma$ is an $E$-manageable legal run of $E^*$.
\end{lemma}

\begin{proof} Assume the conditions of the lemma. They imply that every finite initial segment of $\Gamma$ is a legal position 
of $E^*$, which (by the definition of ``legal run'') means that $\Gamma$ is a legal run of $E^*$. Also, obviously $\Gamma$ satisfies conditions 1, 2, 3  and 4 of Definition \ref{qint} because it has arbitrarily long initial segments that satisfy   those conditions. So, what remains to show is that $\Gamma$ also satisfies condition 5 of Definition \ref{qint}.

Suppose, for a contradiction, that this is not the case. In particular, let $\Sigma^+$, $\Sigma^-$ and $P_q$ be as described in the antecedent of condition 5, and 
assume that $\Sigma^+$ is not a $\pp$-delay of $\rneg\Sigma^-$. This means that at least one of the following two statements is true:

\begin{description}
\item[(i)] For one of the players $\xx$, the subsequence of the $\xx$-labeled moves of $\Sigma^+$ 
(i.e. the result of deleting in $\Sigma^+$ all $\pneg \xx$-labeled moves) is not the same as that 
of $\rneg\Sigma^-$, or
\item[(ii)] For some $k,n$, in $\rneg\Sigma^-$ the $n$th $\pp$-labeled move is made later than the $k$th $\oo$-labeled move, but in $\Sigma^+$ the $n$th $\pp$-labeled move is made earlier than the $k$th $\oo$-labeled move.
\end{description}

Whether (i) or (ii) is the case, it is not hard to see that, beginning from some (finite) $m$,
every initial segment $\Psi$ of $\Gamma$ of length $\geq m$ will satisfy the same  (i) or (ii) in the role 
of $\Gamma$, and hence $\Psi$ will not be an $E$-manageable position of $E^*$. This contradicts the assumptions of our
lemma. \end{proof}

\begin{lemma}\label{may19}
Assume $A$ is a constant static game, $\xx$ is either player, and $\Gamma,\Delta$ are runs such that $\Delta$ is a $\xx$-delay of $\Gamma$. Then:
\begin{enumerate}
\item If $\Delta$ is a $\xx$-illegal run of $A$, then so is $\Gamma$.
\item If $\Gamma$ is a $\pneg\xx$-illegal run of $A$, then so is $\Delta$. 
\end{enumerate}
\end{lemma}
\begin{proof} The above is a fact known from \cite{Jap03} (Lemma 4.7).
\end{proof}

Below and later, by saying ``$\xx$ makes move $\alpha$ in position $\Omega$'' we mean that such a move is appended to $\Omega$ as a new move. This creates the new position $\seq{\Omega,\xx\alpha}$, which serves as the contextual run when talking about the effects of $\alpha$ or other, earlier moves of the position.

\begin{lemma}\label{splm}
In each of the following clauses, we assume that $E$ is a $\propseqc$-formula, $^*$ is a perfect interpretation,  and
$\Omega$ is an $E$-manageable legal position of $E^*$.\vspace{5pt}

1. Suppose $\pp$ makes a move $\alpha$ in position $\Omega$, 
whose effect is choosing the $i${\em th} disjunct in an active surface occurrence of a subformula  $G_1\add\ldots\add G_n$ in $E$. Of course, such a move is legal.\hspace{1pt}\footnote{Here and in the subsequent clauses, ``legal'' should be understand with respect to $E^*$. That is, ``$\alpha$ is a legal move made by $\xx$ (in position $\Omega$)'' means nothing but that  $\seq{\Omega,\xx\alpha}$ is a legal position of $E^*$.}
Let  $H$ be the result of replacing in $E$ the above occurrence by $G_i$. Then
$\seq{\Omega}$ is an $H$-manageable legal position of $H^*$, and $\seq{\Omega,\pp\alpha}E^*=\seq{\Omega}H^*$.\vspace{5pt}

2. Suppose $\pp$ makes a move $\alpha$ in position $\Omega$,  whose effect is making a (leading) switch in an active surface occurrence\footnote{Here and in the subsequent clauses, $E$ is the contextual formula. Specifically, by simply saying ``occurrence'', we mean occurrence in $E$.} of a subformula  $G_0\sqd\ldots\sqd \underline{G_i}\sqd G_{i+1}\sqd\ldots\sqd G_n$. Of course, such a move is legal.
Let  $H$ be the result of replacing in $E$ the above occurrence by $G_0\sqd\ldots\sqd G_i\sqd \underline{G_{i+1}}\sqd\ldots\sqd G_n$. Then 
$\seq{\Omega,\pp\alpha}$ is an $H$-manageable legal position of $H^*$, and 
$\seq{\Omega,\pp\alpha}E^*=\seq{\Omega,\pp\alpha}H^*$.\vspace{5pt}

3. Suppose $H$ is a $\propseqc$-formula, and it results 
from $E$ by replacing in it a positive active surface occurrence $O^+$ and a negative active surface occurrence $O^-$ of a general atom $P$ by a hybrid atom $P_q$. 
Further assume that $\pi_1,\ldots,\pi_n$ and $\nu_1,\ldots,\nu_m$ are the sequences of moves made so far (during playing $\Omega$, that is) by $\oo$ within $O^+$ and $O^-$, respectively. Let $\Omega'$ be the result of adding to $\Omega$ $m$ moves by $\pp$ whose effects are making the moves $\nu_1,\ldots,\nu_m$ in $O^+$, and further adding to the resulting position $n$ moves by $\pp$ whose effects are making the moves $\pi_1,\ldots,\pi_n$ in $O^-$.
Then $\Omega'$ is an $H$-manageable legal position of $H^*=E^*$.\vspace{5pt}

4. Suppose $\oo$ makes a legal move $\alpha$ in position $\Omega$, 
whose effect is moving within some abandoned occurrence of a subformula or a widowed occurrence of a hybrid literal.  Then $\seq{\Omega,\oo\alpha}$ remains 
an $E$-manageable legal position of $E^*$.\vspace{5pt}

5. Suppose $\oo$ makes a legal move $\alpha$ in position $\Omega$, whose effect is moving in some active surface occurrence of a general atom.  Then $\seq{\Omega,\oo\alpha}$ remains 
an $E$-manageable legal position of $E^*$.\vspace{5pt}

6. Suppose $\oo$ makes a legal move $\alpha$ in position $\Omega$, whose effect is making a (catch-up) switch in an active surface occurrence of a subformula  $G_0\sqd\ldots\sqd \underline{G_i}\sqd G_{i+1}\sqd\ldots\sqd G_n$.  Then
$\seq{\Omega,\oo\alpha}$ remains an $E$-manageable legal position of $E^*$.\vspace{5pt}

7. Suppose $\oo$ makes a legal move $\alpha$ in position $\Omega$, whose effect is making a move $\gamma$ within an active surface non-widowed occurrence of a hybrid atom. Let $\beta$ be the move by $\pp$ whose effect is making the same move $\gamma$ within the other active surface occurrence of the same hybrid atom. Then 
$\seq{\Omega,\oo\alpha,\pp\beta}$ remains an $E$-manageable legal position of $E^*$.\vspace{5pt}

8. Suppose $\oo$ makes a legal move $\alpha$ in position $\Omega$, whose effect is choosing the $i${\em th} conjunct in an active surface occurrence of a subformula  $G_1\adc\ldots\adc G_n$. 
Let  $H$ be the result of replacing in $E$ the above occurrence by $G_i$. Then
$\seq{\Omega}$ is an $H$-manageable legal position of $H^*$, and 
$\seq{\Omega,\oo\alpha}E^*=\seq{\Omega}H^*$.\vspace{5pt}

9. Suppose $\oo$ makes a legal move $\alpha$ in position $\Omega$, whose effect is making a (leading) switch in an active surface occurrence of a subformula  $G_0\sqc\ldots\sqc \underline{G_i}\sqc G_{i+1}\sqc\ldots\sqc G_n$. 
Let  $H$ be the result of replacing in $E$ the above occurrence by $G_0\sqc\ldots\sqc G_i\sqc \underline{G_{i+1}}\sqc\ldots\sqc G_n$. Then 
$\seq{\Omega,\oo\alpha,\pp\alpha}$ is an $H$-manageable legal position of $H^*$, and 
$\seq{\Omega,\oo\alpha,\pp\alpha}E^*=\seq{\Omega,\oo\alpha,\pp\alpha}H^*$.\vspace{5pt}

10. If $\oo$ makes a legal move $\alpha$ in position $\Omega$, then it satisfies the conditions of one of the clauses 4-9.

\end{lemma}

\begin{proof} Assume the conditions of the lemma and, in addition, when discussing each clause of the lemma,  assume the additional conditions of that clause.\vspace{5pt} 

{\em Clause 1}. It is not hard to see that each of the five conditions of Definition \ref{qint} is inherited by $H$ and $\Omega$ from $E$ and $\Omega$.
This means that $\seq{\Omega}$ is $H$-manageable. The fact $\seq{\Omega,\pp\alpha}E^*=\seq{\Omega}H^*$ is also obvious, as the effect of (the clearly legal) move $\pp\alpha$ in $\seq{\Omega}E^*$ is turning the 
$(G_1\add\ldots\add G_n)^*$ component of the latter into $G_{i}^{*}$, i.e., turning $E^*$ into $H^*$.\vspace{5pt}

{\em Clause 2}. As in the previous clause, with a little thought one can  see that each of the five conditions of Definition 
\ref{qint} is inherited by $H$ and $\seq{\Omega,\pp\alpha}$ from $E$ and $\Omega$, so $\seq{\Omega,\pp\alpha}$ is $H$-manageable.  
As for $\seq{\Omega,\pp\alpha}$'s being a legal position of $H^*$ and the fact $\seq{\Omega,\pp\alpha}E^*=\seq{\Omega,\pp\alpha}H^*$, it is sufficient to note that we simply have $E^*=H^*$. \vspace{5pt}

{\em Clause 3}. Let 
\[\Phi^+=\seq{\oo\pi_1,\ldots,\oo\pi_n,\pp\nu_1,\ldots,\pp\nu_m}\]
and 
\[\Phi^-=\seq{\oo\nu_1,\ldots,\oo\nu_m,\pp\pi_1,\ldots,\pp\pi_n}.\]
Thus, $\Phi^+$ and $\Phi^-$ are the sequences of all moves made by the two players within $O^+$ and $O^-$, respectively, while playing $\Omega'$ (here the contextual run). 
Note that 
\begin{equation}\label{feb25}
\mbox{$\Phi^+$ is a $\pp$-delay of $\rneg\Phi^-$.}
\end{equation}
This implies that $\Omega'$ is $H$-manageable, because conditions 1-4 of Definition \ref{qint} are obviously inherited by $H$ and $\Omega'$ from $E$ and $\Omega$, and so is condition 5 for any active hybrid atom $R_s$ different from $P_q$.

We, of course, have $H^*=E^*$. So, what remains to show is that $\Omega'$ is a legal position of $E^*$. The assumption that $\Omega$ is a legal position of $E^*$ obviously implies that: 
\begin{eqnarray}
& & \mbox{\em $\seq{\oo\pi_1,\ldots,\oo\pi_n}$ is a legal position of $P^{*}$;}\label{eqq1}\\
& & \mbox{\em $\seq{\oo\nu_1,\ldots,\oo\nu_m}$ is a legal position of $\gneg P^{*}$.}\label{eqq2}
\end{eqnarray}

Assume, for a contradiction, that $\Omega'$ is not a legal position of ${E^*}$. Evidently this can be the case only if either $\Phi^+$ is not a legal position of $P^*$ or $\Phi^-$ is not a legal position of $\gneg P^{*}$ (or both).

Suppose $\Phi^+$ is not a legal position of $P^{*}$. In view of (\ref{eqq1}), $\Phi^+$ cannot be a $\oo$-illegal position of $P^{*}$. So, it must be $\pp$-illegal. But then, by  (\ref{feb25}) and clause 1 of Lemma \ref{may19}, 
$\rneg\Phi^-$ is a $\pp$-illegal position of $P^{*}$, meaning that $\Phi^-$ is a $\oo$-illegal position of $\gneg P^*$. This, however, is in obvious contradiction with (\ref{eqq2}).

Suppose now $\Phi^-$ is not a legal position of $\gneg P^{*}$.  This case is similar/symmetric to the previous one. In view of (\ref{eqq2}), $\Phi^-$ cannot be a $\oo$-illegal position of $\gneg P^{*}$. So, it must be $\pp$-illegal, which means that $\rneg \Phi^-$ is a $\oo$-illegal position of $P^*$. But then, by  (\ref{feb25}) and clause 2 of Lemma \ref{may19}, 
$\Phi^+$ is a $\oo$-illegal position of $P^{*}$. This, however, is in contradiction with (\ref{eqq1}).\vspace{5pt}

{\em Clauses 4,5,6,7} are obvious.\vspace{5pt}

{\em Clause 8} is symmetric to Clause 1. \vspace{5pt}

{\em Clause 9} is similar to Clause 2.\vspace{5pt}

{\em Clause 10} is obvious.
 \end{proof}

\subsection{Finalization}
Note that when a formula $E$ is elementary and an interpretation $^*$
is perfect, $E^*$ is one of the two constant elementary games $\twg$ or $\tlg$. Here we define the relation $\leq$ on constant elementary games by stipulating that $A\leq B$ 
iff $A=\tlg$ or $B=\twg$. In other words, $A\leq B$ iff $(A\mli B)=\twg$. 

\begin{lemma}\label{feb10}
Suppose $^*$ is a perfect interpretation, $F_1$ is an elementary formula, and $F_2$ is the result of replacing 
in $F_1$ an (elementary) literal $L_1$ by an elementary (logical or nonlogical) literal $L_2$, such that $L_{1}^{*}\leq L_{2}^{*}$. Then  $F_{1}^{*}\leq F_{2}^{*}$.
\end{lemma}
\begin{proof} The above 
lemma does nothing but rephrases, in our terms, a known fact from classical logic, according to which, if in an
interpreted  formula $F_1$ we replace a positive occurrence of a subformula $L_1$ by a formula $L_2$ whose 
Boolean value is not less than that of $L_1$ ($L_2$ is ``at least as true'' as $L_1$), then the value of the resulting formula $F_2$ will not be less
than that of $F_1$ ($F_2$ will not be ``less true'' than $F_1$). \end{proof}

Now we introduce the operation $\elzz{\Gamma}{A}$ of the type \[\mbox{\{{\em runs}\}$\times$\{{\em constant games}\} $\rightarrow$ \{{\em constant elementary games}\},}\] which is rather similar to prefixation. Intuitively $\elzz{\Gamma}{A}$, that we call the {\bf $\Gamma$-finalization} of $A$, is
the proposition  ``$\Gamma$ is a $\pp$-won run of $A$". 
This operation is only defined when $\Gamma$ is a legal run of $A$. 
We agree that, every time we make a statement that applies 
$\elzz{\Gamma}{}$ \ to a constant game $A$, we imply that $\Gamma$ is a legal run of $A$ and hence $\elzz{\Gamma}{A}$ is defined.
Here is the formal definition of the operation of finalization:
\begin{definition}\label{elzz}
Assume $A$ is a constant game and $\Gamma$ a legal run of $A$. Then $\elzz{\Gamma}{A}$\label{z45} is the constant elementary game 
defined by $\win{\elzz{\Gamma}{A}}{}\emptyrun=\win{A}{}\seq{\Gamma}$.
\end{definition}

\begin{lemma}\label{feb7}
Assume  $E$ is a stable $\propseqc$-formula, $^*$ is a perfect interpretation, and $\Gamma$ is an $E$-manageable legal run of $E^*$. Then $\Gamma$ is a $\pp$-won run of $E^*$. 
\end{lemma}

\begin{proof} 
Assume the conditions of the lemma. Throughout this proof, $\Gamma$ is the contextual run.

For each (positive) active surface occurrence of each general or hybrid literal in $E$, let us fix an elementary nonlogical atom that we call the {\bf surrogate} for that occurrence. We assume that all surrogates are pairwise distinct, and none of them occurs in  $E$ (either directly or as the elementary component of a hybrid atom). Since these atoms do not occur in $E$, we may make an arbitrary assumption regarding 
how they are interpreted by $^*$ (otherwise replace $^*$ by an appropriate interpretation). In particular, we assume that,
whenever $r$ is the surrogate for an active occurrence $O$ of a general or hybrid literal $L$,  we have $r^{*}=\elzz{\Sigma}{L^*}$, where $\Sigma$ is the sequence of (lab)moves made within that occurrence.

Let $E_1$ denote the result of replacing in the capitalization of $E$:
\begin{itemize}
\item every surface occurrence of a general or hybrid literal by its surrogate; \vspace{-3pt}
\item every surface occurrence of a subformula of the form $H_1\adc\ldots\adc H_n$ by $\twg$;
\item every surface occurrence of a subformula of the form $H_1\add\ldots\add H_n$ by $\tlg$.
\end{itemize} 

With some thought, we can see that 
\begin{equation}\label{feb28}
\elzz{\Gamma}{E^*}=E^{*}_{1}.
\end{equation}

Assume $E$ has $k$ active, positive, non-widowed hybrid atoms, where $q^1,\ldots,q^k$ are the elementary components of those atoms and 
$P^1,\ldots,P^k$ are the corresponding general components. Let 
$r_1,\ldots,r_k$ be the surrogates for the positive occurrences of $P^{1}_{q^1},\ldots, P^{k}_{q^k}$, respectively, and let $\Sigma^{+}_{1},\ldots,\Sigma^{+}_{k}$ be the sequences of moves that have been made (while playing $\Gamma$) within these $k$ occurrences. Next, let 
$s_1,\ldots,s_k$ be the surrogates for the occurrences of $\gneg P^{1}_{q^1},\ldots, \gneg P^{k}_{q^k}$, respectively, and let $\Sigma_{1}^{-},\ldots,\Sigma_{k}^{-}$ be the sequences of moves that have been made  within these $k$ occurrences.
 Let $E_2$ be the result of replacing in $E_1$ each atom $s_i$ ($1\leq i\leq k$) by $\gneg r_i$. According to clause 5 of Definition \ref{qint}, for each $1\leq i\leq k$, $\Sigma_{i}^{+}$ is a $\pp$-delay of $\rneg\Sigma_{i}^{-}$.
Hence, as $P_{i}^{*}$ is a static game, we have $\elzz{\rneg\Sigma_{i}^{-}}{P_{i}^{*}}\leq \elzz{\Sigma_{i}^{+}}{P_{i}^{*}}$, which is the same as to say that  
 $\gneg(\elzz{\Sigma_{i}^{+}}{P_{i}^{*}})  \leq \gneg (\elzz{\rneg\Sigma_{i}^{-}}{P_{i}^{*}})$. But, by the definition of $\gneg$,  \ $\gneg (\elzz{\rneg\Sigma_{i}^{-}}{P_{i}^{*}})=\elzz{\Sigma_{i}^{-}}{\gneg P_{i}^{*}}$. So,  $\gneg(\elzz{\Sigma_{i}^{+}}{P_{i}^{*}})  \leq \elzz{\Sigma_{i}^{-}}{\gneg P_{i}^{*}}$. Now remember that $\elzz{\Sigma_{i}^{+}}{P_{i}^{*}}=r_{i}^{*}$ and $\elzz{\Sigma_{i}^{-}}{\gneg P_{i}^{*}}=s_{i}^{*}$. Thus, $\gneg r^{*}_{i}\leq s^{*}_{i}$. 
Then, applying Lemma \ref{feb10} $k$ times, we get  
\begin{equation}\label{iii}
E^{*}_{2}\leq E^{*}_{1}.
\end{equation} 

Next, let $E_3$ be the result of replacing in $E_2$ each surrogate for a general literal by $\tlg$. Applying Lemma \ref{feb10} as many times as the number of such surrogates, we get 
\begin{equation}\label{iv}
E^{*}_{3}\leq E^{*}_{2}.
\end{equation} 

Now compare $E_3$ with $\elz{E}$. An analysis of how these two formulas have been obtained from $E$ can reveal that 
$E_3$ is just the result of replacing in $\elz{E}$ all (both) occurrences of each atom $q_i$ from the earlier-mentioned list  $q_1,\ldots,q_k$ by $r_{i}$.
That is, $E_3$ is a substitutional instance of $\elz{E}$. The latter is classically
valid because, by our assumptions, $E$ is stable. Therefore $E_3$ is also classically valid, and hence $E_{3}^{*}=\twg$. Then statements (\ref{iv}), (\ref{iii}) and (\ref{feb28}) yield $\elzz{\Gamma}{E^*}=\twg$. In other words, $\Gamma$ is a $\pp$-won run of $E^*$.  \end{proof}

\section{The soundness of $\propseq$}\label{ssnd}

\begin{lemma}\label{sound}
 If $\propseq\vdash F$, then $F$ is valid (any formula $F$). 

Moreover, there is an effective procedure that takes a $\propseq$-proof of an arbitrary formula $F$ and returns an HPM $\cal H$ such that, for every interpretation $^*$, \  ${\cal H}$ computes $F^*$.
\end{lemma}

\begin{proof} According to Theorem \ref{static}, our sequential operations preserve the static property of games. The same is known to be true for all other operations studied in computability logic (Theorem 24 of \cite{Japfin}). So, for any formula $F$ and interpretation $^*$, the game $F^*$ is static. Further, it is known that, for static games, HPMs and EPMs have the same computing power, and that, moreover, every EPM can be effectively converted into an HPM such that the latter wins every static game won by the former (Theorem 28 of \cite{Japfin}). Therefore, it would be sufficient to prove the above lemma --- in particular, the `Moreover' clause of it --- with 
``EPM $\cal S$" instead of ``HPM $\cal H$". 

Furthermore, it would be sufficient to restrict interpretations to perfect ones. Indeed, suppose a machine $\cal M$ (whether it be an EPM or an HPM) wins $F^\dagger$ for every perfect interpretation $^\dagger$, and let $^*$ be a not-necessarily-perfect interpretation. We want to see that the same machine $\cal M$ 
also wins $F^*$. Suppose this is not the case, i.e. $\cal M$ loses $F^*$ on some valuation $e$. This means that, where $\Gamma$ is the run spelled by some $e$-computation branch of $\cal M$, we have 
$\win{e[F^*]}{}\seq{\Gamma}=\oo$. Now, let $^\dagger$ be the perfect interpretation induced by $(^*,e)$. According to Lemma \ref{perf},   $e[F^*]=F^\dagger$. Thus,  $\win{F^\dagger}{}\seq{\Gamma}=\oo$, so that $\cal M$ does not win $F^\dagger$, which is a contradiction.

Finally, Lemma \ref{fdc} allows us to safely replace ``$\propseq$" by ``$\propseqc$" in our present lemma.

In view of the above observations, Lemma \ref{sound} is an immediate consequence of the following Lemma \ref{soundd}. \end{proof}

\begin{lemma}\label{soundd}
There is an effective procedure that takes a $\propseqc$-proof of an arbitrary formula $F$ and returns an EPM $\cal S$ such that, for every perfect interpretation $^*$, \  ${\cal S}$ computes $F^*$.
\end{lemma}

\begin{proof} Every $\propseqc$-proof, in fact, encodes a valuation- and interpretation-independent winning strategy for $\pp$, and the EPM $\cal S$ that we are going to describe just follows such a strategy. 

We provide only a semiformal description of how our strategy/machine $\cal S$ works for a $\propseqc$-provable formula $F$. Restoring suppressed technical details, if necessary, does not present a problem.\footnote{A similar proof for {\bf CL2} given in \cite{Japtocl1} was not lazy to go into all technical details, 
for which reason it was much longer than the present proof.}   Fix an arbitrary valuation $e$ and a perfect interpretation $^*$. Since $^*$ is perfect, the $e$ parameter is irrelevant and it can always be safely omitted. Furthermore, in concordance with our earlier agreements, we will often omit $^*$ as well and write $E$ instead of $E^*$. That is, we continue abusing terminology and notation by (often) identifying game $E^*$ with formula $E$. 

The strategy that our $\cal S$ follows is a recursive one, at every step dealing with $\seq{\Omega}E^*$, where $E$ is a $\propseqc$-provable formula and $\Omega$ is an $E$-manageable legal position of $E^*$. Initially $E=F$ and $\Omega=\emptyrun$. How $\cal S$ acts on $\seq{\Omega}E^*$ depends on by which of the four rules $E$ is derived in $\propseqc$ (some $\propseqc$-proof is assumed to be fixed). 

If $E$ is derived  by Choose$^\circ$ from $H$ as described in Definition \ref{nov23}, $\cal S$ makes the move $\alpha$ whose effect is choosing $G_{i}$ in the $G_1\add\ldots \add G_n$ subformula of $E$. For example, if $E=(G_1\add G_2)\mlc(G_3\adc G_4)$ and $H=G_2\mlc(G_3\adc G_4)$, then `$1.2$' is such a move $\alpha$. Clause 1 of Lemma \ref{splm} tells us that $\Omega$ remains an $H$-manageable legal position of $H^*$ and that $\pp\alpha$ brings $\seq{\Omega}E^*$ down to $\seq{\Omega}H^*$. So, after making move $\alpha$, $\cal S$ switches to its 
strategy for $\seq{\Omega}H^*$. This, by the induction hypothesis, guarantees success.

If $E$ is derived  by Switch$^\circ$ from $H$ as described in Definition \ref{nov23}, then $\cal S$ makes the move $\alpha$ whose effect is making a switch in the \(G_0\sqd\ldots\sqd \underline{G_m}\sqd G_{m+1}\sqd\ldots\sqd G_{n}\) subformula.  
For example, if $E=(G_0\sqd \underline{G_1}\sqd G_2)\mlc(\underline{G_3}\sqd G_4)$ and $H=(G_0\sqd G_1\sqd\underline{G_2})\mlc(\underline{G_3}\sqd G_4)$, then `$1.\S$' is such a move. Clause 2 of Lemma \ref{splm} tells us that $\seq{\Omega,\pp\alpha}$ remains an $H$-manageable legal position of $H^*$ and that $\pp\alpha$ brings $\seq{\Omega}E^*$ down to $\seq{\Omega,\pp\alpha}H^*$. So, after making move $\alpha$, the machine switches to its 
strategy for $\seq{\Omega,\pp\alpha}H^*$ and, by the induction hypothesis, wins.

If $E$ is derived by Match$^\circ$ from $H$ through replacing the two (active surface) occurrences of a hybrid atom $P_q$ in $H$ by $P$, 
then the machine finds within $\Omega$ and copies, in the positive occurrence of $P_q$, all of the moves made so far by the environment in the negative occurrence of $P_q$ (or rather in the corresponding occurrence of $P$), and vice versa. This series of moves brings the game down 
to $\seq{\Omega'}E^*=\seq{\Omega'}H^*$, where $\Omega'$ is result of adding those moves to $\Omega$. Clause 3 of Lemma \ref{splm} guarantees that   $\Omega'$ is an $H$-manageable legal position of $H^*$. So, now $\cal S$ switches to its successful strategy for $\seq{\Omega'}H^*$ and eventually wins.

Finally, suppose $E$ is derived by Wait$^\circ$. Our machine keeps
 granting permission (``waiting''). In view of Lemma \ref{feb7}, if $\oo$ never makes a move, $\cal S$ wins the game. Suppose now $\oo$ makes a move $\alpha$. This should be a legal move, or else $\cal S$ automatically wins. Then, according to clause 10 of Lemma \ref{splm}, $\alpha$ should satisfy the conditions of one of the following 6 cases. 

{\em Case 1}. $\alpha$ is a move whose effect is moving in some abandoned subformula or a widowed hybrid literal of $E$. According to clause 4 of Lemma \ref{splm}, $\seq{\Omega,\oo\alpha}$ remains an $E$-manageable legal position of $E^*$. In this case, $\cal S$ calls itself on $\seq{\Omega,\oo\alpha}E^*$.

{\em Case 2}. $\alpha$ is a move whose effect is moving in some active surface occurrence of a general atom in $E$. According to clause 5 of Lemma \ref{splm}, $\seq{\Omega,\oo\alpha}$ remains an $E$-manageable legal position of $E^*$. Again, in this case, $\cal S$ calls itself on $\seq{\Omega,\oo\alpha}E^*$.

{\em Case 3}. $\alpha$ is a move whose effect is making a catch-up switch in some active surface occurrence of a $\sqd$-subformula. According to clause 6 of Lemma \ref{splm}, $\seq{\Omega,\oo\alpha}$ remains an $E$-manageable legal position of $E^*$. In this case, as in the previous two cases, $\cal S$ calls itself on $\seq{\Omega,\oo\alpha}E^*$.

{\em Case 4}. $\alpha$ is a move whose effect is making a move $\gamma$ in some active surface occurrence of a non-widowed hybrid atom. Let $\beta$ be the move whose effect is making the same move $\gamma$ within the other active surface occurrence of the same hybrid atom.  According to clause 7 of Lemma \ref{splm}, $\seq{\Omega,\oo\alpha,\pp\beta}$ remains an $E$-manageable legal position of $E^*$. In this case, $\cal S$ makes the move $\beta$ and calls itself on $\seq{\Omega,\oo\alpha,\pp\beta}E^*$.

{\em Case 5}: $\alpha$ is a move whose effect is a choice of the $i$th component in an active surface occurrence of a subformula $G_1\adc\ldots\adc G_n$. Then $\cal S$
switches to its winning strategy for $\seq{\Omega}H^*$, where $H$ is the result of replacing the above subformula by $G_i$ in $E$.  Clause 8 of  
Lemma \ref{splm} guarantees that 
$\seq{\Omega}H^*$ is indeed the game to which $\seq{\Omega}E^*$ has evolved and that $\Omega$ is an $H$-manageable legal position of $H^*$, so that, by the
 induction hypothesis, $\cal S$ knows how to win $\seq{\Omega}H^*$.

{\em Case 6}: $\alpha$ signifies a (leading) switch move within an active surface occurrence of a subformula \[G_0\sqc\ldots\sqc \underline{G_m}\sqc G_{m+1}\sqc \ldots\sqc G_n.\] Then $\cal S$ makes the same move $\alpha$ (signifying making a catch-up switch within the same subformula), and calls itself on 
$\seq{\Omega,\oo\alpha,\pp\alpha} H^*$, where $H$ is the result of replacing the above subformula by \[G_0\sqc\ldots\sqc G_m\sqc \underline{G_{m+1}}\sqc \ldots\sqc G_n.\] Clause  of  9 
Lemma \ref{splm} guarantees that 
$\seq{\Omega,\oo\alpha,\pp\alpha}$ is an $H$-manageable legal position of $H^*$, so that, by the induction hypothesis, $\cal S$ will win $\seq{\Omega,\oo\alpha,\pp\alpha}H^*$.

Looking back at $\cal S$'s strategy, the value of $E$ keeps changing from conclusion to a premise, starting from the original formula $F$. It is therefore obvious that such a value should stabilize at some $E_{\mbox{\em final}}$ and never change afterwards. It is also obvious that this $E_{\mbox{\em final}}$ should be derived by Wait$^\circ$, or else it would further change. For the same reason, while $\cal S$ acts following the prescriptions for the case of Wait$^\circ$ with $E=E_{\mbox{\em final}}$, Cases 5 and 6 never occur. But in all other four cases, as well as in the case when $\oo$ does not make any moves, (the perhaps continuously updated) $\Omega$ remains 
$E_{\mbox{\em final}}$-manageable, and --- according to Lemma \ref{feb21} --- $\Omega$ will remain so even if it grows infinite. Also, as a conclusion of Wait, 
$E_{\mbox{\em final}}$ is stable. This, by Lemma \ref{feb7}, implies that the overall game will be won by $\cal S$. 

To officially complete our proof of the lemma, it remains to note that $\cal S$, of course, can be constructed effectively from a given $\propseqc$-proof of $F$. 
\end{proof}

\section{Preliminaries for the completeness proof}

\subsection{Machines against machines}

This subsection borrows a discussion from \cite{Japtocl1}, providing certain background information necessary for our completeness proof but missing in \cite{Japfin}, the only external source on computability logic on which the present paper was promised to rely. 

 For a run $\Gamma$ and a computation branch $B$ of an HPM or EPM, we say that $B$ {\bf cospells} $\Gamma$ iff
$B$ spells $\rneg\Gamma$ ($\Gamma$ with all labels reversed) in the sense of Section 6 of \cite{Japfin}. 
Intuitively, when a machine $\cal M$ plays as  $\oo$ (rather than $\pp$), then the run that is generated by  a given computation branch $B$ of $\cal M$ is the run cospelled (rather than spelled) by $B$, for the moves that $\cal M$ makes 
get the label $\oo$, and the moves that its adversary makes get the label $\pp$.

We say that an EPM $\cal E$ is {\bf fair} iff, for every valuation $e$, every $e$-computation branch 
of $\cal E$ is fair in the sense of Section 6 of \cite{Japfin}. 
 
\begin{lemma}\label{lem}
Assume $\cal E$ is a fair EPM, $\cal H$ is any HPM, and $e$ is any valuation. There are a uniquely defined 
 $e$-computation branch $B_{\cal E}$ of $\cal E$ and a uniquely defined $e$-computation branch $B_{\cal H}$ of $\cal H$
--- which we respectively call {\bf the $({\cal E},e,{\cal H})$-branch} and {\bf the $({\cal H},e,{\cal E})$-branch}
 --- such that the run spelled by $B_{\cal H}$, called  {\bf the $\cal H$ vs. $\cal E$ run on $e$}, 
 is the run cospelled by $B_{\cal E}$.\end{lemma}
 
When ${\cal H},{\cal E},e$ are as above, $\Gamma$ is the $\cal H$ vs. $\cal E$ run on $e$ and $A$ is a game with 
$\win{A}{e}\seq{\Gamma}=\pp$ (resp. $\win{A}{e}\seq{\Gamma}=\oo$), we say that $\cal H$ {\bf wins}
(resp. {\bf loses}) $A$ {\bf against $\cal E$ on $e$}. 

A strict proof of the above lemma can be found in \cite{Jap03} (Lemma 20.4), and we will not reproduce  
the formal proof here.  Instead, the following intuitive explanation should suffice:\vspace{7pt}

{\bf Proof idea.} Assume $\cal H$, $\cal E$, $e$ are as in Lemma \ref{lem}. The play that we are going to describe is the unique 
play generated when the two machines play against each other, with $\cal H$ in the role of $\pp,$  $\cal E$ in the role of $\oo$, and $e$ spelled on the valuation tapes of both machines. 
We can visualize this play as follows.
Most of the time during the process $\cal H$ remains inactive (sleeping); it is woken up only when $\cal E$ enters a permission state, on which event $\cal H$ makes a (one single) transition to its next computation step --- that may or may not result in making a move --- and goes back into a sleep that will continue until $\cal E$ enters  a permission state again, and so on. From ${\cal E}$'s perspective, $\cal H$ acts as a patient adversary who makes one or zero move only when granted permission, just as the EPM-model assumes.  And from $\cal H$'s perspective, which, like a person in a coma,  has no sense of time during its sleep and hence can think that the wake-up events that it calls the beginning of a clock cycle happen at a constant rate, $\cal E$ acts as an adversary who can make any finite number of moves during a clock cycle (i.e. while $\cal H$ was sleeping), just as the HPM-model assumes. This scenario uniquely determines an $e$-computation branch $B_{\cal E}$ of $\cal E$ that we call
the $({\cal E},e,{\cal H})$-branch, and an $e$-computation branch $B_{\cal H}$ of $\cal H$ that we call
the $({\cal H},e,{\cal E})$-branch. What we call the $\cal H$ vs. $\cal E$ run on $e$ is the run generated 
in this play. In particular --- since we let $\cal H$ play in the role of $\pp$ --- this is the run spelled by $B_{\cal H}$. $\cal E$, who plays in the role of $\oo$, sees the same run, only it sees the labels of the moves of that run in negative colors. 
That is, $B_{\cal E}$ cospells rather than spells that run. This is exactly what Lemma \ref{lem} asserts.

\subsection{Logics $\propseqe$, $\propseqec$  and $\propseqecc$}

Our proof of the completeness part of Theorem \ref{thcl2} employs the conservative, elementary-base fragment $\propseqe$ of $\propseq$, obtained by restricting the language of the latter to elementary-base formulas --- we refer to formulas of this restricted language as {\bf $\propseqe$-formulas} --- and (correspondingly) deleting the rule of Match.  Logic {\bf CL1}, historically the first system for computability logic proven (in \cite{Japtocl1}) to be sound and complete, is a $\sqc,\sqd$-free counterpart of $\propseqe$. Of course, $\propseqe$ inherits soundness from $\propseq$. In this section we are going to prove the completeness of $\propseqe$. 

Our completeness proof for $\propseqe$, in turn, employs the modification $\propseqec$ of $\propseqe$. {\bf $\propseqec$-formulas}
are nothing but elementary-base $\propseqc$-formulas, i.e., $\propseqc$-formulas that contain no general or hybrid atoms. Thus, the only difference between 
$\propseqe$-formulas and $\propseqec$-formulas is that, in the latter one of the components of each sequential subformula is underlined. 

Logic $\propseqec$ is a conservative fragment of logic $\propseqc$ in the same sense as $\propseqe$ is a fragment of $\propseq$. Namely, $\propseqec$ is obtained from $\propseqc$ by restricting its language to elementary-base formulas, and deleting the rule of Match$^\circ$. Here we (re)produce the rules of 
$\propseqec$ for the convenience of later references:

\begin{definition}\label{nov29}
The inference rules of $\propseqec$ are:
 
\begin{description}
\item[Wait$^\circ$:]  $\vec{H}\vdash F$, where $F$ is stable and $\vec{H}$ is the smallest set of formulas satisfying the following two conditions: 
\begin{itemize} 
\item whenever $F$ has an active surface occurrence of a subformula $G_1\adc\ldots\adc G_n$, for each 
$i\in\{1,\ldots,n\}$, $\vec{H}$ contains the result of replacing that occurrence in $F$ by $G_i$;
\item whenever $F$ has an active surface occurrence of a subformula $G_0\sqc \ldots\sqc \underline{G_m}\sqc G_{m+1}\sqc\ldots\sqc G_n$, $\vec{H}$ contains the result of replacing that occurrence in $F$ by $G_0\sqc \ldots\sqc G_m\sqc \underline{G_{m+1}}\sqc\ldots\sqc G_n$.
\end{itemize}
\item[Choose$^\circ$:]  $H\vdash F$, where $H$ is the result of replacing in $F$ an active surface occurrence of a subformula $G_1\add\ldots\add G_n$  by $G_i$ for some $i\in\{1,\ldots, n\}$.
\item[Switch$^\circ$:]  $H\vdash F$, where $H$ is the result of replacing in $F$ an active surface occurrence of a subformula \(G_0\sqd\ldots\sqd \underline{G_m}\sqd G_{m+1}\sqd\ldots\sqd G_{n}\)  by \(G_0\sqd\ldots\sqd {G_m}\sqd \underline{G_{m+1}}\sqd\ldots\sqd G_{n}.\)
\end{description}
\end{definition}

The following lemma establishes equivalence between $\propseqe$ and $\propseqec$: 

\begin{lemma}\label{fdcc}
For any formula $G$,  $\propseqec\vdash G$ iff $\propseqe\vdash G$.
\end{lemma}

\begin{proof} This lemma is pretty straightforward: a $\propseqec$-proof of $G$ turns into a $\propseqe$-proof of $G$ after replacing, in each formula of the former, every subformula $E_1\sqc\ldots \sqc\underline{E_m}\sqc \ldots\sqc E_n$ by   $E_m\sqc\ldots\sqc E_n$ and 
every subformula $E_1\sqd\ldots \sqd\underline{E_m}\sqd \ldots\sqd E_n$ by   $E_m\sqd\ldots\sqd E_n$.  And vice versa: a $\propseqe$-proof of $G$ turns into a $\propseqec$-proof of $G$ after underlining the head of each sequential subformula of $G$. Then, in the bottom up view of the proof, every time Switch is used, it should simply move the underline to the next sequential component instead of deleting the head. Similarly for the premises of Wait that are associated with sequential subformulas. \end{proof}

Next, we define Logic $\propseqecc$ which is a ``dual'' of $\propseqec$:

\begin{definition}\label{nov27}
The language of $\propseqecc$ is the same as that of $\propseqec$, and the rules of inference are:

\begin{description}
\item[$\overline{\mbox{Wait$^\circ$}}$:]  $\vec{H}\vdash F$, where $F$ is instable and $\vec{H}$ is the smallest set of formulas satisfying the following two conditions: 
\begin{itemize} 
\item whenever $F$ has an active surface occurrence of a subformula $G_1\add\ldots\add G_n$, for each 
$i\in\{1,\ldots,n\}$, $\vec{H}$ contains the result of replacing that occurrence in $F$ by $G_i$;
\item whenever $F$ has an active surface occurrence of a subformula $G_0\sqd \ldots\sqd \underline{G_m}\sqd G_{m+1}\sqd\ldots\sqd G_n$, $\vec{H}$ contains the result of replacing that occurrence in $F$ by $G_0\sqd \ldots\sqd G_m\sqd \underline{G_{m+1}}\sqd\ldots\sqd G_n$.
\end{itemize}
\item[$\overline{\mbox{Choose$^\circ$}}$:]  $H\vdash F$, where $H$ is the result of replacing in $F$ an active surface occurrence of a subformula $G_1\adc\ldots\adc G_n$  by $G_i$ for some $i\in\{1,\ldots, n\}$.
\item[$\overline{\mbox{Switch$^\circ$}}$:]  $H\vdash F$, where $H$ is the result of replacing in $F$ an active surface occurrence of a subformula \(G_0\sqc\ldots\sqc \underline{G_m}\sqc G_{m+1}\sqc\ldots\sqc G_{n}\)  by \(G_0\sqc\ldots\sqc {G_m}\sqc \underline{G_{m+1}}\sqc\ldots\sqc G_{n}.\)
\end{description}
\end{definition}

\begin{lemma}\label{nov4}
$\propseqec\not\vdash F$ iff $\propseqecc\vdash F$ (any $\propseqec$-formula $F$).
\end{lemma}

\begin{proof} We prove this lemma by induction on the complexity of $F$. It would be sufficient to verify the `only if' part, as the `if' part (which we do not need anyway) can be handled in a fully symmetric way. So, assume $\propseqec\not\vdash F$ and let us see that then $\propseqecc\vdash F$. There are two cases to consider:

{\em Case 1}: $F$ is stable. Then there must be a $\propseqec$-unprovable formula $H$ satisfying one of the following two conditions, for otherwise $F$ would be $\propseqec$-derivable by Wait$^\circ$:
\begin{itemize}
\item $H$ is the result of replacing in $F$ an active surface occurrence of a subformula  $G_1\adc\ldots\adc G_n$  by $G_i$ for some $i\in\{1,\ldots, n\}$.
\item $H$ is the result of replacing in $F$ an active surface occurrence of a subformula \(G_0\sqc\ldots\sqc \underline{G_m}\sqc G_{m+1}\sqc\ldots\sqc G_{n}\)   by \(G_0\sqc\ldots\sqc {G_m}\sqc \underline{G_{m+1}}\sqc\ldots\sqc G_{n}.\)
\end{itemize}
In either case, by the induction hypothesis, $\propseqecc\vdash H$, whence, by $\overline{\mbox{Choose$^\circ$}}$ (if the first condition is satisfied) or $\overline{\mbox{Switch$^\circ$}}$ (if the second condition is satisfied), $\propseqecc\vdash F$.

{\em Case 2}: $F$ is instable. Let $\vec{H}$ be the smallest set of formulas such that the following two conditions are 
satisfied: 
\begin{itemize}
\item whenever $F$ has an active surface occurrence of a subformula $G_1\add\ldots\add G_n$, for each 
$i\in\{1,\ldots,n\}$, $\vec{H}$ contains the result of replacing that occurrence in $F$ by $G_i$;
\item whenever $F$ has an active surface occurrence of a subformula $G_0\sqd \ldots\sqd \underline{G_m}\sqd G_{m+1}\sqd\ldots\sqd G_n$, $\vec{H}$ contains the result of replacing that occurrence in $F$ by $G_0\sqd \ldots\sqd G_m\sqd \underline{G_{m+1}}\sqd\ldots\sqd G_n$.
\end{itemize}
None of the elements of $\vec{H}$ is provable in $\propseqec$, for otherwise $F$ would also be derivable in $\propseqec$ by Choose$^\circ$ or Switch$^\circ$. Therefore, by the induction hypothesis, each element of $\vec{H}$ is $\propseqecc$-provable, whence, by  $\overline{\mbox{Wait$^\circ$}}$, we have  $\propseqecc\vdash F$. 
\end{proof}

\subsection{The completeness of $\propseqe$}

\begin{lemma}\label{oct5}
If $\propseqe\not\vdash F$, then $F$ is not valid (any $\propseqe$-formula $F$).

In particular, if $\propseqe\not\vdash F$, then $F^*$ is not computable for some interpretation $^*$ that interprets all atoms as finitary predicates of complexity $\Delta_2$.
\end{lemma}

\begin{proof} Assume $\propseqe\not\vdash F$, which, by Lemmas \ref{fdcc} and \ref{nov4}, means that $\propseqecc\vdash F$. Fix a $\propseqecc$-proof of $F$. From such a proof, we can extract an environment's interpretation- and valuation-independent EPM-counterstrategy $\cal C$ for $F$ in a way fully symmetric to the way we extracted the machine's strategy $\cal S$ from a  $\propseqc$-proof in Section \ref{ssnd}. $\cal C$ is a {\em counterstrategy} in the sense that $\cal C$ plays in the role of $\oo$ rather than $\pp$.\footnote{If we want to see $\cal C$ as a strategy in the ordinary sense, then it is a strategy for $\gneg F$.}   In fact, $\cal C$ is much simpler than $\cal S$, because in the present case we only deal with elementary-base formulas. Since the work of $\cal C$ depends neither on $^*$ nor on $e$, in our description of it we can safely omit these parameters,  and write $E$ instead of $e[E^*]$.

The strategy that $\cal C$ follows is a recursive one which, just like its counterpart from Section \ref{ssnd}, at every step deals with   $\seq{\Omega}E$, where $E$ is some formula from the $\propseqecc$-proof of $F$, and $\Omega$ is a certain legal position of $E$. Initially $E=F$ and $\Omega=\emptyrun$. How $\cal C$ acts on $\seq{\Omega}E$ depends on by which of the three rules $E$ is derived in $\propseqecc$. 

If $E$ is derived  by $\overline{\mbox{Choose}^\circ}$ from $H$ as described in Definition \ref{nov27}, $\cal C$ makes the move $\alpha$ whose effect is choosing $G_{i}$ in the $G_1\adc\ldots \adc G_n$ subformula of $E$. Then it updates $E$ to $H$ (without changing $\Omega$), and calls (repeats) itself.

If $E$ is derived  by $\overline{\mbox{Switch}^\circ}$ from $H$ as described in Definition \ref{nov27}, then $\cal C$ makes the move $\alpha$ whose effect is making a switch in the \(G_0\sqc\ldots\sqc \underline{G_m}\sqc G_{m+1}\sqc\ldots\sqc G_{n}\) component, updates $E$ to $H$ (without changing $\Omega$), and calls itself.    
  
Finally, suppose $E$ is derived by $\overline{\mbox{Wait}^\circ}$. $\cal C$ keeps
 granting permission. Now and then the adversary may be making moves in abandoned components, or catch-up switch moves within sequential subformulas.  To such moves  $\cal C$ reacts by adding them to its internal record of $\Omega$ (updating $\Omega$), but otherwise $\cal C$ does not move. 
However, what will typically happen
 during this stage (except one --- the last --- case) is that sooner or later the adversary makes a legal move $\alpha$\footnote{And if $\alpha$ is illegal, then $\cal C$'s job is done.} that causes  $\cal C$ to update $E$ to one of its premises. In particular, one of the following will be the case: 

{\em Case 1}: $\alpha$ is a move signifying a choice of the $i$th component in an active surface occurrence of a subformula $G_1\add\ldots\add G_n$. Then $\cal C$
updates $E$ to the result of replacing in $E$ the above subformula by $G_i$ (without changing $\Omega$), and calls itself.  

{\em Case 2}: $\alpha$ is a move signifying a switch in an active surface occurrence of a subformula \[G_0\sqd\ldots\sqd \underline{G_m}\sqd G_{m+1}\sqd \ldots\sqd G_n.\] Then $\cal C$ makes the same move $\alpha$, updates $E$ to the result of replacing in it the above subformula by $G_0\sqd\ldots\sqd G_m\sqd \underline{G_{m+1}}\sqd \ldots\sqd G_n$, updates $\Omega$ to $\seq{\Omega,\oo\alpha,\pp\alpha}$, and calls itself.

As we can see from the above description, the value of $E$ starts with $F$ and, moving up along one of the branches of the $\propseqecc$-proof of $F$, eventually stabilizes at $E=E_{\mbox{\em final}}$ for one of the instable formulas of the proof ($E_{\mbox{\em final}}$  is instable because it is the conclusion of $\overline{\mbox{Wait}^\circ}$). Call such a formula $E_{\mbox{\em final}}$  the {\bf limit formula} of the given play. Of course, $\cal C$ is a fair EPM because it will grant permission infinitely many times after reaching the limit formula. 

For reasons fully symmetric to those that we relied upon in Section \ref{ssnd}, at each step of the work of $\cal C$, $\seq{\Omega}E$ can be seen to be exactly the game to which the original game $F$ has been brought down in the play. So, where $\Omega_{\mbox{\em final}}$ is the final value of $\Omega$, $\cal C$ will be the winner in the overall play over $F$ iff $\Omega_{\mbox{\em final}}$ is a $\oo$-won run of $E_{\mbox{\em final}}$ . More precisely, 
for any given valuation $e$ and interpretation $^*$ (the parameters that we have been suppressing so far), the overall play over $e[F^*]$ is won by $\cal C$ iff
$\Omega_{\mbox{\em final}}$ is a $\oo$-won run of $e[E^{*}_{\mbox{\em final}}]$.
Next, taking into account that $\Omega_{\mbox{\em final}}$ only contains (the meaningless) moves in abandoned components and catch-up switches, with some thought one can see that $\Omega_{\mbox{\em final}}$ is a $\oo$-won run of $e[E^{*}_{\mbox{\em final}}]$ if and only if $\elz{E_{\mbox{\em final}}}^*$ is false at $e$. 

Of course, different $\pp$'s strategies (HPMs) $\cal H$ and different valuations $e$ may yield different runs  and hence induce different limit formulas. So, our goal now is    
 to select an interpretation $^*$ such that, for any HPM $\cal H$, there is a valuation $e$ at which $\elz{E_{\mbox{\em final}}}^*$ is false, where 
$E_{\mbox{\em final}}$ is the limit formula of the play of $\cal H$ against $\cal C$ on valuation $e$. This would mean that no HPM can win $F^*$ against $\cal C$. 

Let us fix some standard way of describing HPMs, and let 
\[{\cal H}_1,\ {\cal H}_2,\ {\cal H}_3,\ \ldots\]
be the list of all HPMs arranged according to the lexicographic order of their descriptions, so that each constant $c$ can be considered the code of ${\cal H}_c$. Next, we fix a variable $x$ and agree that, for each constant $c$, \[e_c\] is the valuation with $e_c(x)=c$ (and, say,  $e_c(y)=1$ for any other variable $y\not=x$). 
Further, \[L_c\] will denote the limit formula of the game over $F$ between ${\cal H}_c$, in the role of $\pp$, and our $\cal C$, in the role of $\oo$, on valuation $e_c$. In more precise terms, $L_c$ is the limit formula induced by the ${\cal H}_c$ vs. $\cal C$ run on $e_c$ (remember Lemma \ref{lem}). 

Finally, let \[G_1,\ldots,G_k\] be all (instable) formulas from the $\propseqecc$-proof of $F$ that are obtained by $\overline{\mbox{Wait}^\circ}$. For each such $G_i$, we fix a classical model (true/false assignment for atoms) $M_i$ such that 
\[\mbox{\em $M_i$ makes the elementarization of $G_i$ false.}\]
 And, for each $i\in\{1,\ldots,k\}$, we define the predicate $T_i$ by 
\[\mbox{\em $T_i$ is true at a valuation $e$ iff $L_{e(x)}=G_i$.}\]  
Now we define the interpretation $^*$ by stipulating that, for each atom $p$, 
\[p^*=\mld\{T_i\ |\ 1\leq i\leq k, \ \mbox{\em $p$ is true in $M_i$}\}\]
($\mld {\cal F}$ means the $\mld$-disjunction of all elements of ${\cal F}$, understood as $\tlg$ when the set $\cal F$ is empty.)

Consider an arbitrary $c\in\{1,2,\ldots\}$. As noted earlier, we must have $L_c=G_j$ for one of the $j\in\{1,\ldots,k\}$. Fix this $j$. Observe that, at valuation $e_c$, $T_j$ is true and all other $T_i$ ($i\not=j$, $1\leq i\leq k$) are false. With this fact in mind, it is easy to see that, for every atom $p$, 
$p$ is true in $M_j$ iff the predicate $p^*$ is true at $e_c$. This obviously extends from atoms to their $\gneg,\mlc,\mld$-combinations, 
so that $\elz{G_j}$ is true in $M_j$ iff the predicate ${\elz{G_j}}^*$ is true at $e_c$. And, by our choice 
of the models $M_i$, the formula $\elz{G_j}$, i.e. $\elz{L_c}$, is false in $M_j$. Consequently, ${\elz{L_c}}^*$ is false at 
$e_c$.  But $L_c$ is the limit formula of the play between $\cal H$ and $\cal C$ on $e_c$ and, as we noted earlier, the fact that ${\elz{L_c}}^*$ is false at 
$e_c$ implies that $\cal H$ loses $F^*$ against $\cal C$ on valuation $e_c$. 

Thus, no ${\cal H}_c$ computes $F^*$, meaning that  $F^*$ is not computable, because every HPM is ${\cal H}_c$ for some $c$.  Note also that, as promised in the lemma, the predicate $p^*$ (any atom $p$) is finitary as only the value assigned to $x$ matters. To officially complete the present proof, it remains to show that 
\begin{equation}\label{delta}
\mbox{\em the complexity of $p^*$ is $\Delta_2$ (any atom $p$).}
\end{equation}

Remember that an arithmetical predicate $A(c)$ (with $c$ here seen as a variable) is said to have complexity $\Sigma_2$ iff it can be written as $\exists x\forall y B(c,x,y)$ for some decidable predicate $B(c,x,y)$; and $A(c)$ is of complexity $\Delta_2$ iff both $A(c)$ and $\gneg A(c)$ are of complexity $\Sigma_2$. 
 
We defined $p^*$ as a disjunction of some $T_i$, that we now think of as unary arithmetical predicates and write as $T_i(c)$. Disjunction is known to preserve  $\Delta_2$ --- as well as $\Sigma_2$ ---
complexity, so, in order to verify (\ref{delta}), it would be sufficient to show that each $T_i(c)$ ($1\leq i\leq k$) 
is of complexity $\Delta_2$. Looking at the meaning of $T_i(c)$, this predicate asserts nothing but that  
the value of ${\cal C}$'s internal record of $E$ in the process of playing the $({\cal C},e_c,{\cal H}_c)$-branch (see Lemma \ref{lem}) --- call this branch $B_c$ --- will stabilize at $G_i$. So, $T_i(c)$ can be 
written as $\exists x\forall y(y\geq x\mli K_i(c,y))$, where $K_i(c,y)$ means ``the value of record $E$ at the $y$th computation step of branch $B_c$ is $G_i$".
Furthermore, we know that the value of  $E$ should indeed stabilize at one of the instable 
formulas $G_1,\ldots,G_n$ of the proof. Hence, $\gneg T_i(c)$ is equivalent to $\mld\{T_j(c)\ |\ 1\leq j\leq n, j\not=i\}$. 
Consequently, in order to show that each $T_i(c)$ is of complexity $\Delta_2$, it would suffice to show that 
each $T_i(c)$ is of complexity $\Sigma_2$. For the latter, in turn, verifying that $K_i(c,y)$ is a decidable 
predicate would be sufficient. But $K_i(c,y)$ is indeed decidable. A decision procedure for it first 
constructs the machine ${\cal H}_c$ from number $c$. Then it lets this machine play against ${\cal C}$ on valuation $e_c$ 
as described in the proof idea for Lemma \ref{lem}. In particular, it traces, in parallel, how the configurations of the two machines evolve up to the $y$th computation step of ${\cal C}$, i.e. its $y$th configuration. Then the procedure looks at the value of record $E$ in that configuration, and says ``yes" or ``no" depending on whether the latter is $G_i$ or not. \end{proof}

\section{The completeness of $\propseq$}\label{s9}

\begin{lemma}\label{compl}
If $\propseq\not\vdash F$, then $F$ is not valid (any $\propseq$-formula $F$).

Moreover, if $\propseq\not\vdash F$, then $F^*$ is not computable  for some  interpretation 
$^*$ that interprets all elementary atoms of $F$ as finitary predicates of arithmetical complexity $\Delta_2$, and interprets all general atoms of $F$ as problems of the form
\((A^{1}_{1}\add\ldots\add A_{m}^{1})\adc\ldots\adc (A_{1}^{m}\add\ldots\add A_{m}^{m}),\)
where each $A_{i}^{j}$ is a finitary predicate of arithmetical complexity $\Delta_2$.
\end{lemma}

{\bf Proof idea}. We are going to show that if $\propseq\not\vdash F$, then there is a $\propseqe$-formula $\lceil F\rceil$ of the same form as $F$ that is not provable in $\propseqe$. Precisely, ``the same form as $F$" here means that $\lceil F\rceil$ is the result of rewriting/expanding in $F$ every general atom $P$ as a certain elementary-base formula $\check{P}_{\add}^{\adc}$. This, in view of the already known completeness of $\propseqe$, immediately yields non-validity for $F$. As it turns out, the above formulas  $\check{P}_{\add}^{\adc}$, that we call {\em molecules}, can be chosen to be 
as simple as sufficiently long $\adc$-conjunctions of sufficiently long $\add$-disjunctions of arbitrary ``neutral" (not occurring in $F$ and pairwise distinct) elementary atoms, with the ``sufficient length" 
of those conjunctions/disjunctions being bounded by the number of occurrences of general atoms in $F$. 

Intuitively, the reason why $\propseqe\not\vdash\lceil F\rceil$, i.e. why $\pp$ cannot win (the game represented by) $\lceil F\rceil$, is that a smart environment may start choosing different conjuncts/disjuncts in different occurrences of $\check{P}_{\add}^{\adc}$. The best that $\pp$ can do in such a play is to match any given positive or negative occurrence of $\check{P}_{\add}^{\adc}$ with one (but not more!) 
negative or positive occurrence of the same subgame --- match in the sense that mimic environment's moves in order to keep the subgames/subformulas at the two occurrences identical. Yet, this 
is insufficient for $\pp$ to achieve a guaranteed success. This is so because $\pp$'s matching decisions for $\lceil F\rceil$ could be modeled by appropriate applications of the rule of Match in an attempted $\propseq$-proof for $F$, and so can be --- through the rules of Wait, Choose and Switch ---
either player's decisions required by choice and sequential connectives in the non-molecule parts of $\lceil F\rceil$. A winning strategy ($\propseqe$-proof) for 
$\lceil F\rceil$ would then translate into a $\propseq$-proof for $F$, which, however, does not exist.  \vspace{10pt}  

\begin{proof} Fix a $\propseq$-formula $F$. Let $\cal P$ be the set of all general atoms occurring in $F$. 
Let us fix $m$ as the total number of occurrences of such atoms in $F$;\footnote{In fact, a much smaller $m$ would be sufficient for our purposes. E.g., $m$ can be chosen to be 
such that no given general atom has more than $m$ occurrences in $F$. But why try to economize?} 
if there are fewer than $2$ of such occurrences, then we take $m=2$. 

For the rest of this section, let us agree that 
\[\mbox{\em $a,b$ always range over $\{1,\ldots,m\}$.}\]

For each $P\in{\cal P}$ and each $a,b$, let us fix an  elementary atom
\begin{itemize} 
\item $\check{P}_{b}^{a}$\label{z53} 
\end {itemize}
not occurring in $F$. We assume that $\check{P}^{a}_{b}\not=\check{Q}^{c}_{d}$ as 
long as either $P\not= Q$ or $a\not=c$ or $b\not=d$. Note that the $\check{P}_{b}^{a}$ are elementary atoms despite our
``tradition" according to which the capital letters $P,Q,\ldots$ stand for general atoms.    

Next, for each $P\in{\cal P}$ and each $a$, we define

\begin{itemize}
\item $\check{P}^{a}_{\add}\ =\ \check{P}^{a}_{1}\add\ldots\add \check{P}^{a}_{m}$.
\end{itemize}

Finally, for each $P\in{\cal P}$, we define 
\begin{itemize}
\item $\check{P}^{\adc}_{\add}\ = \check{P}^{1}_{\add}\adc\ldots\adc\check{P}^{m}_{\add}$, \ i.e. \ 
\(\check{P}^{\adc}_{\add}\ =\ (\check{P}^{1}_{1}\add\ldots\add \check{P}^{1}_{m})\adc\ldots\adc (\check{P}^{m}_{1}\add\ldots\add \check{P}^{m}_{m}).\)
\end{itemize}

We refer to the above formulas $\check{P}^{a}_{b}$, $\check{P}^{a}_{\add}$ and $\check{P}^{\adc}_{\add}$ as 
{\bf molecules},
in particular, {\bf $P$-based molecules}. 
To differentiate between the three sorts of molecules, we call the molecules of the type $\check{P}^{a}_{b}$ 
{\bf small}, call the molecules of the type $\check{P}^{a}_{\add}$ 
{\bf medium}, and call the molecules of the type $\check{P}^{\adc}_{\add}$ 
{\bf large}. Thus, where $k$ is the cardinality of $\cal P$, altogether there are $k$ large molecules, $k\times m$ medium molecules and $k\times m\times m$ small molecules.

For simplicity, for the rest of this section we assume/pretend that the languages of  
$\propseq$ and $\propseqe$ have no nonlogical atoms other than those occurring in $F$ plus the atoms $\check{P}_{a}^{b}$ 
($P\in {\cal P}$, \ $a,b\in\{1,\ldots,m\}$). This way the scope of the term ``formula" is correspondingly redefined. 
 
An occurrence of a molecule $M$ in a formula can be {\bf positive} or {\bf negative}. While a positive occurrence literally means $M$, a negative occurrence looks 
like $\gneg M$, which --- unless $M$ is a small molecule --- should be considered a standard abbreviation. For example, a negative occurrence of  the medium 
molecule $\check{P}^{a}_{1}\add\ldots\add \check{P}^{a}_{m}$ is nothing but an (``ordinary'', positive) occurrence of $\gneg\check{P}^{a}_{1}\adc\ldots\adc\gneg \check{P}^{a}_{m}$. One should be especially careful when applying the terms ``positive occurrence'' and ``negative occurrence'' to small molecules, as here the meaning of our terminology somewhat diverges from its earlier-used meaning for atoms. Specifically, a positive occurrence of a small molecule $\check{P}^{a}_{b}$ means --- as expected --- an occurrence that comes without $\gneg$ in the formula. As for a negative occurrence of the {\em molecule} $\check{P}^{a}_{b}$ (as opposed to the {\em atom} $\check{P}^{a}_{b}$), it means an occurrence of the subformula 
$\gneg \check{P}^{a}_{b}$ rather than just the $\check{P}^{a}_{b}$ part of it under $\gneg$. So, for example, the result of replacing the negative occurrence of 
$\check{P}^{a}_{b}$ by $Q$ in the formula $E\mld \gneg \check{P}^{a}_{b}$ is the formula $E\mld Q$ rather than $E\mld\gneg Q$, as $\gneg$ was a part of what we call 
a ``negative occurrence of $\check{P}^{a}_{b}$''.   

Let us say that a (positive or negative) occurrence of a molecule in a given $\propseqe$-formula is {\bf independent} iff it is not a part of another (``larger") molecule. For example, the negative occurrence of the medium molecule $\check{P}_{1}^{1}\add\ldots\add\check{P}_{m}^{1}$     in the following formula is independent while its positive occurrence is not:
\[(\gneg\check{P}_{1}^{1}\adc\ldots\adc\gneg\check{P}_{m}^{1}) \mld \bigl((\check{P}_{1}^{1}\add\ldots\add\check{P}_{m}^{1})\adc\ldots\adc(\check{P}_{1}^{m}\add\ldots\add\check{P}_{m}^{m})\bigr).\] 
Of course, surface occurrences of molecules are always independent, and so 
are any --- surface or non-surface --- occurrences of large molecules.\vspace{5pt}

We say that a $\propseqe$-formula $E$ is {\bf good} iff the following conditions are satisfied:
\begin{description}
\item[Condition (i):] $E$ contains at most $m$ independent occurrences of molecules.
\item[Condition (ii):] Only large molecules (may) have independent non-surface occurrences in $E$.  
\item[Condition (iii):] Each small molecule has at most one positive and at most one negative 
independent occurrence in $E$. 
\item[Condition (iv):] For each medium molecule $\check{P}^{a}_{\add}$, $E$ has at most one positive independent occurrence of 
$\check{P}^{a}_{\add}$, and when $E$ has such an occurrence, then for no $b$ does $E$ have a positive independent occurrence of the small molecule $\check{P}^{a}_{b}$.\vspace{2pt}
\end{description}

Let $E$ be a $\propseqe$-formula. By an {\bf isolated} small molecule of $E$ (or {\bf $E$-isolated} small molecule, or a small molecule {\bf isolated in $E$}) 
we will mean a small molecule that has exactly one independent occurrence in $E$. We will say that such a molecule is {\bf positive} or {\bf negative} depending on whether its independent occurrence in $E$ is positive or negative. 

Next, the {\bf floorification} of $E$, denoted 
  \[\lfloor E\rfloor,\]
is the result of replacing in $E$ every positive (resp. negative) independent occurrence of every $P$-based (each $P\in{\cal P}$) large, medium and $E$-isolated small molecule\footnote{Remember what was said earlier about the meaning of ``negative occurrence'' for small molecules.} by the general literal $P$ (resp. $\gneg P$).\vspace{10pt}

\noindent {\bf Claim 1.} {\em For any good $\propseqe$-formula $E$, if $\propseqe\vdash E$, then $\propseq\vdash\lfloor E\rfloor$.}\vspace{8pt}

To prove this claim, assume $E$ is a good $\propseqe$-formula, and $\propseqe\vdash E$. By induction on the length of the 
$\propseqe$-proof of $E$, we want to show that $\propseq\vdash\lfloor E\rfloor$. We need to consider the following three cases, depending on which of the three rules of $\propseqe$ was used (last) to derive $E$.\vspace{8pt}

{\em CASE 1:} $E$ is derived by Wait. Let us fix the set $\vec{H}$ of premises of $E$. Each formula $H\in\vec{H}$ is provable in $\propseqe$. Hence, by the induction hypothesis, we have:

\begin{equation}\label{may25}
\mbox{\em For any $H\in\vec{H}$, if $H$ is good, then $\propseq\vdash\lfloor H\rfloor$.}
\end{equation}

We consider the following three subcases. The first two subcases are not mutually exclusive, and either one can be chosen when both of them apply. Specifically, Subcase 1.1 (resp. 1.2) is about when $E$ has a positive (resp. negative) surface occurrence of a large (resp. medium) molecule. Then, as we are going to see, replacing that molecule by a ``safe'' $\adc$-conjunct of it, corresponding to a smart environment's possible move, yields a good formula $H$ from $\vec{H}$ such that $\lfloor E\rfloor=\lfloor H\rfloor$. This,  by (\ref{may25}), automatically means the $\propseq$-provability of $\lfloor E\rfloor$. The remaining Subcase 1.3 is about when all surface occurrences of large (resp. medium) molecules in $E$ are negative (resp. positive). This will be shown to imply that $\lfloor E\rfloor$ follows from the floorifications of some elements of $\vec{H}$ by Wait for ``the same reasons as'' $E$ follows from $\vec{H}$.\vspace{5pt}  

{\em Subcase 1.1:} $E$ has a positive surface occurrence of a large molecule $\check{P}_{\add}^{\adc}$, i.e., an occurrence of 
\[\check{P}_{\add}^{1}\adc\ldots\adc \check{P}_{\add}^{m}.\]
Pick any $a\in\{1,\ldots,m\}$ such that neither the medium molecule 
$\check{P}_{\add}^{a}$ nor any small molecule $\check{P}_{b}^{a}$ (whatever $b$) have independent occurrences in $E$. Such an $a$ exists, for otherwise we would have at least $m+1$ independent occurrences of molecules  in $E$ (including the occurrence of $\check{P}_{\add}^{\adc}$), which violates Condition (i) of  the definition of ``goodness''. Let $H$ be the result of replacing in $E$
the above occurrence of $\check{P}_{\add}^{\adc}$ by $\check{P}_{\add}^{a}$. Clearly $H\in\vec{H}$. Observe that when transferring from $E$ to $H$, we just ``downsize" $\check{P}_{\add}^{\adc}$ and otherwise do not create any additional independent occurrences of molecules, so Condition (i) continues to be satisfied for $H$. Neither do we introduce any new non-surface occurrences of molecules or any new independent occurrences of small molecules, so Conditions (ii) 
 and (iii) also continue to hold for $H$. And 
our choice of $a$ obviously guarantees that so does Condition (iv). To summarize, $H$ is good. Therefore, by (\ref{may25}),  $\propseq\vdash \lfloor H\rfloor$. 
Finally, note that, when floorifying a given formula, both $\check{P}_{\add}^{\adc}$ and $\check{P}_{\add}^{a}$ get replaced by the same atom $P$; and, as the only difference between $E$ and $H$ is that $H$ has $\check{P}_{\add}^{a}$ where $E$ has $\check{P}_{\add}^{\adc}$, 
obviously $\lfloor H \rfloor=\lfloor E\rfloor$. Thus, $\propseq\vdash \lfloor E\rfloor$.\vspace{5pt} 

{\em Subcase 1.2:} $E$ has a negative surface occurrence of a medium molecule $\check{P}_{\add}^{a}$ --- that is, an occurrence of 
\[\gneg\check{P}_{a}^{1}\adc\ldots\adc \gneg\check{P}_{a}^{m}.\] Pick any $b$ such 
that $E$ does not have an independent occurrence of $\check{P}_{b}^{a}$. Again, in view of Condition (i), such a $b$ exists. Let $H$ be the result of replacing in $E$
the above occurrence of $\gneg\check{P}_{a}^{1}\adc\ldots\adc \gneg\check{P}_{a}^{m}$ by $\gneg\check{P}_{b}^{a}$. Certainly $H\in\vec{H}$. Conditions 
(i) and (ii) continue to hold for $H$ for the same reasons as in Subcase 1.1. In view of our choice of $b$, Condition (iii) is also inherited by $H$ from $E$. And so is Condition (iv), because $H$ has the same positive occurrences of (the same) molecules as $E$ does.
Thus, $H$ is good. Therefore, by (\ref{may25}),  $\propseq\vdash \lfloor H\rfloor$. 
It remains to show that $\lfloor H\rfloor=\lfloor E\rfloor$.  Note that when floorifying $E$, $\check{P}_{\add}^{a}$ gets replaced by $P$. But so does $\check{P}_{b}^{a}$ when floorifying $H$ because, by our choice of $b$, $\check{P}_{b}^{a}$ is an isolated small molecule of $H$.
Since the only difference between $H$ and $E$ is that $H$ has $\check{P}_{b}^{a}$ where $E$ has $\check{P}_{\add}^{a}$, it is then obvious that indeed $\lfloor H\rfloor=\lfloor E\rfloor$.\vspace{5pt}

{\em Subcase 1.3:} Neither of the above two conditions is satisfied. This means that in $E$ all surface occurrences of large molecules are negative, and all surface occurrences of medium molecules  are positive. Every such occurrence is an occurrence of a $\add$-formula whose surface occurrences, as we remember, get replaced by $\tlg$ when transferring from $E$ to $\elz{E}$; but the same happens to the corresponding occurrences of $\gneg P$ or $P$ in $\lfloor E\rfloor$ when transferring from $\lfloor E\rfloor$ to $\elz{\lfloor E\rfloor}$. Based on this observation, with a little thought we can see that 
$\elz{\lfloor E\rfloor}$ is ``almost the same" as $\elz{E}$; specifically, the only difference between these two formulas 
is that $\elz{\lfloor E\rfloor}$ has $\tlg$ where $\elz{E}$ has isolated small molecules (positive or negative).  Obviously this means that $\elz{\lfloor E\rfloor}$ is a substitutional instance of $\elz{E}$ --- the result of substituting, in the latter, each positive isolated small molecule by $\tlg$ and the atomic part (the part under $\gneg$) of each negative isolated small molecule by $\twg$.    
As $E$ is derived by Wait, $\elz{E}$ is classically valid.  Therefore $\elz{\lfloor E\rfloor}$, as a substitutional instance of $\elz{E}$, is also classically valid. So, we have:

\begin{equation}\label{mar2}
\mbox{\em $\lfloor E\rfloor$ is stable.}
\end{equation} 

Now consider an arbitrary formula $H'$ that is the result of replacing in $\lfloor E\rfloor$ a surface occurrence of a subformula 
$G'_1\adc\ldots\adc G'_n$ by $G'_i$ for some $i\in \{1,\ldots,n\}$. Our goal is to show that 
\begin{equation}\label{oct28a}
\mbox{\em $\propseq\vdash H'$ (arbitrary $H'$ satisfying the above condition)}.
\end{equation}
The logical structure of $E$ is the same as that of $\lfloor  E\rfloor$, with the only difference that, wherever $\lfloor  E\rfloor$ has general literals, $E$ has  molecules. Hence $E$ has an  
occurrence of a subformula $G_1\adc\ldots\adc G_n$ where  $\lfloor E\rfloor$ has the above occurrence of $G'_1\adc\ldots\adc G'_n$. Let then 
$H$ be the result of replacing $G_1\adc\ldots\adc G_n$ by $G_i$ in $E$. Of course $H\in\vec{H}$. 
 So, in view of (\ref{may25}), it would suffice to show (in order to verify (\ref{oct28a}))  that $H$ is good and 
$H'=\lfloor H\rfloor$. Let us first see that $H$ is good. When transferring from $E$ to $H$, Condition (i) is inherited by $H$ for the same or a similar reasons as in all of the previous cases. So is Condition (ii) because we are not creating any new non-surface occurrences. Furthermore, notice that $G_1\adc\ldots\adc G_n$ 
is not a molecule, for otherwise in $\lfloor E\rfloor$ we would have a general literal rather than $G'_1\adc\ldots\adc G'_n$.
Hence, in view of Condition (ii), $G_i$ is not a small or medium molecule. This means that, when transferring from $E$ to $H$, we are not creating new 
independent/surface occurrences of any small or medium molecules, so that 
Conditions (iii) and (iv) are also inherited by $H$ from $E$. To summarize, $H$ is indeed good. 
Finally, it is also rather obvious that $H'=\lfloor H\rfloor$. The only case when we might have $H'\not=\lfloor H\rfloor$
would be if there was a small molecule $\check{P}_{b}^{a}$ isolated in $E$ but not in $H$, or vice versa (so that the independent occurrence of that molecule in $E$ 
would become $P$ in  $\lfloor E\rfloor$ and hence in $H'$ but stay $\check{P}_{b}^{a}$ in $\lfloor H\rfloor$, or vice versa). But, as we   observed just a little  while ago, $E$ and $H$ do not differ in what independent/surface occurrences of what small molecules they have.

Next, consider an arbitrary formula $H''$ that is the result of replacing in $\lfloor E\rfloor$ a surface occurrence of a subformula 
$G''_0\sqc\ldots\sqc G''_n$ by $G''_1\sqc\ldots\sqc G''_n$. Our goal is to show that 
\begin{equation}\label{oct28b}
\mbox{\em $\propseq\vdash H''$ (arbitrary $H''$ satisfying the above condition)}.
\end{equation}
This case is very similar to the case handled in the previous paragraph. The logical structure of  $E$ the same as that of $\lfloor  E\rfloor$, so
$E$ has an  
occurrence of a subformula $G_0\sqc\ldots\sqc G_n$ where  $\lfloor E\rfloor$ has the above occurrence of $G''_0\sqc\ldots\sqc G''_n$. Let then 
$H$ be the result of replacing $G_0\sqc\ldots\sqc G_n$ by $G_1\sqc\ldots\sqc G_n$ in $E$. Of course $H\in\vec{H}$. Continuing arguing as in the previous paragraph, we find that $H$ is good and that  $H''=\lfloor H\rfloor$, which, by (\ref{may25}), implies the desired (\ref{oct28b}).

Based on (\ref{mar2}), (\ref{oct28a}) and (\ref{oct28b}), we find that $\lfloor E\rfloor$ is derivable in $\propseq$ by Wait.\vspace{8pt}

The remaining two CASES 2 and 3 are about when $E$ is derived by Choose or Switch from a premise $H$. Such an $H$ turns out to be good and hence (by the induction hypothesis) its floorification $\propseq$-provable. And, ``almost always'' $\lfloor E\rfloor$ follows from $\lfloor H\rfloor$ by Choose or Switch for the same reasons as $E$ follows from $H$. An exception is the special case of Choose when $H$ is the result of replacing in $E$ a positive occurrence of a medium molecule 
$\check{P}_{\add}^{a}$ by one of its disjuncts $\check{P}_{b}^{a}$ such that $E$ has a negative independent occurrence of $\check{P}_{b}^{a}$. Using our earlier terms, this is a step signifying $\pp$'s (final) decision to ``match'' the two $P$-based molecules. In this case, while $\lfloor E\rfloor$ does not follow from $\lfloor H\rfloor$ by Choose, it does so by Match. The secret is that the two $P$-based molecules are non-isolated small molecules in $H$ and hence remain elementary literals in $\lfloor H\rfloor$, while they turn into general literals in $\lfloor E\rfloor$.\vspace{8pt} 
 
{\em CASE 2:} $E$ is derived by Choose. That is, we have $\propseqe\vdash H$, where $H$ is the result of replacing in
$E$ a  surface occurrence of a subformula $G =G_1\add\ldots\add G_n$ by $G_i$ for some $i\in\{1,\ldots,n\}$. Fix these formulas and this number $i$. 
Just as in CASE 1 (statement (\ref{may25})), based on the induction hypothesis, we have:
\begin{equation}\label{may26}
\mbox{\em If $H$ is good, then $\propseq\vdash\lfloor H\rfloor$.}
\end{equation}

We need to consider the following three subcases that cover all possibilities:\vspace{5pt}

{\em Subcase 2.1:} $G$ is not a molecule. Reasoning (almost) exactly as we did at the end of our discussion of Subcase 1.3, 
we find that $H$ is good. Therefore, by (\ref{may26}), $\propseq\vdash\lfloor H\rfloor$. Now, a little thought can 
convince us that $\lfloor E\rfloor$ follows from $\lfloor H\rfloor$ by Choose, so that $\propseq\vdash\lfloor E\rfloor$.\vspace{3pt}

{\em Subcase 2.2:} $G$ is a negative large molecule $\gneg\check{P}_{\add}^{1}\add\ldots\add \gneg\check{P}_{\add}^{m}$. So, 
$G_i=\gneg\check{P}_{\add}^{i}$. A (now already routine for us) examination of Conditions (i)-(iv) reveals that 
each of these four conditions are inherited by $H$ from $E$, so that $H$ is good. Therefore, by (\ref{may26}), 
$\propseq\vdash\lfloor H\rfloor$. Now, $\lfloor H\rfloor$ can be easily seen to be the same as $\lfloor E\rfloor$,
and thus $\propseq\vdash\lfloor E\rfloor$.\vspace{3pt}

{\em Subcase 2.3:} $G$ is a positive medium molecule $\check{P}_{1}^{a}\add\ldots\add \check{P}_{m}^{a}$. So, $G_i=\check{P}_{i}^{a}$. There are two subsubcases to consider:

{\em Subsubcase 2.3.1:} $E$ contains no independent occurrence of $\check{P}_{i}^{a}$. One can easily verify that $H$ is good and that
$\lfloor H\rfloor =\lfloor E\rfloor$. By (\ref{may26}), we then get the desired $\propseq\vdash\lfloor E\rfloor$.

{\em Subsubcase 2.3.2:} $E$ has an independent occurrence of $\check{P}_{i}^{a}$.  Since $E$ also has a positive independent occurrence of $\check{P}_{\add}^{a}$, 
Condition (iv) implies that the above occurrence of $\check{P}_{i}^{a}$ in $E$ is negative. This, in conjunction with Condition (iii), means that $E$ does not have any other independent occurrences of $\check{P}_{i}^{a}$, and thus $H$ has exactly two --- one negative and one positive --- independent occurrences of $\check{P}_{i}^{a}$. 
This guarantees that Condition (iii) is satisfied for $H$, because $H$ and $E$ only differ in that $H$ has $\check{P}_{i}^{a}$ where $E$ has $\check{P}_{\add}^{a}$. Conditions (i) and (ii) are straightforwardly inherited by $H$ from $E$. Finally, Condition (iv) also transfers from $E$ to $H$ because, even though $H$ --- unlike $E$ --- has a positive independent occurrence of $\check{P}_{i}^{a}$, it no longer has a positive independent occurrence of $\check{P}_{\add}^{a}$ (which, by the same Condition (iv) for $E$, was unique in $E$). Thus, $H$ is good and, by (\ref{may26}), $\propseq\vdash\lfloor H\rfloor$. Note that since $H$ is good, by 
Condition (ii), both of the independent occurrences of $\check{P}_{i}^{a}$ in it 
are surface occurrences. The same, of course, is true for the corresponding occurrences of $\check{P}_{i}^{a}$ and 
$\check{P}_{\add}^{a}$ in $E$. Let us now compare $\lfloor E\rfloor$ with $\lfloor H\rfloor$. According to our earlier observation, $\check{P}_{i}^{a}$ only has one independent occurrence in $E$, i.e. $\check{P}_{i}^{a}$ is $E$-isolated. 
Hence the independent occurrence of $\check{P}_{i}^{a}$, just as that of $\check{P}_{\add}^{a}$, gets replaced by $P$ when floorifying $E$. On the other hand, $\check{P}_{i}^{a}$ is no longer isolated in $H$, so the two 
independent occurrences of it stay as they are when floorifying $H$. Based on this observation, we can easily see that the only difference between $\lfloor E\rfloor$ and $\lfloor H\rfloor$ is that 
$\lfloor E\rfloor$ has the general atom $P$ where $\lfloor H\rfloor$ has the (two occurrences of) elementary atom $\check{P}_{i}^{a}$. Since 
 $\lfloor E\rfloor$ does not contain $\check{P}_{i}^{a}$ (because the only independent occurrence of it in $E$, as well as all  large and medium $P$-based molecules, got replaced by $P$ when floorifying $E$), and since we are talking about two --- one positive and one negative --- surface occurrences of $P$ in $\lfloor E\rfloor$, we find that $\lfloor E\rfloor$ follows
from $\lfloor H\rfloor$ by Match. We already know that $\propseq\vdash\lfloor H\rfloor$. Hence 
 $\propseq\vdash\lfloor E\rfloor$.\vspace{5pt}

{\em CASE 3:} $E$ is derived by Switch. That is, we have $\propseqe\vdash H$, where $H$ is the result of replacing in
$E$ a  surface occurrence of a subformula \(G_0\sqd  \ldots\sqd  G_k\) by 
\(G_1\sqd \ldots \sqd G_k.\)
Just as in CASES 1 and 2, based on the induction hypothesis, we have:
\begin{equation}\label{may264}
\mbox{\em If $H$ is good, then $\propseq\vdash\lfloor H\rfloor$.}
\end{equation}
Reasoning as in the previous cases, we further find that  
 $H$ is good, and thus,  by (\ref{may264}), $\propseq\vdash\lfloor H\rfloor$. Now, a moment's thought convinces us that $\lfloor E\rfloor$ follows from $\lfloor H\rfloor$ by Switch, so that $\propseq\vdash\lfloor E\rfloor$.\vspace{5pt}

Claim 1 is proven.\vspace{8pt}

Now we are very close to finishing our proof of Lemma \ref{compl}. Assume $\propseq\not\vdash F$. 
Let $\lceil F\rceil$ be the result of replacing in $F$ all occurrences of each general atom $P\in{\cal P}$ by $\check{P}_{\add}^{\adc}$. Obviously $\lceil F\rceil$ is good. Clearly we also have $\lfloor\lceil 
F\rceil\rfloor=F$, so that $\propseq\not\vdash \lfloor\lceil 
F\rceil\rfloor$. Therefore, by Claim 1, $\propseqe\not \vdash\lceil F\rceil$. Hence, by 
Lemma \ref{oct5}, there is an interpretation $^\dagger$ that interprets every elementary atom as a finitary predicate 
of arithmetical complexity $\Delta_2$, such that 
\begin{equation}\label{K}
\mbox{\em $\lceil F\rceil^\dagger$ is not computable.}
\end{equation} 

Let $^*$ be an interpretation such that:
\begin{itemize}
\item $^*$ agrees with $^\dagger$ on all elementary atoms;
\item $^*$ interprets each atom $P\in{\cal P}$ as $(\check{P}_{\add}^{\add})^{\dagger}$.\vspace{2pt}
\end{itemize}
 
Clearly $^*$ interprets atoms as promised in our Lemma \ref{compl}. It is also obvious that 
$F^* = \lceil F\rceil^\dagger$. Therefore, by 
(\ref{K}), \ $F^*$ is not computable, and the lemma is proven. 
\end{proof}

\section{A first-order extension of $\propseq$}

Here we introduce a first-order extension of $\propseq$ called $\predseq$. The latter is also a conservative extension of logic {\bf CL4} (proven to be sound and complete in \cite{Japtcs2}), obtained by augmenting its language with $\sqc,\sqd$. 

For each arity $n$, the language of $\predseq$ has infinitely many nonlogical $n$-ary {\em elementary letters} $p$, $q$, \ldots and {\em general letters} $P$, $Q$, \ldots. An {\em elementary atom} is $p(t_1,\ldots,t_n)$, where $p$ is an $n$-ary elementary letter and each $t_i$ is a {\em term} (a variable or a constant). {\em General atoms} are defined similarly. As in the case of $\propseq$, each interpretation $^*$ is required to interpret elementary atoms as elementary games, and general atoms as any static games. Such an interpretation $^*$ then extends to all formulas by commuting with the operation of substitution of variables by terms, and seeing all logical operators as the corresponding operations on games. There are some straightforward additional ``admissibility'' conditions on interpretations to avoid collisions of variables and some other unpleasant effects. We refer for details to \cite{Japfin} or \cite{Japtcs2}.

The logical vocabulary of $\predseq$ is that of $\propseq$ plus the four quantifiers $\ada,\ade,\cla,\cle$. Formulas, that we refer to as {\bf $\predseq$-formulas}, are built from atoms, variables, constants and logical operators in the standard way. As before, negation is officially allowed to be applied only to nonlogical atoms. For safety, we also require that no variable should have both free and bound occurrences in the same formula. The concepts of validity and uniform validity straightforwardly extend from $\propseq$-formulas to $\predseq$-formulas. So do most of the technical concepts defined earlier for the language of $\propseq$. Two of those still need to be slightly redefined. Namely, a {\bf surface occurrence} now is an occurrence that is not in the scope of a choice connective or a choice quantifier and is not in the tail of any sequential subformula. And
the {\bf elementarization} of a formula $F$ now means the result of replacing in the capitalization of $F$ every surface occurrence of the form $G_1\adc\ldots\adc G_n$ or $\ada x G$ by $\twg$, every surface occurrence of the form $G_1\add\ldots\add G_n$ or $\ade x G$ by $\tlg$, and every surface occurrence of each general literal 
 by $\tlg$. 
Finally,  a formula is said to be {\bf stable} iff its elementarization is a valid formula of classical first-order logic; otherwise it is {\bf instable}.\vspace{10pt}

In the above language, the logic is axiomatized as follows:

\begin{definition}\label{defcl9a}
The rules of inference of $\predseq$ are: 

\begin{description}
\item[Wait:]  $\vec{H}\vdash F$, where $F$ is stable and $\vec{H}$ is a set of formulas satisfying the following three conditions: 
\begin{itemize} 
\item whenever $F$ has a surface occurrence of a subformula $G_1\adc\ldots\adc G_n$, for each 
$i\in\{1,\ldots,n\}$, $\vec{H}$ contains the result of replacing that occurrence in $F$ by $G_i$;
\item whenever $F$ has a surface occurrence of a subformula $G_0\sqc G_1\sqc\ldots\sqc G_n$, $\vec{H}$ contains the result of replacing that occurrence in $F$ by $G_1\sqc\ldots\sqc G_n$;
\item whenever $F$ has a surface occurrence of a subformula $\ada xG(x)$, $\vec{H}$ contains the result of replacing that occurrence in $F$ by $G(y)$, where $y$ is a variable not occurring in $F$. 
\end{itemize}
\item[$\add$-Choose:]  $H\vdash F$, where $H$ is the result of replacing in $F$ a surface occurrence of a subformula $G_1\add\ldots\add G_n$  by $G_i$ for some $i\in\{1,\ldots, n\}$.
\item[$\ade$-Choose:]  $H\vdash F$, where $H$ is the result of replacing in $F$ a surface occurrence of a subformula $\ade xG(x)$ by $G(t)$, where $t$ is a term with no bound occurrence in $F$.  
\item[Switch:]  $H\vdash F$, where $H$ is the result of replacing in $F$ a surface occurrence of a subformula $G_0\sqd G_1\sqd\ldots\sqd G_n$  by 
$G_1\sqd\ldots\sqd G_n$.
\item[Match:] $H\vdash F$, where $H$ is the result of replacing in $F$ two --- one positive and one negative ---
surface occurrences of some $n$-ary general letter $P$ by a nonlogical elementary letter $p$ that does not occur in $F$.
\end{description}
\end{definition}

There is every reason to expect that the already known soundness and completeness theorems for $\propseq$ and {\bf CL4} extend to their common extension $\predseq$. A proof of this fact can be obtained by combining the techniques and ideas employed in the above two soundness/completeness proofs. The author does not see any reasons why such an (almost mechanical) combination would not work. This job has to be actually done though, and until then the following statement should be officially considered only  a conjecture rather than a theorem:

  \begin{conjecture}\label{thcl2new}
$\predseq\vdash F$ iff $F$ is valid $($any $\predseq$-formula $F$$)$.
Furthermore:

a) There is an effective procedure that takes a $\predseq$-proof of an arbitrary formula $F$ and 
constructs an HPM $\cal H$ such that,  for every interpretation $^*$, \ $\cal H$ computes $F^*$.

b) If $\predseq\not\vdash F$, then $F^*$ is not computable  for some  interpretation 
$^*$ that interprets all elementary atoms of $F$ as finitary predicates of arithmetical complexity $\Delta_2$, and interprets all general atoms of $F$ as problems of the form
\((A^{1}_{1}\add\ldots\add A_{m}^{1})\adc\ldots\adc (A_{1}^{m}\add\ldots\add A_{m}^{m}),\)
where each $A_{i}^{j}$ is a finitary predicate of arithmetical complexity $\Delta_2$.
\end{conjecture}

 \begin{theorem}\label{nw}
The $\cla,\cle$-free fragment of $\predseq$ (i.e., the set of all $\cla,\cle$-free theorems of $\predseq$) is decidable in polynomial space. 
\end{theorem}

\begin{proof} This is similar to the corresponding theorem for {\bf CL4} proven in \cite{Japtcs2} --- the presence of $\sqc,\sqd$ in formulas hardly creates any differences. So, it would be sufficient to give just a schematic outline of the proof idea for our present theorem. A polynomial-space decision algorithm for $\predseq$-provability of a $\cla,\cle$-free formula $F$ is a recursive one. At each level of recursion, it tests all possible premises (for any of the four possible rules) for $F$, calling itself on those premises. In each case, there is only a finite number of premises to test. Strictly speaking, there are infinitely many possible premises for Match. However, those premises only differ from each other in selecting a fresh elementary letter to replace two occurrences of a general letter. Of course, one choice of such a letter can yield a provable premise iff any other choice can, so considering only one premise would be sufficient. Similarly for the rules associated with $\ada$ and $\ade$ ($\ada$-Choose and Wait). Each time the algorithm deals with Wait, it has to test whether the conclusion is stable. While stability of $\predseq$-formulas is generally undecidable, for $\cla,\cle$-free $\predseq$-formulas it can be easily seen to be decidable in linear space. Each level of recursion thus only takes polynomial space. And the depth of recursion is limited by the size of the formula. So, the whole algorithm runs in polynomial space. 
\end{proof}

We close this section with two examples that can help us get some syntactic feel of $\predseq$ and appreciate its value as a problem-solving tool.

\begin{example}\label{n29a}
Remember formula (\ref{nov16d}), claimed to be valid in Section \ref{s2seq}. Below is a $\predseq$-proof of it:\vspace{10pt}

1. $\gneg q(y)\mld P(y)\mld q(y)$ (from $\{\}$  by Wait)\vspace{5pt}

2. $\gneg q(y)\mld \bigl(\gneg p(y)\sqc  P(y)\bigr)\mld q(y)$ (from $\{1\}$  by Wait)\vspace{5pt}

3. $\gneg P(y)\mld \bigl(\gneg p(y)\sqc  P(y)\bigr)\mld P(y)$ (from 2 by Match)\vspace{5pt}

4. $\gneg P(y)\mld \bigl(\gneg p(y)\sqc  P(y)\bigr)\mld \bigl(P(y)\add\gneg P(y)\bigr)$ (from 3 by $\add$-choose)\vspace{5pt}

5. $\gneg P(y)\mld q(y)\mld  \gneg q(y)$ (from $\{\}$ by Wait)\vspace{5pt}

6. $\bigl(p(y)\sqc\gneg P(y)\bigr)\mld q(y)\mld  \gneg q(y)$ (from $\{5\}$ by Wait)\vspace{5pt}

7. $\bigl(p(y)\sqc\gneg P(y)\bigr)\mld P(y)\mld \gneg P(y)$ (from 6 by Match)\vspace{5pt}

8. $\bigl(p(y)\sqc\gneg P(y)\bigr)\mld P(y)\mld \bigl(P(y)\add\gneg P(y)\bigr)$ (from 7 by $\add$-Choose)\vspace{5pt}

9. $\bigl(p(y)\sqc\gneg P(y)\bigr)\mld \bigl(\gneg p(y)\sqc  P(y)\bigr)\mld \bigl(P(y)\add\gneg P(y)\bigr)$ (from $\{4,8\}$  by Wait)\vspace{5pt}

10. $\bigl(P(y)\sqc\gneg P(y)\bigr)\mld \bigl(\gneg P(y)\sqc P(y)\bigr)\mld \bigl(P(y)\add\gneg P(y)\bigr)$ (from 9 by Match)\vspace{5pt}

11. $\bigl(P(y)\sqc\gneg P(y)\bigr)\mld\ade x\bigl(\gneg P(x)\sqc  P(x)\bigr)\mld \bigl(P(y)\add\gneg P(y)\bigr)$ (from 10 by $\ade$-Choose)\vspace{5pt}

12. $\ade x\bigl(P(x)\sqc\gneg P(x)\bigr)\mld\ade x\bigl(\gneg P(x)\sqc P(x)\bigr)\mld \bigl(P(y)\add\gneg P(y)\bigr)$ (from 11 by $\ade$-Choose)\vspace{5pt}

13. $\ade x\bigl(P(x)\sqc\gneg P(x)\bigr)\mld\ade x\bigl(\gneg P(x)\sqc P(x)\bigr)\mld \ada x\bigl(P(x)\add\gneg P(x)\bigr)$ (from $\{12\}$ by Wait)

\end{example}
 
\begin{example}\label{n29b}
Let us now see the $\predseq$-provability of formula  (\ref{nov16e}) from Section \ref{s2seq}. Below $E\not\leftrightarrow F$ is an abbreviation of 
$\gneg (E\leftrightarrow F)$, i.e., of $(E\mlc \gneg F)\mld (F\mlc \gneg E)$:\vspace{10pt}

1. $\bigl(q(z)\not\leftrightarrow p(v)\bigr)\mld \gneg p(v)\mld q(z)$ (from $\{\}$ by Wait)\vspace{5pt}

2. $\bigl(q(z)\not\leftrightarrow p(v)\bigr)\mld \gneg p(v)\mld \bigl(\gneg q(z)\sqd q(z)\bigr)$ (from  1 by Choose)\vspace{5pt}

3. $\bigl(q(z)\not\leftrightarrow p(v)\bigr)\mld \bigl(p(v)\sqc\gneg p(v)\bigr)\mld \bigl(\gneg q(z)\sqd q(z)\bigr)$ (from $\{2\}$ by Wait)\vspace{5pt}

4. $\bigl(q(z)\not\leftrightarrow p(v)\bigr)\mld \ade x\bigl(p(x)\sqc\gneg p(x)\bigr)\mld \bigl(\gneg q(z)\sqd q(z)\bigr)$ (from 3 by $\ade$-Choose)\vspace{5pt}

5. $\ada y\bigl(q(z)\not\leftrightarrow p(y)\bigr)\mld \ade x\bigl(p(x)\sqc\gneg p(x)\bigr)\mld \bigl(\gneg q(z)\sqd q(z)\bigr)$ (from $\{4\}$ by Wait)\vspace{5pt}

6. $\ade x\ada y\bigl(q(x)\not\leftrightarrow p(y)\bigr)\mld \ade x\bigl(p(x)\sqc\gneg p(x)\bigr)\mld \bigl(\gneg q(z)\sqd q(z)\bigr)$ (from 5 by $\ade$-Choose)\vspace{5pt}

7. $\ade x\ada y\bigl(q(x)\not\leftrightarrow p(y)\bigr)\mld \ade x\bigl(p(x)\sqc\gneg p(x)\bigr)\mld \ada x\bigl(\gneg q(x)\sqd q(x)\bigr)$ (from $\{6\}$ by Wait)

\end{example}

\section{Appendix A: On abstract resource semantics} Abstract resource semantics, introduced in \cite{Cirq}, is a companion  of computability logic. One could probably characterize it as a ``lazy'' and naive form of the semantics of CL. Here we outline it in very informal terms.

All rules of our systems --- $\propseq$, $\propseq^{\circ}$, $\predseq$ --- pretty much look like (the effects of) moves in associated games. The only exception is Match, which has no direct counterpart in games. The central idea of abstract resource semantics is to make Match a legitimate move (by $\pp$) in its own right, called {\em allocation}. \cite{Cirq} provided plenty of intuitive explanations, showing how this sort of an approach yields a direct materialization of the resource intuitions traditionally (and somewhat wrongly) associated with linear logic, in the ``can you get both a candy and an ice cream for one dollar?'' style. 

The language that abstract resource semantics deals with is the same as the language of computability logic (possibly with just minor differences such as considering hyperformulas instead of formulas), with two sorts --- elementary and general --- of atoms. And, as in computability logic, formulas are understood as games. There are three main differences. 

One, rather minor, difference is that abstract resource semantics prefers to see each position not as a sequence of moves but rather the formula --- more precisely, the hyperformula --- representing the game to which such a sequence brings the original game down. This is an approach taken in all pre-CL papers by the present author, most notably in \cite{Jap02}. For example, the position $\seq{\pp 1.2}$ of game $\bigl(P\add(\underline{Q}\sqc R)\bigr)\mlc S$ will be simply seen as the hyperformula 
$(\underline{Q}\sqc R)\mlc S$, and the move $1.2$ by $\pp$ as the action of turning  $\bigl(P\add(\underline{Q}\sqc R)\bigr)\mlc S$ into $(\underline{Q}\sqc R)\mlc S$. The further move $1.\S$ by $\oo$ will be seen as turning $(\underline{Q}\sqc R)\mlc S$ into $(Q\sqc\underline{R})\mlc S$, the further move $1.\S$ by $\pp$ will be seen as turning 
$(Q\sqc\underline{R})\mlc S$ into $\underline{R}\mlc S$, etc.\footnote{Note the minor difference from how we treated catch-up switch moves in Section \ref{shf}: such moves had no effect there, and  $(Q\sqc\underline{R})\mlc S$ would remain $(Q\sqc\underline{R})\mlc S$ instead of turning into $\underline{R}\mlc S$.} 

The second difference is that, while computability logic treats formulas as schemata of games --- that is, syntactic expressions that become games only after an interpretation $^*$ is applied to them --- abstract resource semantics simply sees formulas as full-fledged games (``{\em abstract resources''}). It does not apply or appeal to interpretations, essentially meaning that it has a concept of validity  but no concept of truth.

The third, most important difference, as already noted, is that, along with all ``ordinary'' moves permitted in computability logic, abstract resource semantics allows an additional sort of a move called {\em (resource) allocation}. It consists of pairing one positive and one negative occurrence of a general atom $P$. The effect of such a move can be stipulated to be the result of replacing, in the original formula, the two occurrences of $P$ by the hybrid atom $P_q$ for some fresh elementary atom $q$.  That is, allocation is a direct counterpart of (or the same as) Match --- or, more precisely, Match$^\circ$. If no recurrence operators and no parallel and sequential quantifiers are present, every game will only last a finite number of steps, and will be considered won by $\pp$ iff the final formula/position is stable. It is almost immediately obvious (only for an expert, of course) that our systems $\propseq$ and $\predseq$ continue to be sound and complete with respect this semantics. In fact, proving such soundness and completeness would be by an order of magnitude easier than proving soundness and completeness with respect to the semantics of CL, because the relevant parts of abstract resource semantics are essentially directly ``read'' from the rules of those systems.  
  
The above approach easily extends to the fragment of the language of CL containing parallel and sequential recurrences, as well as all sorts of  quantifiers. The only difference will be that now the ``final'' (``limit'') formula, whose stability determines the outcome of the game, may be infinite, containing  infinite parallel conjunctions and/or disjunctions. Furthermore, a sequential subformula of the ``final'' formula may have no particular underlined component due to an infinite number of leading switches made in it. Such a subformula should be replaced by $\tlg$ (if it was a $\sqd$, $\sqe$ or $\scost$-subformula) or $\twg$ (if it was a $\sqc$, $\sqa$ or $\sst$-subformula). Then extending the concept of stability to such limit formulas presents no problem.   

A relatively nontrivial step is further extending the approach to branching recurrences as well. Those familiar with the relevant pieces of literature (such as Section 4.6 of \cite{Japfin}) would remember that the effect of making a ``replicative move'' in $\st E$ or $\cost E$ is turning it into $\st(E\circ E)$ or $\cost(E\circ E)$, respectively. Adding $\circ$ to the language that we consider is not necessary though, as $\st(A\circ B)$ is equivalent to $\st A\mlc\st B$ and 
 $\cost(A\circ B)$ is equivalent to $\cost A\mld\cost B$. So, we can stipulate that a replicative move turns $\st E$ into $\st E\mlc\st E$ and $\cost E$ into $\cost E\mld \cost E$. And the effect of a non-replicative move $\alpha$ within $\st E$ or $\cost E$ is simply replacing $E$ by the effect of $\alpha$ on $E$. In this respect, branching operators do not differ from any other operators. What makes the case of branching operators special is that general atoms that are in the scope of such operators will have to be reallocated over and over again (as the associated replicative moves create two copies of the argument of $\st,\cost$). Reallocations here should follow the same constraint as all allocations do --- specifically, (re)allocation can only take place between a positive and a negative occurrence of the {\em same} atom, whether such an atom is an original general atom $P$ or a previously already allocated atom which now looks like $P_{\vec{q}}$. When two occurrences of such a $P_{\vec{q}}$ are allocated to each other, they become $P_{\vec{q},r}$ for some fresh $r$. Thus, here we may get an infinite ``final'' formula not only because of infinitely many subformulas, but also because of hybrid atoms $P_{\vec{s}}$ with infinite subscripts $\vec{s}$. Two such atoms $P_{\vec{p}}$ and $Q_{\vec{q}}$ counting as the same iff $P=Q$ and $\vec{p}=\vec{q}$, the concept of stability then painlessly extends to the present case as well.

The approach called the {\em logic of tasks}, introduced and elaborated in \cite{Jap02}, was essentially nothing but the above-outlined abstract resource semantics limited to the language of (not yet officially born) CL without general atoms and without branching and sequential operators. In the absence of general atoms, the logic of tasks did not employ allocation. On the other hand, \cite{Cirq} dealt with a language with general atoms and used allocation as a basic semantical concept. But the logical vocabulary for which abstract resource semantics was fully defined (and, most importantly, well-motivated intuitively) was limited to $\{\gneg,\mlc,\mld\}$. The extension of abstract resource semantics to the full language of CL outlined in the present appendix is a mechanical combination of the approaches of \cite{Cirq} and \cite{Jap02}, 
extended further to the sequential and branching operators that were present in neither \cite{Cirq} nor \cite{Jap02}. 
 
Both \cite{Jap02} and \cite{Cirq} outlined potential applications of abstract resource semantics in resource-based planning systems. It has been also argued that planning systems based on such a semantics are immune to the notorious frame problem and the knowledge preconditions problem. That outline still needs a further materialization and elaboration though. Until that is done, abstract resource semantics (unlike the semantics of CL) should probably be treated as a technical tool rather than a semantics in its own right. As such, it may be very useful in proving various theorems about computability logic.

\section{Appendix B: Proof of Theorem \ref{static}}

\begin{lemma}\label{l14} \ 

1. Assume $A_0,\ldots,A_n$ are constant static games, $\Sigma$ is a $\xx$-delay of $\Gamma$, and $\Sigma$ is a $\xx$-illegal run of $A_0\sqc\ldots\sqc A_n$. Then $\Gamma$ is also a $\xx$-illegal run of $A_0\sqc\ldots\sqc A_n$.

2. Similarly for $A_0\sqc A_1\sqc A_2\sqc\ldots$.
\end{lemma}
\begin{proof} We will prove this lemma by induction on the length of the shortest illegal initial segment 
of $\Sigma$. 

{\em CLAUSE 1}. Assume the conditions of clause 1 of the lemma. We want to show that $\Gamma$ is a $\xx$-illegal run of $A_0\sqc\ldots\sqc A_n$. Let $\seq{\Psi,\xx\alpha}$ be the shortest ($\xx$-) illegal initial segment of $\Sigma$. Let $\seq{\Phi,\xx\alpha}$ be the shortest initial segment of $\Gamma$ containing all the $\xx$-labeled moves\footnote{In this context, different occurrences of the same labmove count as different labmoves. So, a more accurate phrasing would be ``as many $\xx$-labeled moves as..." instead ``all the $\xx$-labeled moves of ...".} of $\seq{\Psi,\xx\alpha}$. If $\Phi$ is a $\xx$-illegal position of $A_0\sqc\ldots\sqc A_n$, then so is $\Gamma$ and we are done. Therefore, for the rest of the proof, we assume that 
\begin{equation}\label{654}
\mbox{\em $\Phi$ is not a $\xx$-illegal position of $A_0\sqc\ldots\sqc A_n$.}
\end{equation}

Let $\Theta$ be the sequence of those $\pneg\xx$-labeled moves of $\Psi$ that are not in $\Phi$. Obviously
\begin{equation}\label{e141}
\mbox{\em  $\seq{\Psi,\xx\alpha}$ is a $\xx$-delay of $\seq{\Phi,\xx\alpha,\Theta}$.}
\end{equation}
We also claim that
\begin{equation}\label{e142}
\mbox{\em $\Phi$ is a legal position of $A_0\sqc\ldots\sqc A_n$.}
\end{equation}
Indeed, suppose this was not the case. Then, by (\ref{654}), $\Phi$ should be $\pneg\xx$-illegal. This would make $\Gamma$  a $\pneg\xx$-illegal run of $A_0\sqc\ldots\sqc A_n$ with $\Phi$ as an illegal initial segment which is shorter than $\seq{\Psi,\xx\alpha}$. Then, by the induction hypothesis, any run for which $\Gamma$ is a $\pneg\xx$-delay, would be $\pneg\xx$-illegal. But, as observed in Lemma 4.6 of \cite{Jap03}, the fact that $\Sigma$ is a $\xx$-delay of $\Gamma$ implies that $\Gamma$ is a $\pneg\xx$-delay of $\Sigma$. So, $\Sigma$ would be $\pneg\xx$-illegal, which is a contradiction because, according to our assumption, $\Sigma$ is $\xx$-illegal.

We are continuing our proof. There are three possible reasons to why $\seq{\Psi,\xx\alpha}$ is an illegal (while 
$\Psi$ being legal)  position of $A_0\sqc\ldots\sqc A_n$:

{\em Reason 1}: $\alpha$ does not have the form $\S$ or $.\beta$. Then, in view of (\ref{e142}), 
$\seq{\Phi,\xx\alpha}$ is a $\xx$-illegal position of $A_0\sqc \ldots\sqc A_n$. As $\seq{\Phi,\xx\alpha}$ happens to be an initial segment of $\Gamma$, the latter then is a $\xx$-illegal run of $A_0\sqc\ldots\sqc A_n$.

{\em Reason 2}: $\alpha=\S$, and either $\xx=\oo$ and the $\oo$-degree of $\Psi$ is $n$, or $\xx=\pp$ and the $\pp$-degree of $\Psi$ equals the $\oo$-degree of $\Psi$. In either case, with (\ref{e142}) in mind, $\seq{\Phi,\xx\alpha}$ can be seen to be a $\xx$-illegal position of $A_0\sqc \ldots\sqc A_n$.  Hence,    as $\seq{\Phi,\xx\alpha}$ is an initial segment of $\Gamma$, the latter is a $\xx$-illegal run of $A_0\sqc\ldots\sqc A_n$. 

{\em Reason 3}: $\alpha=.\beta$, the $\xx$-degree of $\xx\alpha$ is $i\in\{0,\ldots,n\}$, and $\seq{\Psi^{\#i},\xx\beta}\not\in \legal{A_i}{}$. That is,
$\seq{\Psi,\xx\alpha}^{\#i}$ is a $\xx$-illegal position of $A_i$. (\ref{e141}) obviously implies that 
 $\seq{\Psi,\xx\alpha}^{\#i}$ is a $\xx$-delay of $\seq{\Phi,\xx\alpha,\Theta}^{\#i}$. Therefore, since 
$A_i$ is static, clause 1 of Lemma \ref{may19} yields that $\seq{\Phi,\xx\alpha,\Theta}^{\#i}$ is a $\xx$-illegal position of $A_i$. Notice that $\seq{\Phi,\xx\alpha,\Theta}^{\#i}=\seq{\Phi^{\#i},\xx\beta,\Theta^{\#i}}$.
A $\xx$-illegal position will remain $\xx$-illegal after removing a block of $\pneg\xx$-labeled moves (in particular, $\Theta^{\#i}$) at the end of it. Hence,  $\seq{\Phi^{\#i},\xx\beta}$ is a $\xx$-illegal position of $A_i$. In view of (\ref{e142}), this implies that  $\seq{\Phi, \xx \alpha=\xx .\beta}\not\in\legal{A_0\sqci \ldots\sqci A_n}{}$, so that  
$\seq{\Phi,\xx\alpha}$ is a $\xx$-illegal position of $A_0\sqc\ldots\sqc A_n$, and then so is $\Gamma$ because $\seq{\Phi,\xx\alpha}$  is an initial segment of it.

{\em CLAUSE 2}. The reasoning here is virtually the same as in the proof of clause 1. The only difference (that makes the present case simpler) is that, in ``{\em Reason 2}'', the condition ``$\xx=\oo$ and the $\oo$-degree of $\Psi$ is $n$'' does not need to be considered. 
\end{proof}

Since $\sqa x,\sqe x,\sst,\scost$  are nothing but sequential conjunctions/disjunctions, there is no need to separately consider them when proving Theorem \ref{static}. Furthermore, since $\sqd$ is expressible in terms of $\sqc$ and $\gneg$ (with $\gneg$ already known to preserve the static property), considering only $\sqc$ would be sufficient. For simplicity, here we restrict ourselves to the $n$-ary case of $\sqc$. Adapting our argument to the infinite case of $\sqc$ does not present a problem. 

Assume that $A_0,\ldots,A_n$ are static constant games, $\win{A_0\sqci\ldots\sqci A_n}{}\seq{\Gamma}=\xx$ and $\Sigma$ is a $\xx$-delay of $\Gamma$. Our goal is to show that $\win{A_0\sqci\ldots\sqci A_n}{}\seq{\Sigma}=\xx$. 

If $\Sigma$ is a $\pneg\xx$-illegal run of $A_0\sqc\ldots\sqc A_n$, then it is won by $\xx$ and we are done. So, assume that 
$\Sigma$ is not $\pneg\xx$-illegal. Lemma 4.6 of \cite{Jap03} asserts that, if $\Sigma$ is a $\xx$-delay of $\Gamma$, then $\Gamma$ is a $\gneg\xx$-delay of $\Sigma$. So, by Lemma \ref{l14}, our $\Gamma$ cannot be $\pneg\xx$-illegal, for otherwise so would be $\Sigma$. 
$\Gamma$ also cannot be $\xx$-illegal, because otherwise it would not be won by $\xx$. Consequently, $\Sigma$ cannot be $\xx$-illegal either, for otherwise, by Lemma \ref{l14}, $\Gamma$ would be $\xx$-illegal. Thus, we have narrowed down our considerations to the case when both $\Gamma$ and $\Sigma$ are legal runs of $A_0\sqc\ldots\sqc A_n$.

$\win{A_0\sqci\ldots\sqci A_n}{}\seq{\Gamma}=\xx$, together with $\Gamma\in\legal{A_0\sqci\ldots\sqci A_n}{}$, implies that, where $i$ ($0\leq i\leq n$) is the $\oo$-degree of $\Gamma$,  
$\Gamma^{\#i}$ is a $\xx$-won run of $A_i$. Taking into account that $\Sigma^{\#i}$ is obviously a $\xx$-delay of $\Gamma^{\#i}$ and that $A_i$ is 
 static, the above, in turn, implies that $\Sigma^{\#i}$ is a $\xx$-won run of $A_i$, which, taking into account that 
$\Sigma\in\legal{A_0\sqci \ldots\sqci A_n}{}$,  means nothing but that $\Sigma$ is a $\xx$-won run of $A_0\sqc\ldots\sqc A_n$.

\end{document}